\documentclass[%
superscriptaddress,
preprint,
 amsmath,amssymb,
 aps,
prc,
]{revtex4-1}

\usepackage{graphicx}
\usepackage{dcolumn}
\usepackage{bm}
\usepackage{enumerate}
\usepackage{adjustbox}
\usepackage{float}

\newcommand{\sign}{\text{sgn}}
\newcommand{\MeV}{\text{MeV}}
\newcommand{\htil}{\tilde{h}(\theta)}
\newcommand{\htilprime}{\tilde{h}^\prime(\theta)}
\newcommand{\htilsec}{\tilde{h}^{\prime\prime}(\theta)}

\newcommand{\g}{g(\theta)}
\newcommand{\gprime}{g^\prime(\theta)}
\newcommand{\gsec}{g^{\prime \prime}(\theta)}

\begin{document}

\preprint{MIT-CTP-5015}

\title{A QCD equation of state matched to lattice data and exhibiting a critical point singularity}

\author{Paolo Parotto}
\email[Corresponding author: ]{\url{parotto@uni-wuppertal.de}}
\affiliation{Department  of  Physics,  University  of  Houston, Houston,  TX  77204,  USA}
\affiliation{University of Wuppertal, Department of Physics, Wuppertal D-42219, Germany}

\author{Marcus Bluhm}
\affiliation{Institute of Theoretical Physics, University of Wroclaw, 50204 Wroclaw, PL}
\affiliation{SUBATECH  UMR  6457  (IMT  Atlantique,  Universit\'e  de  Nantes, IN2P3/CNRS),  4  rue  Alfred  Kastler,  44307  Nantes,  France}

\author{Debora Mroczek}
\affiliation{Department  of  Physics,  University  of  Houston, Houston,  TX  77204,  USA}

\author{Marlene Nahrgang}
\affiliation{SUBATECH  UMR  6457  (IMT  Atlantique,  Universit\'e  de  Nantes, IN2P3/CNRS),  4  rue  Alfred  Kastler,  44307  Nantes,  France}

\author{J. Noronha-Hostler}
\affiliation{Department of Physics, 
University of Illinois at Urbana-Champaign, Urbana, IL 61801, USA}

\author{Krishna Rajagopal}
\affiliation{Center for Theoretical Physics, Massachusetts Institute of Technology, Cambridge, MA 02139, USA}

\author{Claudia Ratti}
\affiliation{Department  of  Physics,  University  of  Houston, Houston,  TX  77204,  USA}

\author{Thomas Sch\"afer}
\affiliation{Department of Physics, North Carolina State University, Raleigh, NC 27695, USA}

\author{Mikhail Stephanov}
\affiliation{Physics Department, University of Illinois at Chicago, Chicago, IL 60607, USA}

\date{\today}

\begin{abstract}
We construct a family of equations of state for QCD in the temperature range 30 MeV $\leq T\leq$ 800 MeV and in the chemical potential range $0\leq \mu_B \leq$ 450 MeV. These equations of state match available lattice QCD results up to $\mathcal{O}(\mu_B^4)$ and in each of them we place a critical point in the 3D Ising model universality class. The position of this critical point can be chosen in the range of chemical potentials covered by the second Beam Energy Scan at RHIC. We discuss possible choices for the free parameters, which arise from mapping the Ising model onto QCD. Our results for the pressure, entropy density, baryon density, energy density and speed of sound can be used as inputs in the hydrodynamical simulations of the fireball created in heavy ion collisions. We also show our result for the second cumulant of the baryon number in thermal equilibrium, displaying its divergence at the critical point. In the future, comparisons between RHIC data and the output of the hydrodynamic simulations, including calculations of fluctuation observables, built upon the model equations of state that we have constructed may be used to locate the critical point in the QCD phase diagram, if there is one to be found.
\end{abstract}

\maketitle


\section{Introduction} \label{sec:Intro}

The search for a possible QCD critical point is receiving increasing attention, which will culminate in the second Beam Energy Scan (BES-II) at the Relativistic Heavy Ion Collider (RHIC) at Brookhaven National Laboratory. The main goal of the BES-II program is to discover a critical point, or constrain its location, on the phase diagram of strongly interacting matter. One of the central questions that these experiments aim to answer is whether the continuous crossover \cite{Aoki:2006we} between quark-gluon plasma and hadronic matter that occurs as a function of decreasing $T$ at $\mu_B=0$ turns into a first order phase transition above some critical point at a nonzero $\mu_B$, corresponding to heavy ion collisions below some collision energy \cite{Luo:2017faz,Busza:2018rrf}.

Lattice QCD simulations cannot currently be performed at finite density. For this reason, a first principle prediction of the existence and position of the critical point is still missing. Several QCD-based models predict its location on the phase diagram, which naturally depends on the model parameters and approximations (for a review see e.g. \cite{Stephanov:2007fk}). This aspect makes the critical point search challenging, and is at the basis of the systematic scan in collision energies performed at RHIC. We anticipate that non-monotonous dependence of specific observables on collision energy will indicate the presence of the critical point as the freezeout point traverses the critical region \cite{Stephanov:1998dy,Stephanov:1999zu}. As the BES-II approaches, it is therefore important to predict the effects of the critical point on several observables.

One of the main theoretical approaches to pursue this goal is represented by hydrodynamical simulations of the evolution of the fireball produced in heavy ion collisions (see e.g. \cite{Jeon:2015dfa} and references therein). While modifications of the hydrodynamical approach itself are needed in the vicinity of the critical point \cite{Stephanov:2009ra,Nahrgang:2011mg,Stephanov:2017ghc,Nahrgang:2018afz}, the Equation of State (EoS) used as an input in these simulations needs to reflect all theoretical knowledge currently available as well as contain the singularity associated with the QCD critical point at a parametrically controllable location. Thus, the purpose of this manuscript is to produce a family of model equations of state for QCD, each of which contains a critical point somewhere in the region of the phase diagram covered by the BES-II at RHIC, and all of which respect what we know from lattice QCD calculations  up to $\mathcal{O}(\mu_B^4)$. Previous hydrodynamical calculations at finite $\mu_B$ have either incorporated models for the contribution from a critical point (for example, see Refs. \cite{Hama:2005dz,Steinheimer:2010ib,Auvinen:2017fjw,Gardim:2017ruc}) or have used the Taylor-expanded equation of state from lattice QCD calculations to describe a crossover \cite{Bellwied:2015rza,Monnai:2016kud,Denicol:2018wdp}. Our study is the first to incorporate both, and in addition correctly captures the universal physics near the hypothesized critical point. 

At chemical potential $\mu_B=0$, the EoS of QCD is known with high precision, in the case of 2+1 \cite{Borsanyi:2010cj,Borsanyi:2013bia,Bazavov:2014pvz} and 2+1+1 \cite{Borsanyi:2016ksw} quark flavors. Extensions to finite chemical potential are usually performed through a Taylor series in powers of $\mu_B/T$ \cite{Allton:2002zi,Allton:2005gk,Gavai:2008zr,Basak:2009uv,Kaczmarek:2011zz} or an analytic continuation from imaginary $\mu_B$ \cite{deForcrand:2002hgr,DElia:2002tig,Wu:2006su,DElia:2007bkz,Conradi:2007be,deForcrand:2008vr,DElia:2009pdy,Moscicki:2009id}.
The Taylor expansion of the pressure in $\mu_B/T$ around $\mu_B=0$ can be written as:
\begin{equation}
P (T,\mu_B) = T^4 \sum_n c_{2n} (T) \left( \frac{\mu_B}{T} \right)^{2n} \, \, ,
\end{equation} \label{eq:PTayl}
where the coefficients of the expansion are the susceptibilities of the baryon number:
\begin{equation}
c_n (T) = \frac{1}{n!} \left. \frac{\partial^n P/T^4}{\partial (\mu_B/T)^n} \right|_{\mu_B=0} = \frac{1}{n!} \chi_n (T) \, \, .
\end{equation}

After the early results for $c_2,~c_4$ and $c_6$ \cite{Allton:2005gk}, the first continuum extrapolated results for $c_2$ were published in Ref. \cite{Borsanyi:2012cr}; in Ref. \cite{Hegde:2014sta} $c_4$ was shown, but only at finite lattice spacing. The continuum limit for $c_6$ was published for the first time in \cite{Gunther:2016vcp}, and later in \cite{Bazavov:2017dus}. In \cite{DElia:2016jqh}, a first determination of $c_8$, at two values of the temperature and $N_t=8$ was presented. More recently, an estimate for $c_8$ as a function of the temperature for $N_t=12$ was presented in \cite{Borsanyi:2018grb}. The advantage of the Taylor expansion method is that all the quantities are calculated at vanishing baryon chemical potential, where lattice QCD simulations do not suffer from the fermion sign problem. Moreover, the knowledge of the expansion coefficients can in principle provide information on the location of the critical point, under the assumption that such point is the closest singularity to $\mu_B=0$ in the complex-$\mu_B$ plane. Unfortunately, the fact that only a few coefficients are known makes this task extremely hard, leading to just an \textit{indication} that the region  corresponding to $\mu_B \lesssim 2 \, T$ in the phase diagram cannot contain the critical point \cite{Bazavov:2017dus}. In addition to this, the knowledge of the equation of state of QCD \textit{beyond} the critical point (i.e. for $\mu_B > \mu_{BC}$) could not come from a Taylor expansion, as a singularity cannot be reproduced in this method. 
 
In this manuscript, we produce an equation of state which matches everything we know from lattice QCD simulations up to $\mathcal{O}(\mu_B^4)$ (for a recent review, see e.g. \cite{Ratti:2018ksb}) in the region where they are applicable, and which shows the correct singular behavior at and around the hypothesized critical point. The latter can be inferred from the fact that the critical point of QCD is expected to be in the same universality class as the one of the 3D Ising model \cite{Rajagopal:1992qz,Berges:1998rc,Halasz:1998qr,Karsch:2001nf,deForcrand:2003vyj}. Our approach is similar to the one presented in Refs. \cite{Nonaka:2004pg,Bluhm:2006av}, but in our case the critical contribution is built on top of a first-principle result for the EoS, instead of relying on models such as the MIT bag or the quasiparticle model.

We will adopt the following strategy:
\begin{enumerate}[i)]
\item Choose a location in the $(\mu_B,T)$ plane at which to put a critical point;
\item Make use of a suitable parametrization to describe the universal scaling behavior of the EoS in the 3D Ising model near the critical point;
\item Map the 3D Ising model phase diagram onto the one of QCD via a parametric, non-universal change of variables;
\item Use the thermodynamics of the Ising model EoS to estimate the critical contribution to the expansion coefficients up to $\mathcal{O}(\mu_B^4)$ from lattice QCD; 
\item Reconstruct the full pressure, matching lattice QCD up to $\mathcal{O}(\mu_B^4)$ at $\mu_B=0$ and including the correct critical behavior.
\end{enumerate} 

It is important to notice that our approach is based on the assumption that the critical point of QCD is the closest singularity to $\mu_B=0$ on the real $\mu_B$ axis. Only in this case we are allowed to merge the contributions of the lattice and 3D Ising approaches in the way detailed below. The result of this procedure will be an equation of state that meets our requirements and depends on the parameters of the non-universal map between Ising variables and QCD coordinates \cite{code:2018}. These parameters include the coordinates of the critical point. The ultimate goal of this project is to provide a family of model equations of state that can be used as inputs to future hydrodynamic calculations and calculations of fluctuation observables that can then be compared to experimental data from the BES-II program, resulting in constraints on the parameters in the map that we have constructed, in particular the parameters representing the location of the critical point. We will follow up on this discussion in the following sections. Note that we place a critical point in each of our equations of state entirely by construction; our calculation {\it in isolation} is therefore not a path toward determining whether the phase diagram of QCD features a critical point and where it lies.  But, by comparing experimental data to predictions obtained by using our family of equations of state within future calculations of hydrodynamics and fluctuations progress toward this goal can be realized.

The family of model equations of state that we construct can also be used directly to illustrate the divergence of quantities in the vicinity of the critical point, in particular the cumulants of the baryon number. We show in Section \ref{sec:thermfull} our result for the second cumulant, which can be related to the variance of the net-proton distribution in heavy-ion collision experiments assuming thermal equilibrium \cite{Hatta:2003wn}. In order to make contact with experiment, investigating out-of-equilibrium physics is important because of critical slowing down in the dynamics near a critical point and because the matter produced in a heavy-ion collision does not spend a long time near the critical point \cite{Berdnikov:1999ph,Mukherjee:2015swa,Mukherjee:2016kyu}. Simulations of hydrodynamics and fluctuations that are built upon the family of equations of state that we have constructed would be a good next step in this direction.

\section{Scaling EoS in the 3D Ising model}
\begin{figure} 
\center
\includegraphics[width=0.48\textwidth]{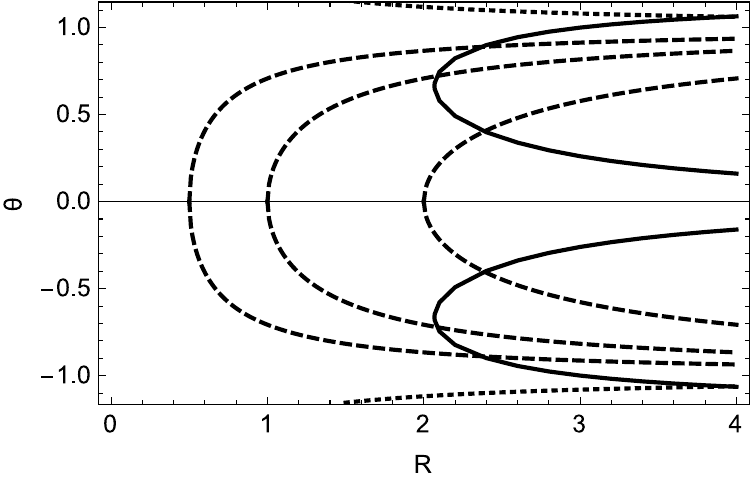}
\includegraphics[width=0.48\textwidth]{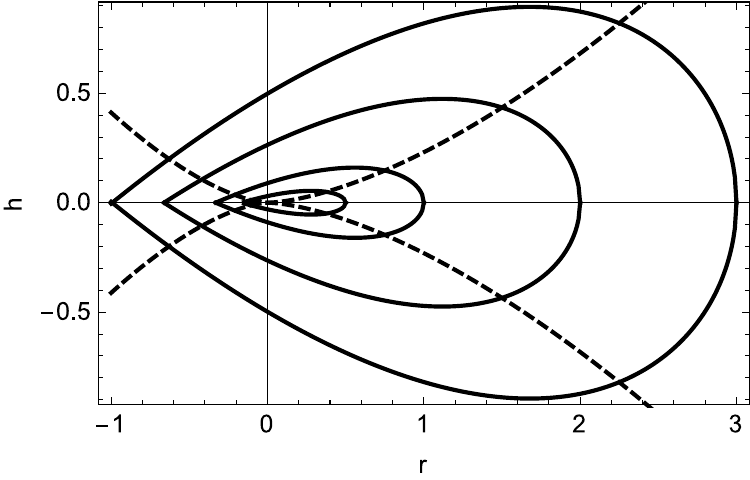}
\caption{Map of the parametrization in Eqs.(\ref{eq:param1})-(\ref{eq:param3}). (Left) Lines of constant $h$ and $r$ are shown in the $\theta-R$ plane with solid and dashed lines respectively. (Right) Lines of constant $\theta$ and $R$ are shown in the $h-r$ plane with dashed and solid lines respectively.}
\label{fig:paramMap}
\end{figure}

The first ingredient for this work is a parametrization of the Ising model equation of state in the vicinity of the critical point, which corresponds to a map between two variables $(R,\theta)$ to Ising variables $(r,h)$, where $r$ is the reduced temperature $r = (T - T_c)/T_c$ and $h$ is the magnetic field. The map needs to accommodate the correct behavior of the order parameter $M$ (the magnetization) as a function of $r$ and $h$ themselves. The following form for the parametrization meets the requirements \cite{Nonaka:2004pg,Guida:1996ep,Schofield:1969zz,Bluhm:2006av}:
\begin{align}
M &= M_0 R^\beta \theta \, \, , \label{eq:param1} \\
h &= h_0 R^{\beta \delta} \tilde{h}(\theta) \, \, , \label{eq:param2} \\
r &= R (1- \theta^2) \, \, . \label{eq:param3}
\end{align}
where $M_0$, $h_0$ are normalization constants, $\tilde{h}(\theta) = \theta (1 + a \theta^2 + b \theta^4)$  with $a=-0.76201$, $b=0.00804$. $\beta \simeq 0.326$ and $\delta \simeq 4.80$ are 3D Ising critical exponents, and the parameters take on the values $R \geq 0$, $\left| \theta \right| \leq \theta_0 \simeq 1.154$, $\theta_0$ being the first non-trivial zero of $\tilde{h}(\theta)$. The values of the normalization constants are such that $M(r=-1,h=0^+) = 1$ and $M(r=0,h) \propto \sign(h) \left| h \right|^{1/\delta}$: this yields $M_0 \simeq 0.605$, $h_0 \simeq 0.394$. Fig.\ref{fig:paramMap} shows a pictorial representation of the parametrization: the lines of constant $h$ and $r$ in the $\theta-R$ plane (left panel) and the lines of constant $\theta$ and $R$ in the $h-r$ plane (right panel).

Starting from this parametrization, it is possible to define the Gibbs free energy density:
\begin{equation}
G(h,r) = F(M,r) - M h \, \, ,
\end{equation}
where $F(M,r)$ is the free energy density, defined as:
\begin{equation}
F(M,h) = h_0 M_0 R^{2-\alpha}g(\theta) \, \, ,
\end{equation}
where $\alpha \simeq 0.11$ is another critical exponent of the 3D Ising model (also, the relation $2 - \alpha = \beta (\delta + 1)$ holds). The function $g(\theta)$ is fixed by noticing that $h = (\partial F/ \partial M)_h$ and solving the following differential equation:
\begin{equation}
\tilde{h}(\theta) (1 - \theta^2 + 2 \beta \theta^2) = 2 (2 - \alpha) \theta g(\theta) + (1 - \theta^2) g^\prime (\theta)
\end{equation}
which results in:
\begin{equation}
g(\theta) = c_0 + c_1 (1 - \theta^2) + c_2 (1 - \theta^2)^2 + c_3 (1 - \theta^2)^3 \, \, ,
\end{equation}
with:
\begin{align*}
c_0 &= \frac{\beta}{2 - \alpha} (1 + a + b) \, \, , \\
c_1 &= - \frac{1}{2} \frac{1}{\alpha - 1} \left\lbrace (1 - 2 \beta) (1 + a + b) - 2 \beta (a + 2 b) \right\rbrace \, \, , \\
c_2 &= - \frac{1}{2 \alpha} \left\lbrace 2 \beta b - (1 - 2 \beta) (a + 2 b) \right\rbrace \, \, , \\
c_3 &= - \frac{1}{2 (\alpha + 1)} b (1 - 2 \beta) \, \, .
\end{align*}

Now everything is determined, and one can build an expression for the pressure in the 3D Ising model in the scaling regime, noticing that the Gibbs free energy density equals the pressure up to a minus sign: $G = - P$, hence:
\begin{equation} \label{eq:PIsing}
P_{\text{Ising}}(R,\theta) = h_0 M_0 R^{2 - \alpha} \left[ \theta \tilde{h}(\theta) - g(\theta) \right] \, \, .
\end{equation}
Notice that this pressure is dimensionless. Besides, this expression is completely analytic in $(R,\theta)$ in the whole range of parameter values. However, the map $(R,\theta) \longmapsto (r,h)$ is not globally invertible. 

\section{Non-universal map from Ising to QCD}
The next step is to build a map from Ising variables to QCD coordinates, so that Eq. (\ref{eq:PIsing}) that we derived for the pressure becomes useful for our purpose. We want to map the phase diagram of the 3D Ising model onto the one of QCD, so that the critical point of the Ising model $r=h=0$ corresponds to the one of QCD, and that the lines of first order phase transition and crossover in the Ising  model are mapped onto those of QCD. \\

\begin{widetext}
\begin{figure}
\center
\includegraphics[width=\textwidth]{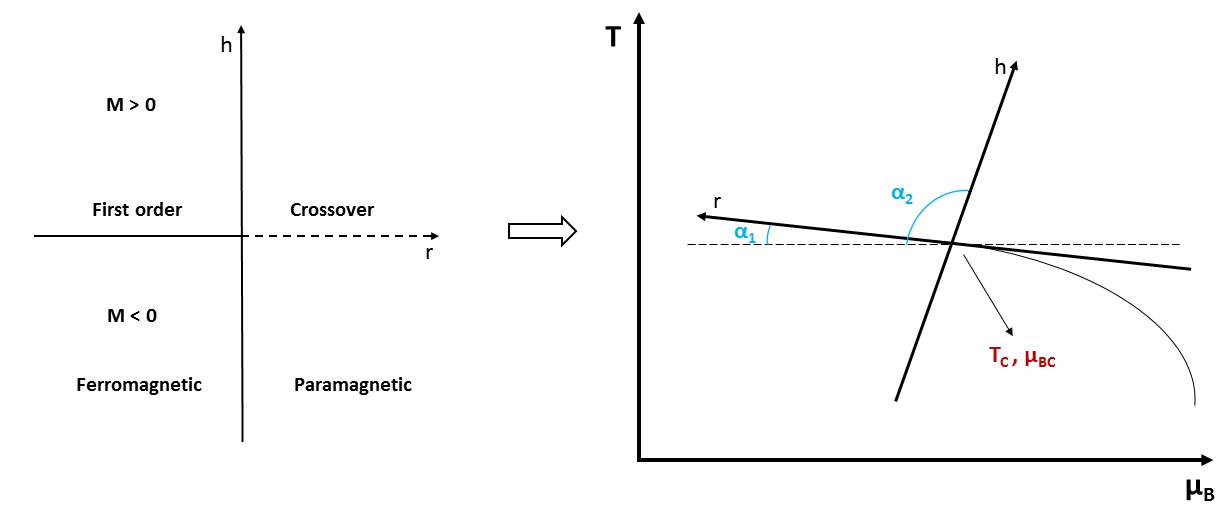}
\caption{Non-universal map from Ising variables $(r,h)$ to QCD coordinates $(T,\mu_B)$.}
\label{fig:IsingQCD}
\end{figure}
\end{widetext}

The simplest way to do so is through a linear map as follows \cite{Rehr:1973zz}:
\begin{align}
\frac{T - T_C}{T_C} &=  w \left( r \rho \,  \sin \alpha_1  + h \, \sin \alpha_2 \right) \, \, , \label{eq:IsQCDmap1} \\ 
\frac{\mu_B - \mu_{BC}}{T_C} &=  w \left( - r \rho \, \cos \alpha_1 - h \, \cos \alpha_2 \right) \, \, , \label{eq:IsQCDmap2}
\end{align} 
which can be visualized in Fig. \ref{fig:IsingQCD}. This map makes use of six parameters, two of which correspond to the location of the critical point on the QCD phase diagram, two are the angles that the $r$ and $h$ axes form with the $T = \text{const.}$ lines, and $(w,\rho)$ are scale factors for the variables $r$ and $h$. While $w$ represents a \textit{global} scaling for the Ising variables, namely determining the size of the critical region, $\rho$ represents a \textit{relative} scaling of $r$ and $h$, thus roughly determining the shape of it. \\

At this point, we have a double map between coordinates:
\begin{equation} \label{eq:map}
(R,\theta) \longmapsto (r,h) \longleftrightarrow (T,\mu_B) \, \, ,
\end{equation}
where the second step is globally invertible. We will now apply the thermodynamics we developed in the previous section for the Ising model, making use of the additional variables $(R,\theta)$, to the QCD phase diagram, in a parametrized form given by Eqs. (\ref{eq:IsQCDmap1}), (\ref{eq:IsQCDmap2}). \\

In order to do this analytically, we would need the map $(R,\theta) \longmapsto (T,\mu_B)$, which unfortunately cannot be globally inverted. Therefore it is necessary to solve the following relations numerically:
\begin{eqnarray}
T (R,\theta) - T_i &= &0 \, \, ,  \label{eq:numMap1} \\
\mu_B (R,\theta) - {\mu_B}_i &=& 0 \, \, , \label{eq:numMap2}
\end{eqnarray}
for each value of $(T,\mu_B)$ needed in the QCD phase diagram. We proceed in the following way: we choose a range of interest for $T$ and $\mu_B$, and given a choice of the parameters in the Ising-QCD map, we solve Eqs. (\ref{eq:numMap1}) and (\ref{eq:numMap2}) numerically for a two-dimensional grid in $T$ and $\mu_B$ in the desired range, thus providing a discrete inverse map $(T,\mu_B) \longmapsto (R,\theta)$. 

With this solution, although not analytic, it is possible to transport the thermodynamics of the Ising model (written in terms of $(R,\theta)$), into the QCD phase diagram, given a choice of parameters for the map.

\section{Thermodynamics}

\subsection{Strategy}

The strategy we wish to pursue in order to produce an equation of state for QCD which meets the requirements stated in Section \ref{sec:Intro} is the following. Starting from the Taylor expansion coefficients up to $\mathcal{O}(\mu_B^4)$ in Eq. (\ref{eq:PTayl}), available from lattice QCD simulations, we re-write them as a sum of an ``Ising'' contribution coming from the critical point of QCD, and a ``Non-Ising'' contribution, which would contain the regular part as well as any other possible criticality present in the region of interest:
\begin{equation} \label{eq:coeffs}
T^4 c_n^{\text{LAT}} (T) = T^4 c_n^{\text{Non-Ising}} (T) + f(T,\mu_B=0) c_n^{\text{Ising}} (T) \, \, ,
\end{equation}
where $f(T,\mu_B)$ is a regular function of the temperature and chemical potential, with dimension of energy to the fourth power. Away from the critical regime, $f$ just reshuffles the regular terms and can be chosen arbitrarily. Near the critical point, the choice for $f$ is almost arbitrary, with the only requirement being that it must not add any leading singular behavior. In general, though, any term in $f$ beyond a constant introduces sub-leading behavior in the vicinity of the critical point. For this reason, the simplest choice is to take $f$ to be a constant, with the appropriate dimension. This also ensures that no sub-leading behavior is introduced near the critical point, which cannot be predicted through universality.
Note that Eq. (\ref{eq:coeffs}) is to be understood as a definition for the $c_n^{\text{Non-Ising}}$ coefficients. 

Once these coefficients are obtained, we will build a Taylor expansion in $\mu_B$ analogous to the lattice one, using the ``Non-Ising'' coefficients. The latter have the advantage that the critical behavior coming from the critical point has been removed, so that the expansion can be pushed to larger values of $\mu_B$. This provides an expression for the ``Non-Ising'' pressure over a broad region of the QCD phase diagram. The assumption here is that the Ising critical point contribution to the Taylor coefficients from lattice QCD can be reproduced upon imposing the correct scaling behavior in the vicinity of the critical point.  

Once this expansion is carried out, the full pressure is then reconstructed simply by adding the critical contribution at any $(T,\mu_B)$ to the Taylor expanded ``Non-Ising'' one:
\begin{equation} \label{eq:Pfull}
P (T, \mu_B) = T^4 \sum_n c_{2n}^{\text{Non-Ising}} (T) \left( \frac{\mu_B}{T} \right)^{2n} +P^{\text{QCD}}_{\text{crit}}(T, \mu_B) \, \, . 
\end{equation}
Note that in Eq. (\ref{eq:Pfull}), the critical pressure is obtained from Eq. (\ref{eq:PIsing}) with the use of the relation in Eq. (\ref{eq:map}) and the multiplication by the regular function $f(T,\mu_B)$ in Eq. (\ref{eq:coeffs}):
\begin{equation} \label{eq:PQCD} 
P_{\text{QCD}}^{\, \text{crit}}(T, \mu_B) = f(T, \mu_B) P^{\text{Ising}} (R(T, \mu_B) ,\theta (T, \mu_B)) \, \, ,
\end{equation} 
which is extremely easy to calculate using the above relations. We will hereafter consider the following choice for the function  $f(T,\mu_B)$:
$$f(T,\mu_B) = T_C^4 \, \, .$$ 

The prescription we follow in order to enforce the matching of our EoS to lattice QCD at vanishing chemical potential should not be understood as a way to force a critical point in the QCD phase diagram without direct relation to the thermodynamics. Although we can, for any choice of the parameters, ensure that the coefficients at $\mu_B=0$ coincide with lattice, this does not guarantee that the resulting EoS is thermodynamically stable. An essential step of our procedure is to check that thermodynamic inequalities are satisfied by the EoS we generate. Thanks to this, given a choice of parameters in the Ising-to-QCD map, we can ensure that the resulting EoS realizes a merging with lattice QCD without leading to pathological behavior of the thermodynamic quantities. In Section \ref{sec:parameters} (see Fig. \ref{fig:paramtable}) we show an example of such an analysis using our EoS when varying two of the parameters in the Ising-to-QCD map, holding the others fixed.

\subsection{Taylor coefficients in the Ising model}

The other quantities we need to calculate from the parametrization of the Ising model thermodynamics are the contributions to the expansion coefficients of the pressure, which are simply the derivatives of the latter with respect to the baryonic chemical potential at fixed temperature:
\begin{equation}
c_n^{\text{Ising}} (T) = \frac{1}{n!} T^n \left. \frac{\partial^n P^{\text{Ising}}}{\partial \mu_B^n} \right|_{\mu_B=0} = \frac{1}{n!} \chi^{\text{Ising}}_n (T) \, \, .
\end{equation} 

Unfortunately, the expression for the critical pressure is given in terms of the additional variables $(R,\theta)$ and not as a function of $(r,h)$ or $(T,\mu_B)$. In order to obtain the derivatives we need, we will have to use the rules for the derivative of the inverse and for the multivariate chain rule in order to be able to express everything analytically as a function of $(R,\theta)$, and convert to QCD coordinates only at the end.  

We have to calculate expressions such as:
\begin{equation} \label{eq:Chis}
\chi_n (T, \mu_B=0) = - T^n \left( \frac{\partial^n G}{\partial \mu_B^n} \right)_T \, \, ,
\end{equation}
which we will have to re-write as ($n=1$ as an example):
\begin{equation}
\frac{\chi_1 (T)}{T} = - \left( \frac{\partial G}{\partial \mu_B} \right)_T = - \left( \frac{\partial G}{\partial r} \right)_h \frac{\partial r}{\partial \mu_B} - \left( \frac{\partial G}{\partial h} \right)_r \frac{\partial h}{\partial \mu_B} \, \, ,
\end{equation}
where:
\begin{align*}
\left( \frac{\partial G}{\partial r} \right)_h = \frac{\partial G}{\partial R} \left( \frac{\partial R}{\partial r} \right)_h + \frac{\partial G}{\partial \theta} \left( \frac{\partial \theta}{\partial r} \right)_h \, \, , \\
\left( \frac{\partial G}{\partial h} \right)_r = \frac{\partial G}{\partial R} \left( \frac{\partial R}{\partial h} \right)_r + \frac{\partial G}{\partial \theta} \left( \frac{\partial \theta}{\partial h} \right)_r \, \, . 
\end{align*}

Since we do not have explicit expressions for the dependence of $(R,\theta)$ on $(r,h)$, we need to proceed in the following way:
\begin{enumerate}[i.]
\item Use the rule for the derivative of the inverse, so that we can express derivatives of $(R,\theta)$ wrt $(r,h)$ as combinations of derivatives of $(r,h)$ wrt $(R,\theta)$;
\item Use the rule for derivatives of a function holding another function constant.
\end{enumerate}

\subsubsection{Derivatives of the inverse}

Naming $Q_n$ the $n^{\text{th}}$ derivative of an invertible function $y = y(x)$:
$$Q_n = \frac{d^n y}{d x^n} \, \, ,$$
we can exploit the recursive relationship:
\begin{equation} \label{eq:derinv}
Q_1^{2n - 1} \frac{d^n x}{d y^n} = P_n  \, \, ,   \qquad \text{with} \qquad P_{n+1} =Q_1 P^\prime_n - (2n - 1) Q_2 P_n \, \, , 
\end{equation}
where the $P_n$ are polynomials in $\left\lbrace Q_k \right\rbrace$ and $P_1 = 1$. $P^\prime$ indicates derivation with respect to $x$. 

For example, one can find:
$$\left(\frac{\partial^2 R}{\partial r^2}\right)_h = - \left(\frac{\partial^2 r}{\partial R^2}\right)_h \left(\frac{\partial r}{\partial R}\right)^{-3}_h \, \, .$$

\subsubsection{Derivatives with functions held constant}

We will have to use the following:
\begin{equation} \label{eq:derconst}
\left( \frac{\partial }{\partial x_1} \right)_{y_1} y_2 = \left( \frac{\partial}{\partial x_1} + \left( \frac{d x_2}{d x_1} \right)_{y_1} \frac{\partial}{\partial x_2} \right) y_2 \, \, ,
\end{equation}
where, in our case $(x_1,x_2) = (R,\theta)$ and $(y_1,y_2) = (r,h)$. 

For example, one has:
$$\left( \frac{\partial h}{\partial R} \right)_r = \left( \frac{\partial }{\partial R} + \left( \frac{d R }{d \theta} \right)_r \frac{\partial }{\partial \theta} \right) h = \frac{h_0 R^{-1 + \beta \delta}}{2\theta} \frac{1 - \theta^2}{2 \beta \delta \theta \tilde{h}(\theta) + (1-\theta^2) \tilde{h^\prime}(\theta)}.$$

The sequential application of Eq. (\ref{eq:derconst}) gives the correct expression for higher order derivatives. The explicit expressions increase in complexity extremely fast when higher order derivatives are considered, but they remain completely analytic in terms of the variables $(R,\theta)$, and allow us to have any of these derivatives defined at any point in the QCD phase diagram, provided a choice of parameters for the transformation map is given, the only step to be performed numerically being the solution of Eqs. (\ref{eq:numMap1}), (\ref{eq:numMap2}). 

\subsection{Critical pressure} \label{sec:CritPress}

Our procedure reduces to the use of Eqs. (\ref{eq:PIsing}) and (\ref{eq:Chis}), because the dependence on $(R,\theta)$ is well defined and the numerical inversion allows us to transport any quantity to any point in the QCD phase diagram. A remark is in order at this point.

\begin{figure}
\center
\includegraphics[width=0.49\textwidth]{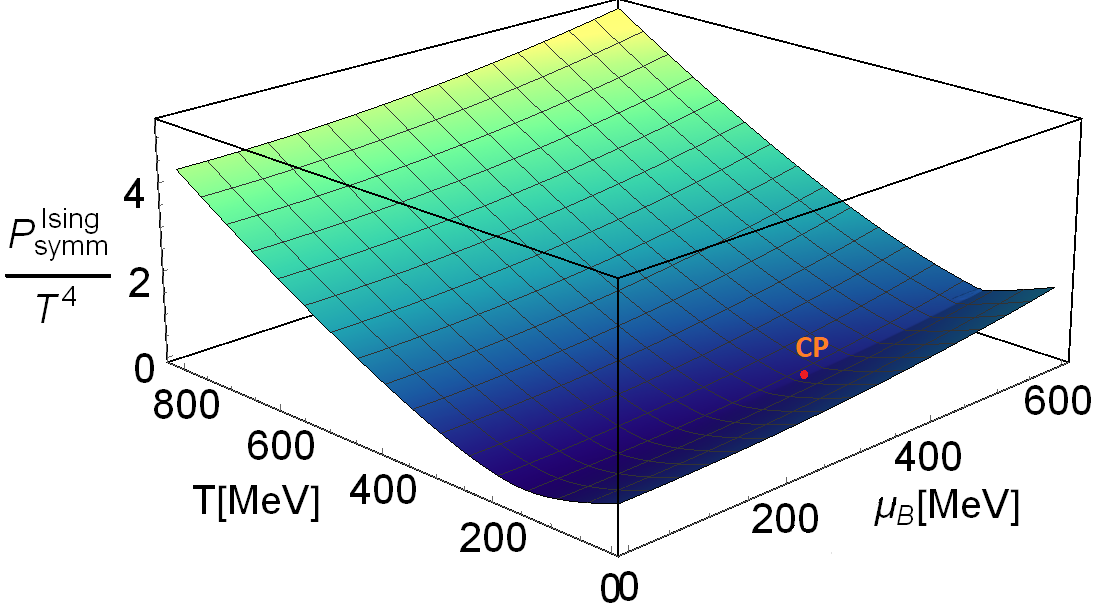} \hfill
\includegraphics[width=0.49\textwidth]{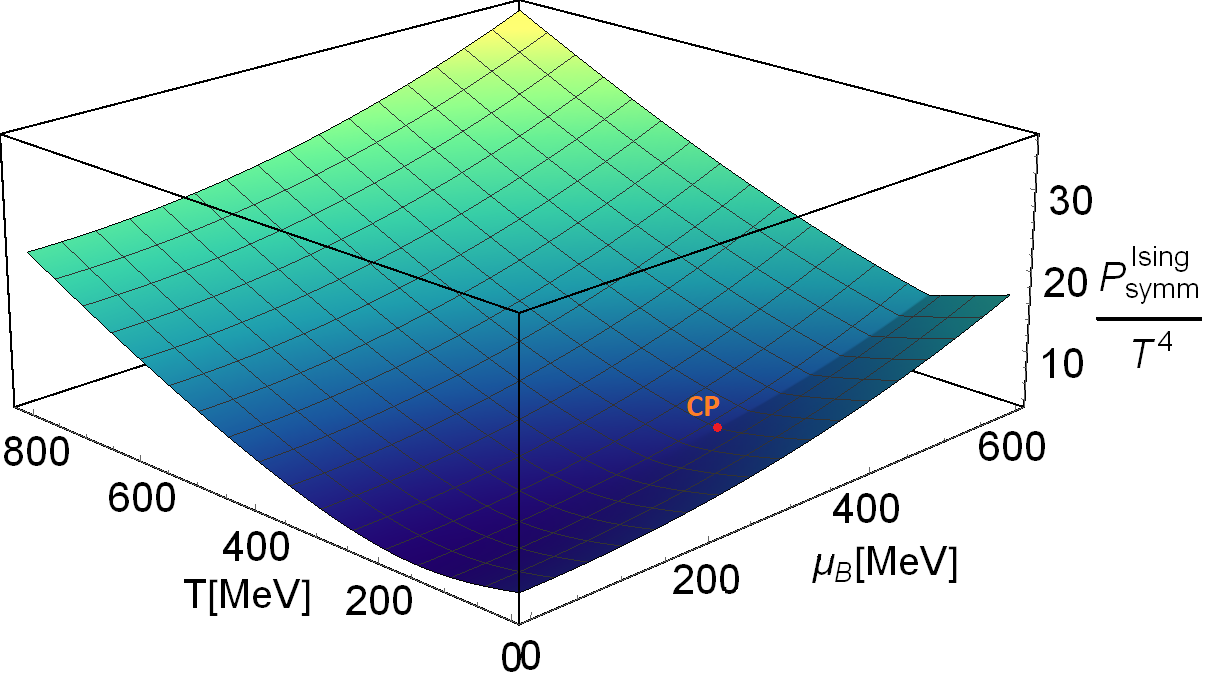}
\caption{The critical pressure for the choice of parameters in Section \ref{sec:paramRed} (left), and for a smaller value of $w=0.25$ (right), obtained with the mapping from the 3D Ising model and symmetrized around $\mu_B=0$. The critical point is located at $\mu_{BC}=350 \, \text{MeV}$ and $T_C  \simeq 143.2 \, \text{MeV}$ in both cases. The singular behavior is evident for $\mu_B > \mu_{BC}$, where the first order transition occurs. We can see that a smaller value of the scaling parameter $w$ corresponds to a larger ``Ising'' contribution to the pressure.}
\label{fig:Pcrit}
\end{figure}

Because of the charge conjugation symmetry, in QCD the partition function needs to be an even function of the baryon chemical potential:
\begin{equation}
{\cal Z} (T,-\mu_B) = {\cal Z} (T,\mu_B) \, \, , 
\end{equation}
as well as the pressure. Thus QCD must possess a critical point at both $\mu_{BC}$ and $-\mu_{BC}$. To achieve this we need to write Eq. (\ref{eq:PsymmIsing}) below. This form does not modify the singular critical behavior at the critical point(s) and automatically ensures that the odd-power coefficients in the Taylor expansion in $\mu_B$ vanish, as they should.
\begin{align} \label{eq:PsymmIsing}
P_{\text{QCD}}^{\, \text{crit}}(T, \mu_B) &= \frac{1}{2} f(T, \mu_B) P_{\, \text{symm}}^{\text{Ising}} (R(T, \mu_B) ,\theta (T, \mu_B))  =
\nonumber \vspace{4mm} \\ 
 &= \frac{1}{2} f(T, \mu_B) \left\lbrace P^{\text{Ising}} (R(T, \mu_B) ,\theta (T, \mu_B)) + P^{\text{Ising}} (R(T, - \mu_B) ,\theta (T, - \mu_B)) \right\rbrace \, \, ,
\end{align} 
which will have the effect of slightly changing the form of the critical pressure (the main one being that now the pressure at the critical point is non-zero, whereas it would be zero in the straightforward definition) but not its singular behavior, leaving all the even order derivatives unchanged. Fig. \ref{fig:Pcrit} shows the symmetrized form of the critical pressure for the choice of parameters in Section \ref{sec:paramRed} (left panel) and for a smaller value of $w=0.25$ (right panel).

\section{Methodology and parameter choice}

With the prescription exposed in Section \ref{sec:CritPress}, we now have a well defined procedure to produce an expression for the pressure that meets all our requirements, needing only to make a choice for the parameters in the Ising-to-QCD map.

\subsection{Acceptable parameter values}

The choice of parameters is a key part of the procedure, because it can provide physical information, e.g. indications on the location of the critical point through comparison with experimental data. In general, the linear map we introduced has a total of six parameters. As we will detail below, most of them are not arbitrary.

Some indication or constraint on this choice comes from our current knowledge of the QCD phase diagram. For example, since the chiral/deconfinement transition temperature is $T \simeq 155 \, \text{MeV}$ at $\mu_B=0$ \cite{Aoki:2006we,Aoki:2009sc,Borsanyi:2010bp,Bhattacharya:2014ara,Bazavov:2011nk}, and the curvature of the transition line is negative \cite{Cea:2015cya,Bonati:2015bha,Bellwied:2015rza}, we can safely expect the temperature of the critical point to be $T_C \lesssim 155 \, \text{MeV}$. Other works have also shown that the presence of the critical point in the region $\mu_B \lesssim 2 T$ appears to be strongly disfavored \cite{Bazavov:2017dus}.

From the fact that the curvature of the transition line is negative and extremely small, we can easily argue that the angle $\alpha_1$ in the map needs to be positive, and very small as well. The choice of the second angle is rather arbitrary, because no argument of symmetry can be made to guide the choice. For simplicity, we will consider parameter sets in which the $r$ and $h$ axes are orthogonal. 

For the chemical potential at the critical point $\mu_{BC}$, although not required by any physical argument, we will restrict ourselves to values that are within reach of the BES-II program, namely $\mu_{BC} \lesssim 450 \, \text{MeV}$.

The choice of the scale factors $w$ and $\rho$ is definitely less intuitive. When we consider the contribution from the pressure and its derivatives at $\mu_B=0$, the effect of changing the scale factor $w$, keeping $\mu_{BC}$ fixed, is the same as moving away or towards the critical point. In particular, since $\partial_{\mu_B} \sim 1/w$, reducing $w$ would result in a larger ``Ising'' contribution to the pressure and its derivatives at $\mu_B=0$, and a larger critical region. This can be seen in Fig. \ref{fig:Pcrit}: a smaller value of $w$ increases the value of the ``Ising'' contribution to the pressure. Moreover, the pressure grows faster both in the $T$ and $\mu_B$ directions. As for the other scale parameter $\rho$, its role is to govern the behavior in the pressure and its derivatives when moving away from the critical point in the temperature direction. Since we do not know what the scaling should look like in the temperature direction relative to the one in the chemical potential direction, the choice of $\rho$ remains mainly arbitrary. 

\subsubsection{Reducing the number of parameters} \label{sec:paramRed}

In practice, although the most general linear map between Ising variables and QCD coordinates requires the use of six parameters, it is possible to impose some constraint in the choice by making use of additional arguments for the location of the critical point. For example, the curvature of the transition line at $\mu_B=0$ has been estimated in lattice simulations \cite{Cea:2015cya,Bonati:2015bha,Bellwied:2015rza}. The shape of such transition line can be approximated with a parabola:
\begin{equation}\label{eq:trline}
T = T_0 + \kappa \, T_0 \left( \frac{\mu_B}{T_0} \right)^2 + {\cal O} (\mu_B^4),
\end{equation}
where $T_0$ and $\kappa$ are the transition temperature and curvature of the transition line at $\mu_B=0$, respectively. The number of parameters is thus reduced to four, the angle $\alpha_1$ also being fixed by:
\begin{equation}
\alpha_1 = \tan^{-1} \left( 2 \frac{\kappa}{T_0} \mu_{BC} \right) \, \, .
\end{equation} 

In the following, remembering that the aim of the EoS is to be employed in hydrodynamic simulations for heavy-ion collisions in the BES-II program, we will consider a choice of the baryonic chemical potential which is $\mu_{BC} = 350 \, \MeV$, resulting in:
\begin{align} 
T_C \simeq 143.2 \, \MeV \, , \qquad \qquad \alpha_1 \simeq 3.85 \, ^\circ \, \, .
\end{align}
In addition, the axes are chosen to be orthogonal, as we already mentioned, so that $\alpha_2 \simeq 93.85 \, ^\circ$. Finally, the scaling parameters are initially chosen as:
\begin{align}
w = 1 \, \, , & \qquad \qquad \rho = 2 \, \, .
\end{align}
Later we will explore different choices for $w$ and $\rho$, trying to reduce their acceptable range on the basis of physical conditions for the thermodynamic quantities. For completeness, in Appendix B we show the results for the EoS obtained with other allowed parameter choices.

\subsection{Lattice results}

Recalling Eq. (\ref{eq:coeffs}), we can see that the other defining ingredients for our procedure, besides the calculation of the Ising model thermodynamics and its ``translation'' to QCD, are the Taylor coefficients from lattice QCD. In the following we will use data from the Wuppertal-Budapest Collaboration \cite{Borsanyi:2013bia,Bellwied:2015lba} for the pressure and its derivatives at $\mu_B=0$. 

Before actually using the lattice results, we need to address a couple of issues:
\begin{enumerate}[i.]
\item The range of temperatures of the available lattice results is not sufficient to provide an equation of state as needed in hydrodynamical simulations, namely for temperature values $ 30 \, \MeV \lesssim T\lesssim 800 \, \MeV$;
\item The dependence of such quantities on the temperature needs to be smooth enough, such that when we take derivatives of the thermodynamic quantities wrt to $T$ and $\mu_B$ (to calculate e.g. entropy density and baryon density) they do not present an unphysical wiggly behavior.
\end{enumerate}

To solve these issues, we took the following steps:
\begin{enumerate}[i.]
\item Generate data for temperatures below the reach of lattice ($T \geq 135 \, \MeV$ in this case) using the HRG model;
\item Provide a parametrization of the temperature dependence of the pressure and its derivatives in the desired temperature range.
\end{enumerate}

The parametrizations of $\chi_0 (T)$ and $\chi_4 (T)$ were performed through a ratio of $5^{\text{th}}$ order polynomials in the inverse temperature:
\begin{equation}
\chi_i(T) = \frac{a^i_0 + a^i_1/t + a^i_2/t^2 + a^i_3/t^3 + a^i_4/t^4 + a^i_5/t^5}{b^i_0 + b^i_1/t + b^i_2/t^2 + b^i_3/t^3 + b^i_4/t^4 + b^i_5/t^5} \, \, ,
\end{equation}
while for $\chi_2 (T)$, a different expression was used:
\begin{equation}
\chi_2(T) = e^{-h_1/t^\prime - h_2/{t^\prime}^2} \cdot f_3 \cdot (1 + \tanh(f_4/t^\prime + f_5))
\end{equation}
where $t=T/154 \, \text{MeV}$ and $t^\prime = T/200 \, \text{MeV}$ (\cite{Borsanyi:2011sw}). The parametrizations were obtained with lattice data/HRG model results in the range $T=5-500 \, \text{MeV}$, but extrapolated to the range $T=5-800 \, \text{MeV}$. \\

The values of the parameters are given in the following tables:

\vspace{4mm}
\small
\begin{adjustbox}{max width=0.95\textwidth} \small
\begin{tabular}{|c|c|c|c|c|c|c|c|c|c|c|c|c|} 
\hline
    & & & & & & & & & & & & \\
     & $a_0$ & $a_1$ & $a_2$ & $a_3$ & $a_4$ & $a_5$ & $b_0$ & $b_1$ & $b_2$ & $b_3$ & $b_4$ & $b_5$  \\
        & & & & & & & & & & & & \\
\hline
        & & & & & & & & & & & & \\
    $\chi_0(T)$ & $7.53891$ & $-6.18858$ & $-5.37961$ & $7.08750$ & $-0.977970$ & $0.0302636$ & 
    $2.24530$ & $-6.02568$ & $15.3737$ & $-19.6331$ & $10.2400$ & $0.799479$ \\
            & & & & & & & & & & & & \\
\hline
        & & & & & & & & & & & & \\
    $\chi_4(T)$ & $0.0148438$ & $-0.0371572$ & $0.0313008$ & $-0.0101907$ & $0.00144661$ & 
	$-0.000159877$ & $0.0673273$ & $3.33723$ & $-13.6747$ & $20.4745$ & $-13.6013$ & $3.39819$ \\ 
            & & & & & & & & & & & & \\
\hline
\end{tabular}
\end{adjustbox}
\vspace{8mm} 

\begin{center}
\begin{adjustbox}{max width=0.45\textwidth} \small
\begin{tabular}{|c|c|c|c|c|c|} 
\hline
    & & & & &  \\
     & $h_1$ & $h_2$ & $f_3$ & $f_4$ & $f_5$   \\
        & & & & &   \\
\hline
      & & & & &  \\
    $\chi_2(T)$ & $-0.325372$ & $0.497729$ & $0.148987$ & $6.66388$ & $-5.07725$ \\    
            & & & & &   \\
\hline
\end{tabular}
\end{adjustbox}
\end{center}
\vspace{4mm} \normalsize

\begin{figure}[h]
\center
\includegraphics[width=.45\textwidth]{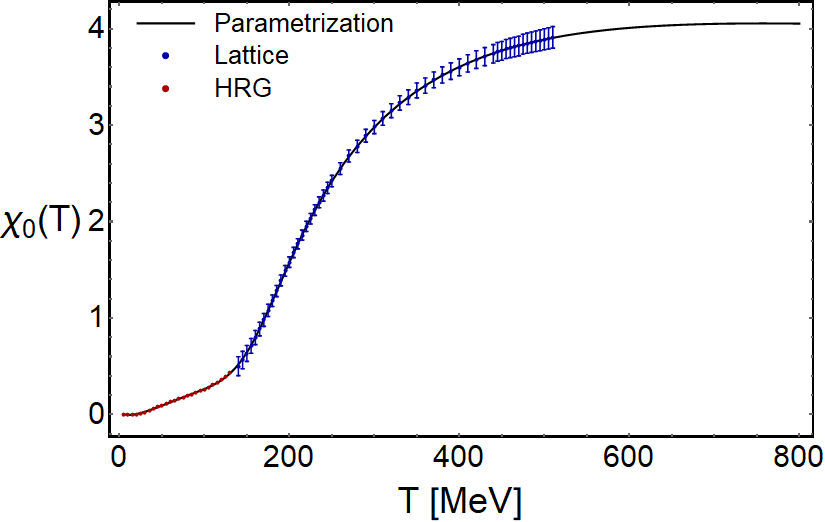}
\includegraphics[width=.465\textwidth]{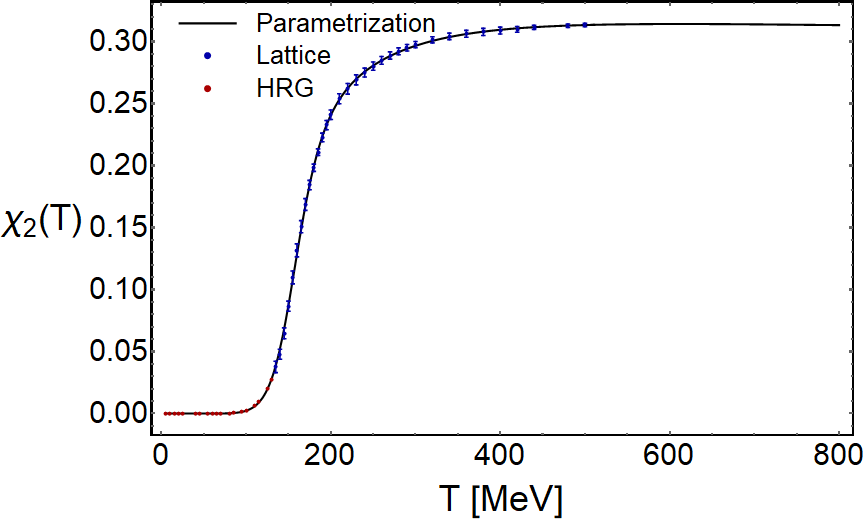} \\
\includegraphics[width=.465\textwidth]{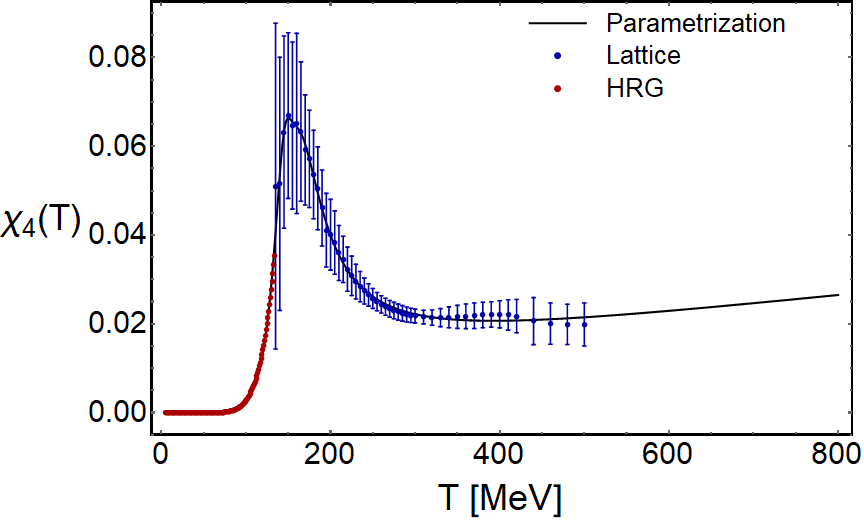}
\caption{Parametrization of baryon susceptibilities from Lattice QCD \cite{Borsanyi:2013bia,Bellwied:2015lba} and HRG model calculations.}
\label{fig:param}
\end{figure}

In Fig. \ref{fig:param} we can see the comparison between the lattice data (and the extension with the HRG model) and the resulting parametrization. The HRG model employed to calculate the pressure does not contain any interaction, and makes use of the most up to date particle list available from the Particle Data Group \cite{Patrignani:2016xqp} (list PDG2016+ in \cite{Alba:2017mqu}).

\begin{figure}[h]
\center
\includegraphics[width=.48\textwidth]{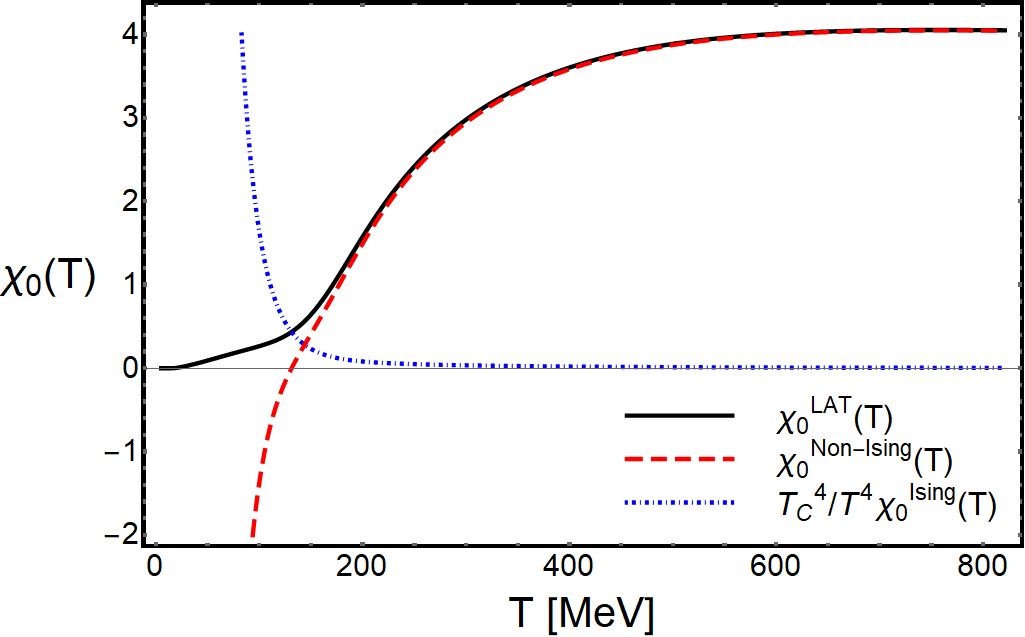} \hfill
\includegraphics[width=.495\textwidth]{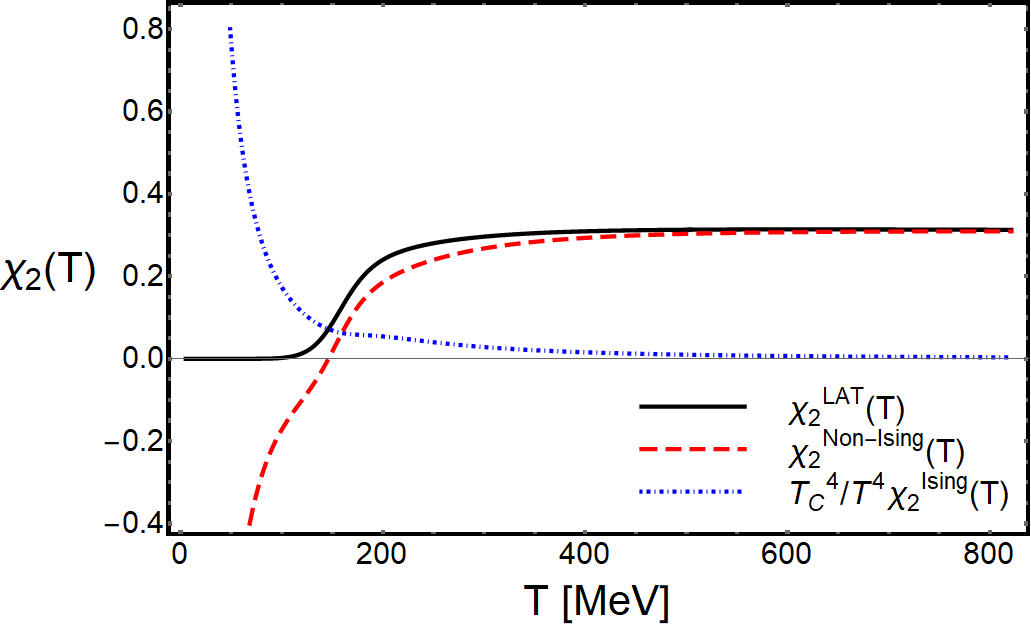} \\
\includegraphics[width=.495\textwidth]{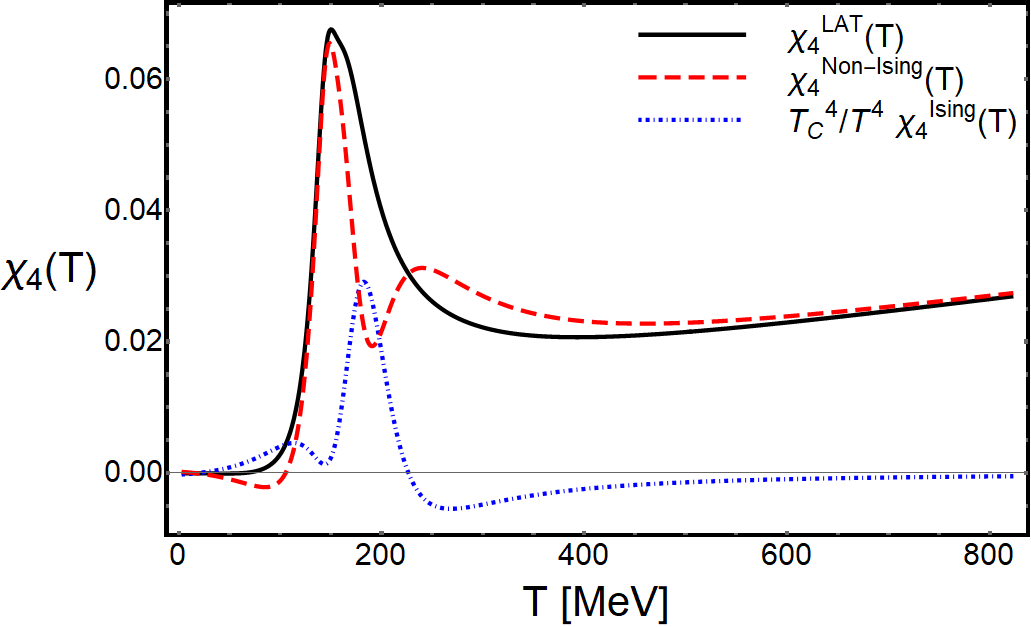} 
\caption{Comparison of critical (blue, dot-dashed) and ``Non-Ising'' (red, dashed) contributions to baryon susceptibilities up to ${\cal O}(\mu_B^4)$ with the parametrized lattice data (black, solid).}
\label{fig:ChisComp}
\end{figure}

The smooth curves obtained from the parametrization will be the $c_n^{\text{LAT}} (T)$ coefficients in Eq. (\ref{eq:coeffs}), thus defining the $c_n^{\text{Non-Ising}} (T)$ coefficients that will be used for the Taylor expansion. Fig. \ref{fig:ChisComp} shows the comparison of the ``Ising'' and ``Non-Ising'' contributions to the parametrized lattice/HRG model results.

\section{Results}


At this point, we have all the ingredients in Eq. (\ref{eq:Pfull}):
\begin{equation}
P (T, \mu_B) = T^4 \sum^2_{n=0} c^{\text{Non-Ising}}_{2n} (T) \left( \frac{\mu_B}{T} \right)^{2n} + T_C^4 \, P_{\text{symm}}^{\, \text{Ising}}(T, \mu_B) \, \, ,
\end{equation}
which is now straightforward. However, although the overall behavior is correct, at low temperatures and in particular in regions where the ratio $\mu_B/T$ is very large, the pressure becomes negative, and so do other observables as well. This is due to the fact that, given our choice of the function $f(T,\mu_B)$ in Eq.  (\ref{eq:Pfull}), the ``Ising'' coefficients at low temperature follow a power law, whereas the full ones from lattice calculations fall off exponentially; hence, there will always be a value of $T$ for which one or more of the $c_n^\text{Non-Ising}(T)$ falls below zero, and thus a value of $\mu_B/T$ large enough that the pressure from the Taylor expansion in Eq. (\ref{eq:Pfull}) is large and negative, resulting in unphysical values for the thermodynamic observables. The recipe to cure this problem  is to make use of the fact that one can reasonably expect the system to find itself in a hadron gas state in that region of the phase diagram, and find a way to smoothly merge the pressure coming from the procedure we developed so far with the one from the HRG model.


The smooth merging can be obtained through a hyperbolic tangent as:
\begin{align}\label{eq:finalpress}
\frac{P_{\text{Final}}(T,\mu_B)}{T^4} &= \frac{P (T,\mu_B)}{T^4} \frac{1}{2} \left(1 + \tanh\left(\frac{T-T^\prime(\mu_B)}{\Delta T^\prime}\right) \right) + \nonumber \\
&+ \frac{P_{\text{HRG}} (T,\mu_B)}{T^4} \frac{1}{2} \left( 1 - \tanh\left(\frac{T - T^\prime(\mu_B)}{\Delta T^\prime}\right) \right) \, ,
\end{align} 
where $T^\prime(\mu_B)$ works as the ``switching temperature'', and $\Delta T^\prime$ is roughly the size of the ``overlap region'' where both pressures contribute to the sum. The dependence on the baryon chemical potential of the ``switching temperature'' is chosen to be parabolic, and parallel to the chiral transition line we assumed in Eq. (\ref{eq:trline}):
$$T^\prime(\mu_B) = T_0 + \frac{\kappa}{T_0} \mu_B^2 - T^* \, \, ,$$
where $T_0$ and $\kappa$ are the transition temperature and curvature of the transition line at $\mu_B=0$, and we choose $T^* = 23 \, \text{MeV}$ and in Eq. (\ref{eq:finalpress}) $\Delta  T^\prime = 17 \, \text{MeV}$. 

\subsection{Full thermodynamic description} \label{sec:thermfull}

\begin{figure}
\center
\includegraphics[width=.8\textwidth]{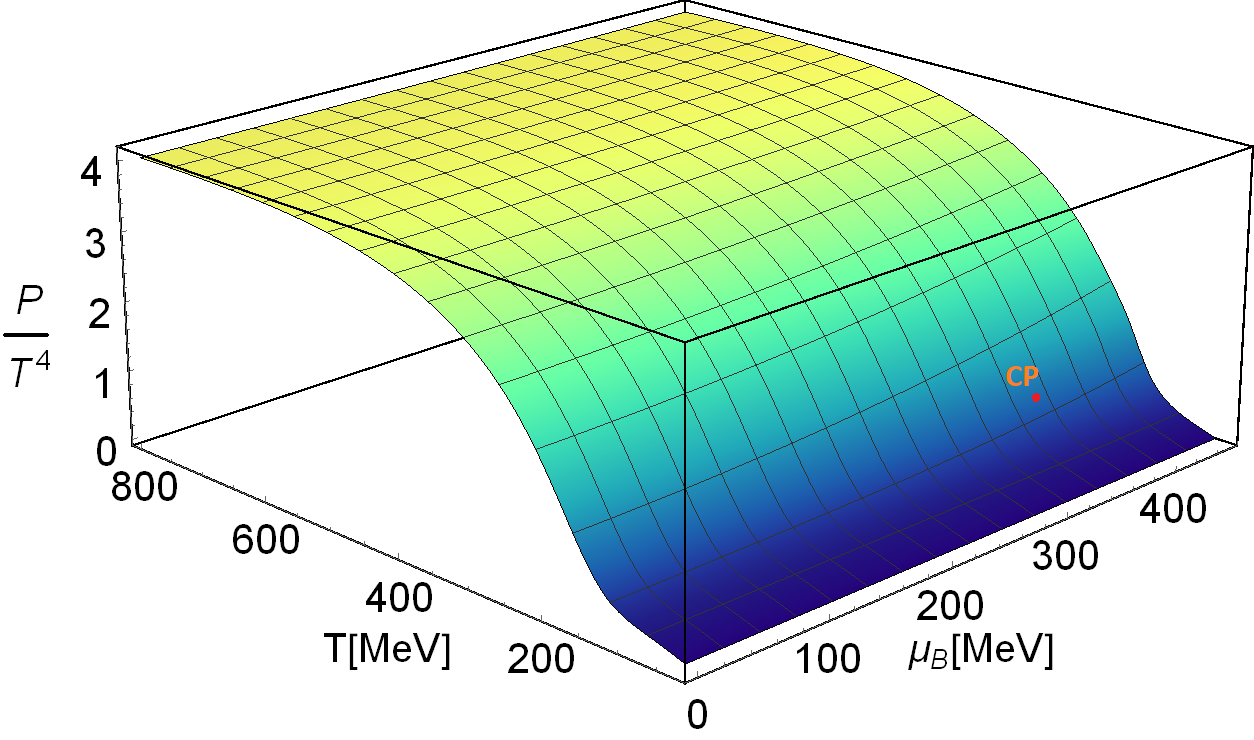}
\caption{Pressure for the choice of parameters in Section \ref{sec:paramRed}, after merging with HRG.}
\label{fig:PressFinal}
\end{figure}
\begin{figure}
\center
\includegraphics[width=.8\textwidth]{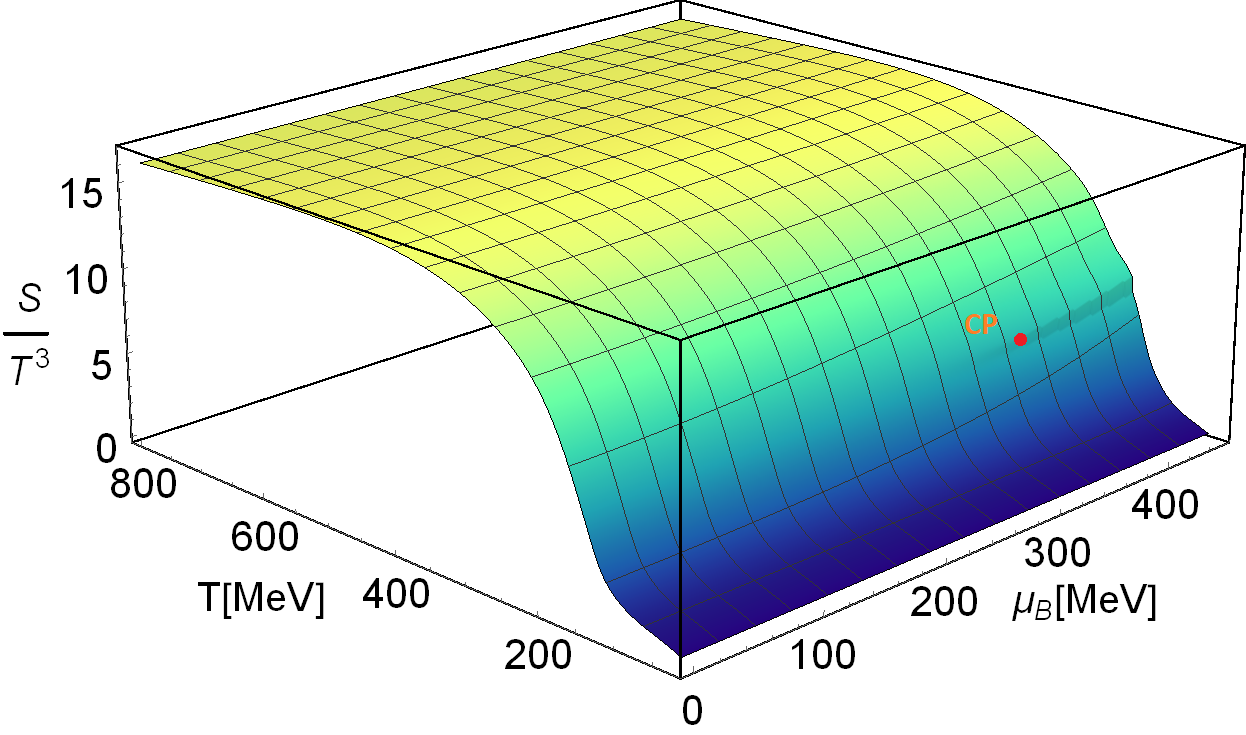}
\caption{Entropy density for the choice of parameters in Section \ref{sec:paramRed}, after merging with HRG.}
\label{fig:EntrFinal}
\end{figure}
\begin{figure}
\center
\includegraphics[width=.8\textwidth]{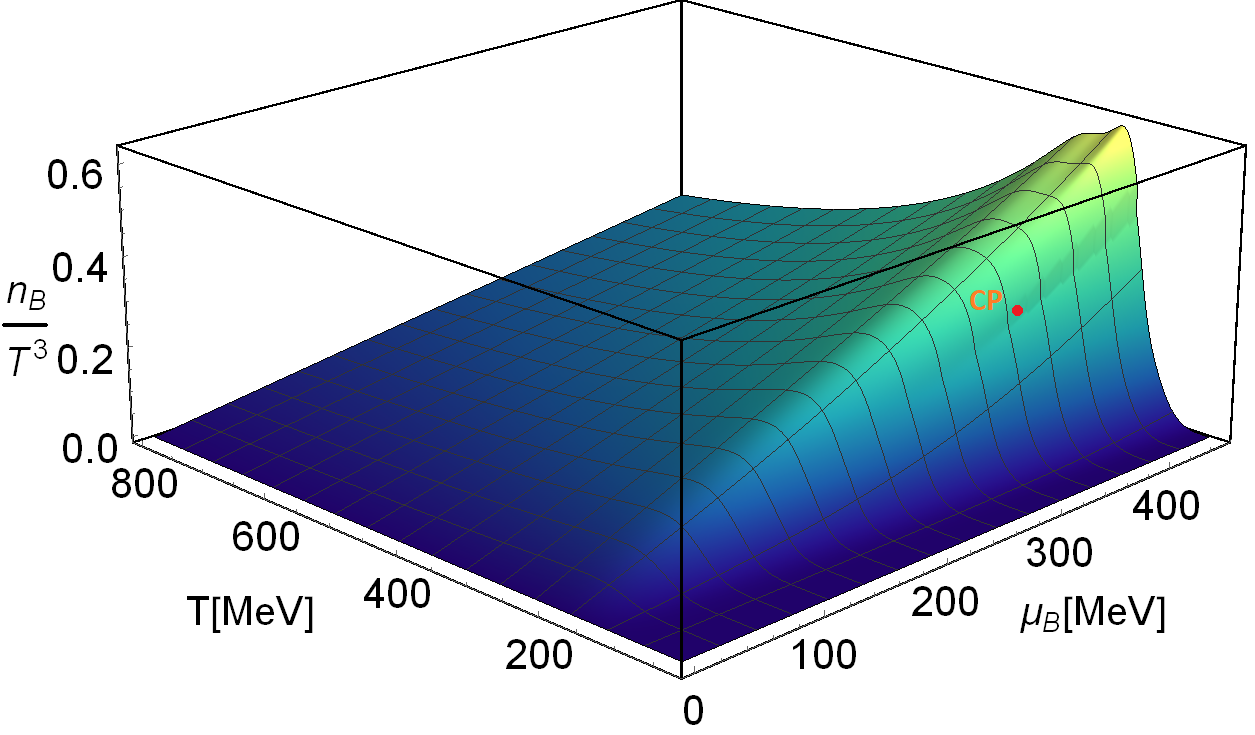}
\caption{Baryon density for the choice of parameters in Section \ref{sec:paramRed}, after merging with HRG.}
\label{fig:BarDensFinal}
\end{figure}
\begin{figure}
\center
\includegraphics[width=.8\textwidth]{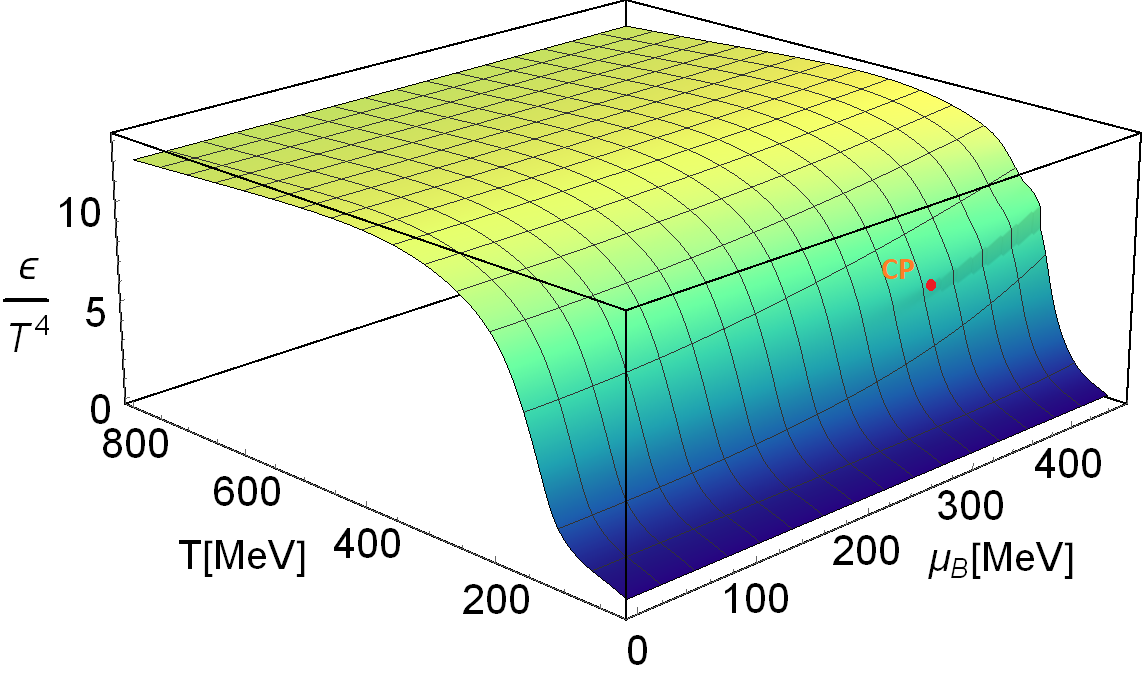}
\caption{Energy density for the choice of parameters in Section \ref{sec:paramRed}, after merging with HRG.}
\label{fig:EnerDensFinal}
\end{figure}
\begin{figure}
\center
\includegraphics[width=.8\textwidth]{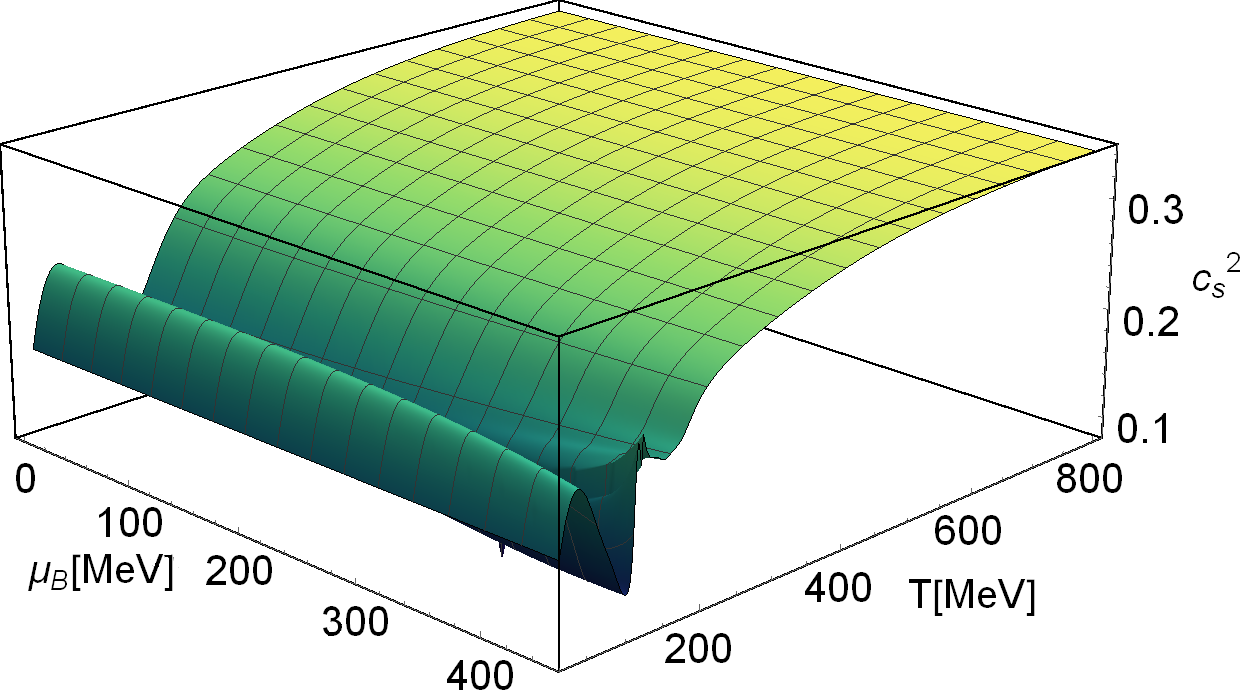}
\caption{Speed of sound for the choice of parameters in Section \ref{sec:paramRed}, after merging with HRG.}
\label{fig:SpSoundFinal}
\end{figure}

In order to complete the thermodynamic description of the finalized equation of state obtained in Eq. (\ref{eq:finalpress}), we can compute various thermodynamic observables of interest. In addition to the pressure, we compute the entropy density, baryon density, energy density and speed of sound normalized by the correct power of the temperature:
\begin{align} 
\frac{P(T,\mu_B)}{T^4} \, \, , & \qquad & \frac{S(T,\mu_B)}{T^3} &= \frac{1}{T^3} \left( \frac{\partial P}{\partial T} \right)_{\mu_B} \, \, , \\
\frac{n_B(T,\mu_B)}{T^3} &= \frac{1}{T^3} \left( \frac{\partial P}{\partial \mu_B} \right)_T \, \, ,  \qquad & \frac{\epsilon(T,\mu_B)}{T^4} &= \frac{S}{T^3} - \frac{P}{T^4} + \frac{\mu_B}{T} \frac{n_B}{T^3} \, \, , \\
c^2_s(T,\mu_B) &= \left( \frac{\partial P}{\partial \epsilon} \right)_{S/n_B} \, \, .
\end{align}

When working in the $(T,\mu_B)$ phase diagram, it is not advantageous to perform the calculation for the speed of sound directly from the definition; however, it is possible to re-write the expression for this observable in terms of derivatives of the pressure with respect to the temperature or the chemical potential only \cite{Floerchinger:2015efa}:
\begin{align}
c^2_s &= \frac{n_B^2 \partial^2_T P - 2 S n_B \partial_T \partial_{\mu_B} P + S^2 \partial^2_{\mu_B} P}{\left( \epsilon + P \right) \left(\partial^2_T P \partial^2_{\mu_B} P - (\partial_T \partial_{\mu_B} P)^2 \right)} \, \, .
\end{align} 
In Figs. \ref{fig:PressFinal}-\ref{fig:SpSoundFinal}, we show the pressure, entropy density, baryon density, energy density and speed of sound in the range of temperatures $T = 30 - 800 \, \MeV$ and chemical potentials $\mu_B = 0 - 450 \, \MeV$ for the EoS that we have constructed, with parameters as specified in Section \ref{sec:paramRed} and merged with the HRG model pressure at low temperatures as described in Eq. (\ref{eq:finalpress}). 

We note that, at large values of $\mu_B$, small wiggles appear in the thermodynamic observables, in particular the speed of sound. This is due to the truncation in the Taylor expansion of the non-Ising contribution to the pressure. In general these wiggles are only appearing at the edge of the allowed $\mu_B$ region and only for some parameter choices (see Appendix B for other choices in which they are not present). Besides, their magnitude is very small so we are confident that they will not affect the results of the hydrodynamic simulations. However, in order to extend our EoS to larger values of $\mu_B$, we need to devote future work to incorporate higher order terms in the Taylor expansion into the construction of model equations of state in order to improve their behavior at higher $\mu_B$.

\begin{figure}[h]
\center
\includegraphics[width=0.55\textwidth]{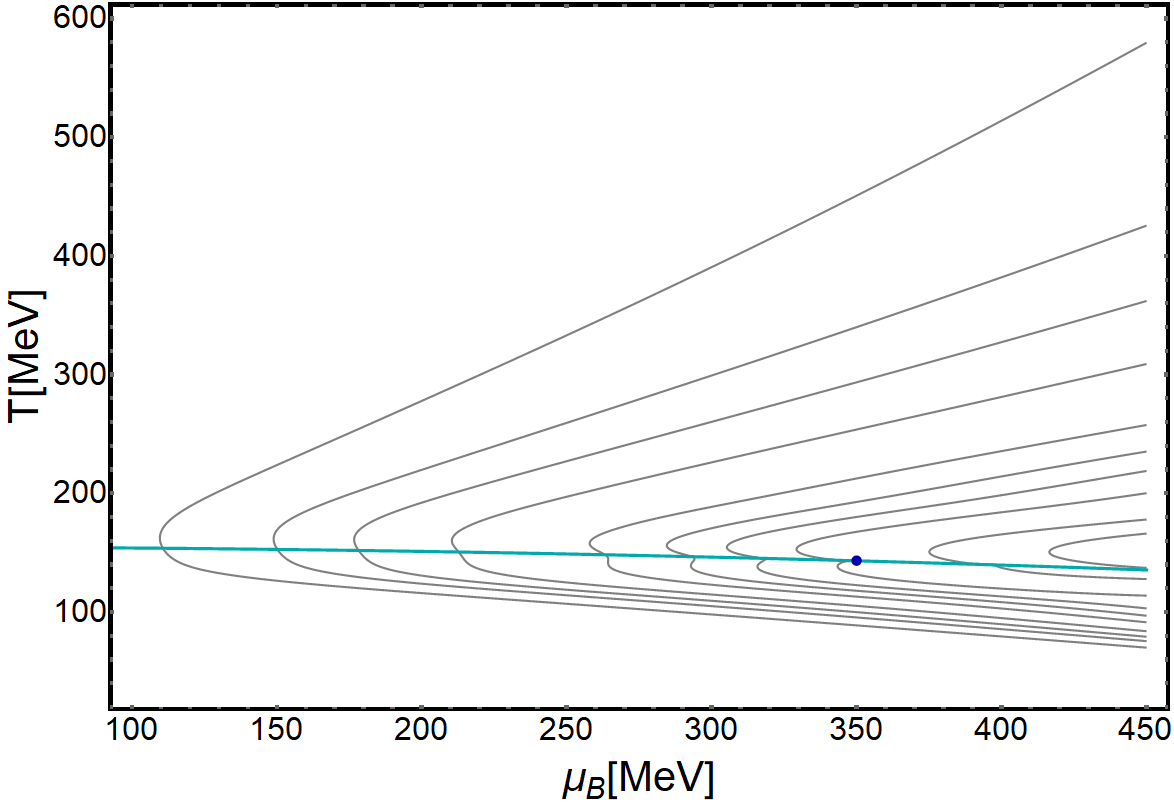}
\caption{Lines of constant $S/n_B$ (from left to right  $S/n_B=68,50,42,35,28,25,23,21,18,16$) in the QCD phase diagram. The blue dot indicates the location of the critical point, whereas in cyan is shown the chiral transition line in Eq. (\ref{eq:trline}).}
\label{isentropes}
\end{figure}

Although not very evident from the pressure, the critical point manifests itself clearly in first order derivatives (entropy, baryon and energy density), where the discontinuity due to a first order phase transition is clearly visible at $\mu_B > \mu_{BC}$. Furthermore, the speed of sound shows a clear dip at the critical point, as well as a (less evident) discontinuity at $\mu_B > \mu_{BC}$.
Figure \ref{isentropes} shows the trajectories at constant $S/n_B$ in the QCD phase diagram.

\begin{figure}[h]
\center
\includegraphics[width=0.8\textwidth]{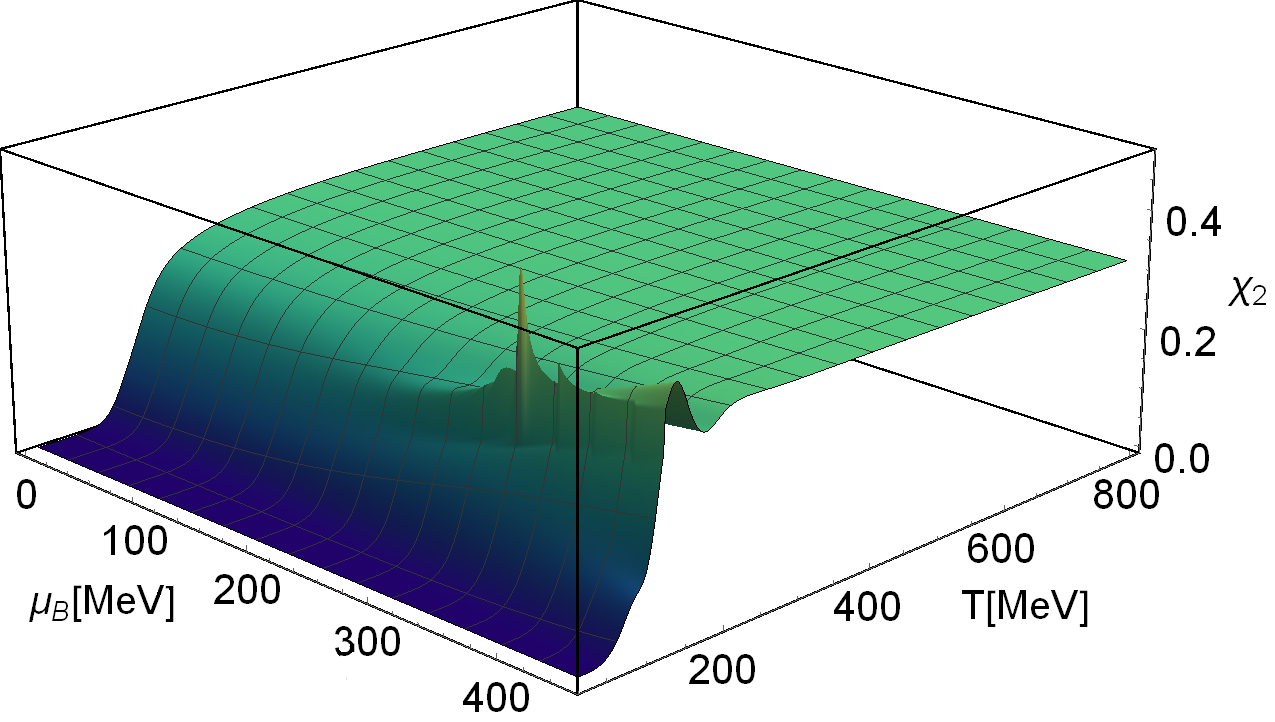}
\caption{The second cumulant of the baryon number $\chi_2^B = T^{-2} \frac{\partial^2 P}{\partial \mu_B^2}$, for the choice of parameters in Section \ref{sec:paramRed}, after merging with HRG (the rather jagged behavior close to the critical point is due to the finite size in the graphics grid used to make the figure).}
\label{fig:chi2}
\end{figure}

In addition to the thermodynamic quantities mentioned above, it is possible to calculate observables that are more sensitive to critical fluctuations: in Fig. \ref{fig:chi2} we show the second cumulant of the baryon number $\chi_2^B = T^{-2 }\frac{\partial^2 P}{\partial \mu_B^2}$ for the choice of parameters in Section \ref{sec:paramRed}. In heavy-ion collision experiments, it is possible to measure related quantities, constructed from moments of the event-by-event distribution of the measured number of protons in a given acceptance. The critical contribution to $\chi^B_2$ diverges at the critical point like $\xi^2$, where $\xi$ is the correlation length of the order parameter fluctuations. Higher, non-Gaussian, moments of the proton number distribution (or the baryon number distribution) receive larger contributions from critical fluctuations, with the third and fourth cumulants diverging like $\xi^{9/2}$ and $\xi^7$ respectively \cite{Stephanov:2008qz}, making these observables more sensitive to the presence of a critical point\cite{Athanasiou:2010kw,Stephanov:2011pb,Ling:2015yau,Brewer:2018abr}, and motivating their experimental measurement in the RHIC Beam Energy Scan \cite{Adamczyk:2013dal,Luo:2015ewa,Luo:2015doi,Akiba:2015jwa,Geesaman:2015fha}, as well as the future investigations of higher moments of the $n_B$ distribution in this model. 

\subsection{Exploration of parameter space} \label{sec:parameters}
\begin{figure}
\center
\includegraphics[width=.5\textwidth]{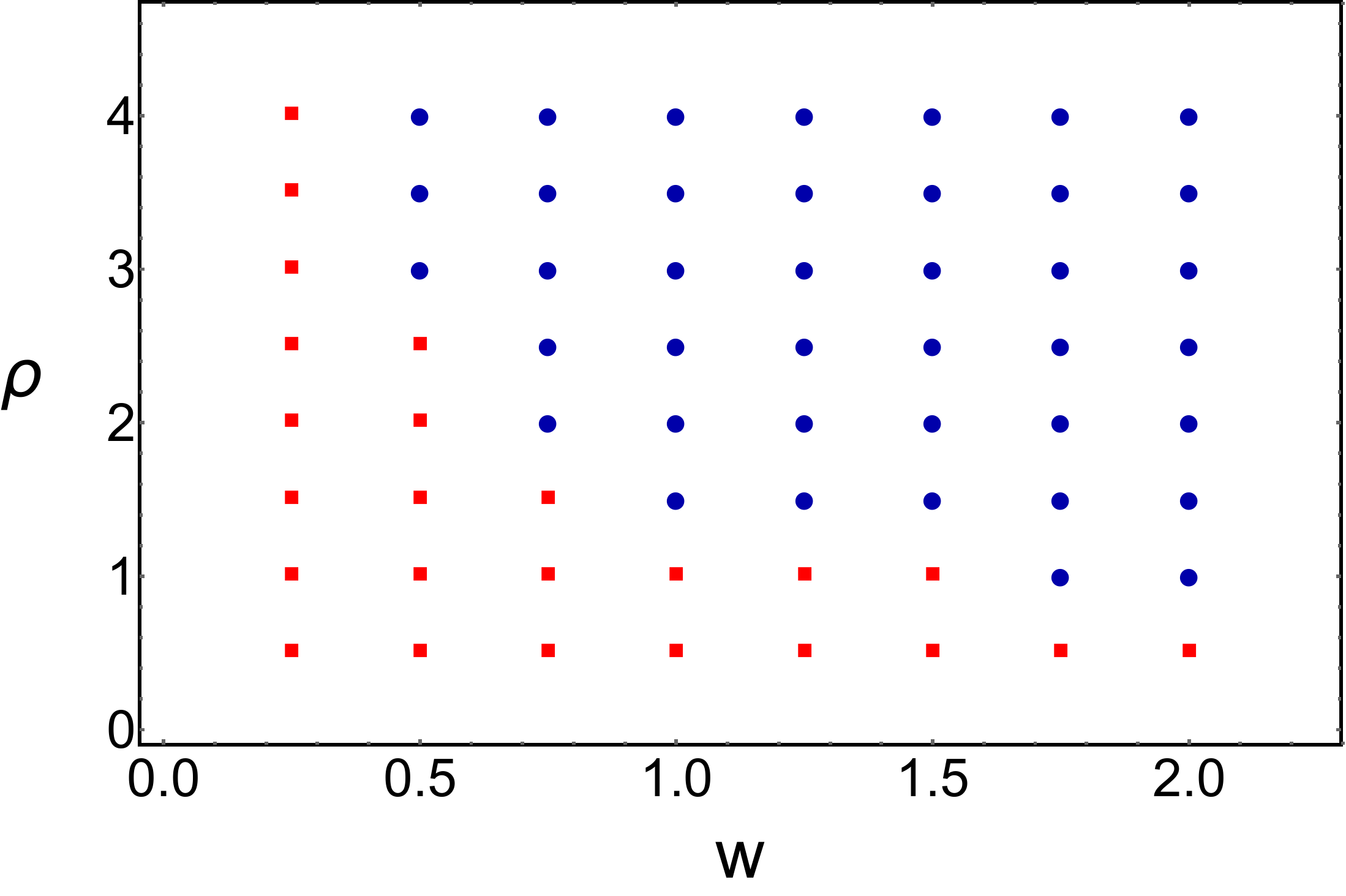}
\caption{Plane of $w$ and $\rho$ parameter values. In red (squares) the points corresponding to pathological choices of parameters, in blue (dots) the acceptable ones.}
\label{fig:paramtable}
\end{figure}

By requiring thermodynamic stability, i.e. positivity of pressure, entropy density, baryon density, energy density and speed of sound, and causality, i.e $c^2_s < 1$, over the whole phase diagram, it is possible to reduce the range of acceptable parameters in the non-universal Ising $\mapsto$ QCD map. By keeping the location of the critical point fixed ($\mu_{BC} = 350 \, \text{MeV}$, $T_C \simeq 143 \, \text{MeV}$), as well as the orientation of the axes ($\alpha_1 \simeq 3.85 ^\circ$, $\alpha_2 - \alpha_1 = 90 ^\circ$), we investigated the role of the scaling parameters $w$, $\rho$. In Fig. \ref{fig:paramtable}, we can see in red the points corresponding to pathological parameter choices, while the blue dots correspond to acceptable ones. We notice that, while most commonly specific parameter choices are unacceptable because of the negativity of $n_B$, for very low $w$ ($w=0.25$) we observe violation of causality as well ($c^2_s > 1$).

\section{Conclusions}
In this manuscript, we presented a procedure to construct a family of model equations of state for QCD, each of which features a critical point in the 3D Ising model universality class. A parametrization of the scaling equation of state in such a model, together with a parametrized map from Ising variables to QCD coordinates, yields explicit expressions for the critical contribution to thermodynamic quantities in QCD.

The comparison with lattice results at vanishing chemical potential allows us to estimate the possible size of the critical contribution, and reconstruct the equation of state in a way that on the one hand contains critical behavior in the correct universality class, and on the other hand matches lattice QCD results up to $\mathcal{O}(\mu_B^4)$ exactly at zero chemical potential. Our result can be readily utilized in hydrodynamic simulations of heavy ion collisions at the energies reached in the BES-II program. We also show the second cumulant of the baryon number, a quantity that diverges at the critical point and can be related to experimentally measurable net-proton fluctuations. 

Higher order cumulants, which are expected to display a more pronounced divergence at the critical point, will be investigated in future work. We also leave to future work the inclusion of strangeness and electric charge chemical potentials, which play an important role in heavy-ion collision physics. The main reason for this is that the same universality arguments we employ in this work cannot be used to answer the question how the critical behavior in the $(T,\mu_B)$ plane would vary with the two additional chemical potentials. Because of this, such an exercise would require a great deal of additional modeling, which is beyond the scope of this work.

The requirement that the resulting equation of state does not violate thermodynamic inequalities (positivity of pressure, entropy density, baryon density, energy density, speed of sound, and causality, i.e. $c^2_s < 1$), together with the comparison between experimental data and results obtained through simulations that employ such an equation of state, can help constrain the values of the parameters in the equation of state, in so doing, we hope, constraining the location of a possible critical point in the QCD phase diagram. 

\section*{Acknowledgements}
The authors acknowledge fruitful discussions within the ``Equation of State and Fluctuations" working group of the BEST Collaboration and with Jorge Noronha, Israel Portillo Vazquez, Greg Ridgway, Ryan Weller, Yi Yin. 
This material is based upon work supported by the National Science Foundation under 
grants no. PHY-1654219 and OAC-1531814 and by the U.S. Department of Energy, Office 
of Science, Office of Nuclear Physics under Contract number DE-SC0011090 and DE-
FG0201ER41195, and Award Number DE-FG02-03ER41260, as well as within the 
framework of the Beam Energy Scan Theory (BEST) Topical Collaboration. P.P., D.M. and 
C.R. also acknowledge the use of the Maxwell Cluster and the advanced support from the 
Center of Advanced Computing and Data Systems at the University of Houston. M.N. 
acknowledges the support of the program ``Etoiles montantes en Pays de la Loire 2017''. 
The work of M.B. is funded by the European Union's Horizon 2020 research and innovation 
program under the Marie Sk\l{}odowska Curie grant agreement No 665778 via the 
National Science Center, Poland, under grant Polonez UMO-2016/21/P/ST2/04035.  P.P. 
also acknowledges support by the DFG grant SFB/TR55.

\newpage
\section*{Appendix A}

The following relations will be used in Eq. (\ref{eq:derconst}):
\begin{align*}
\left. \frac{dR}{d \theta} \right|_r  &=  \frac{2 \theta R}{(1 - Th^2)} \, \, , & \left. \frac{dR}{d \theta} \right|_h &= - \frac{R \tilde{h}^\prime (\theta)} {\beta \delta \tilde{h}(\theta)} \, \, , 
\end{align*}
and hence:
\begin{align*}
\left. \frac{d \theta}{dR} \right|_r  &=  \frac{(1 - Th^2)}{2 \theta R} \, \, , & \left. \frac{d \theta}{dR} \right|_h &= - \frac{\beta \delta \tilde{h}(\theta)}{R \tilde{h}^\prime (\theta)}  \, \, . 
\end{align*}

\subsection*{First order}

\noindent
Derivatives of $G(R,\theta)$ wrt $(R,\theta)$:
\begin{align*}
\frac{\partial G}{\partial R} &= (2-\alpha) h_0 M_0 R^{1-\alpha} \left [ g(\theta) - \theta \htil \right] \, \, , & \frac{\partial G}{\partial \theta} &= h_0 M_0 R^{2-\alpha} \left [ g^\prime(\theta) - \htil - \theta \htilprime \right] \, \, .
\end{align*}

\noindent
Derivatives of $(r,h)$ wrt $(R,\theta)$:
\begin{align*}
\left( \frac{\partial r}{\partial R} \right)_h &= (1 -\theta^2) + 2 \beta \delta \theta \frac{\htil}{\htilprime} \, \, , & \left( \frac{\partial h}{\partial R} \right)_r &= h_0 R^{\beta \delta -1} \left( \beta \delta \htil + \frac{(1 -\theta^2)}{2 \theta} \htilprime \right)  \, \, , \\
\left( \frac{\partial r}{\partial \theta} \right)_h &= - \frac{R}{\beta \delta} \left( 2 \beta \delta \theta + (1 - \theta^2) \frac{\htilprime}{\htil} \right) \, \, , & \left( \frac{\partial h}{\partial \theta} \right)_r &= h_0 R^{\beta \delta} \left( \frac{2 \beta \delta \theta}{1 - \theta^2} \htil + \htilprime \right)  \, \, ,
\end{align*}
hence, derivatives of $(R,\theta)$ wrt $(r,h)$:
\begin{align*}
\left( \frac{\partial R}{\partial r} \right)_h &= \frac{\htilprime}{2 \beta \delta \theta \htil + (1 -\theta^2) \htilprime} \, \, , & \left( \frac{\partial R}{\partial h} \right)_r &= \frac{1}{h_0 R^{\beta \delta -1}} \frac{2 \theta}{2 \beta \delta \theta \htil + (1 -\theta^2) \htilprime} \, \, , \\
\left( \frac{\partial \theta}{\partial r} \right)_h &= - \frac{\beta \delta}{R} \frac{\htil}{2 \beta \delta \theta \htil + (1 -\theta^2) \htilprime} \, \, , & \left( \frac{\partial \theta}{\partial h} \right)_r &= \frac{1}{h_0 R^{\beta \delta}} \frac{1 - \theta^2}{2 \beta \delta \theta \htil + (1 -\theta^2) \htilprime} \, \, ,
\end{align*}

\noindent
Derivatives of $G(R,\theta)$ wrt $(r,h)$:
\begin{align*}
\left( \frac{\partial G}{\partial r} \right)_h &= \frac{\partial G}{\partial R} \left( \frac{\partial R}{\partial r} \right)_h + \frac{\partial G}{\partial \theta} \left( \frac{\partial \theta}{\partial r} \right)_h = \\
& = \frac{h_0 M_0 R^{1-\alpha}}{2 \beta \delta \theta \htil + (1 -\theta^2) \htilprime} \left\lbrace (2 - \alpha) \htilprime \left( g(\theta) - \theta \htil \right) - \beta \delta \htil \left( g^\prime(\theta) - \htil - \theta \htilprime \right) \right\rbrace
 \, \, , \\  \\
 \left( \frac{\partial G}{\partial h} \right)_r &= \frac{\partial G}{\partial R} \left( \frac{\partial R}{\partial h} \right)_r + \frac{\partial G}{\partial \theta} \left( \frac{\partial \theta}{\partial h} \right)_r = \\
 & = \frac{M_0 R^\beta}{2 \beta \delta \theta \htil + (1 -\theta^2) \htilprime} \left\lbrace 2 \theta (2 - \alpha) \left( g(\theta) - \theta \htil \right) + (1 - \theta^2) \left( g^\prime(\theta) - \htil - \theta \htilprime \right) \right\rbrace
 \, \, , \\ 
\end{align*}

\noindent
Finally, the derivatives of $G(R,\theta)$ wrt $(T,\mu_B)$ are constructed as:
\begin{align*}
\left( \frac{\partial G}{\partial \mu_B} \right)_T &= \frac{\partial h}{\partial \mu_B} \left( \frac{\partial G}{\partial h} \right)_r + \frac{\partial r}{\partial \mu_B} \left( \frac{\partial G}{\partial r} \right)_h \\
 \left( \frac{\partial G}{\partial T} \right)_{\mu_B} &= \frac{\partial h}{\partial T} \left( \frac{\partial G}{\partial h} \right)_r + \frac{\partial r}{\partial T} \left( \frac{\partial G}{\partial r} \right)_h
\end{align*}

\subsection*{Second order}

The relationships start to become more complicated at the second order. \\

\noindent
Derivatives of $G(R,\theta)$ wrt $(R,\theta)$:
\begin{align*}
\frac{\partial^2 G}{\partial R^2} &= (2-\alpha) (1 - \alpha) h_0 M_0 R^{-\alpha} \left [ \g - \theta \htil \right] \, \, , \\
\frac{\partial^2 G}{\partial \theta^2} &= h_0 M_0 R^{2-\alpha} \left [ \gsec -2 \htilprime - \theta \htilsec \right] \, \, , \\
\frac{\partial^2 G}{\partial R \partial \theta} &= (2-\alpha) (1 - \alpha) h_0 M_0 R^{-\alpha} \left [ \gprime - \htil - \theta \htilprime \right] \, \, .
\end{align*}

\noindent
Derivatives of $G(R,\theta)$ wrt $(T,\mu_B)$ are constructed as:
\begin{align*}
\left( \frac{\partial^2 G}{\partial \mu_B^2} \right)_T &= \left( \frac{\partial h}{\partial \mu_B} \right)^2 \left( \frac{\partial^2 G}{\partial h^2} \right)_r + \left( \frac{\partial r}{\partial \mu_B} \right)^2 \left( \frac{\partial^2 G}{\partial r^2} \right)_h + 2 \frac{\partial h}{\partial \mu_B} \frac{\partial r}{\partial \mu_B} \frac{\partial^2 G}{\partial r \partial h} \\
 \left( \frac{\partial^2 G}{\partial T^2} \right)_{\mu_B} &= \left( \frac{\partial h}{\partial T} \right)^2 \left( \frac{\partial^2 G}{\partial h^2} \right)_r + \left( \frac{\partial r}{\partial T} \right)^2 \left( \frac{\partial^2 G}{\partial r^2} \right)_h + 2 \frac{\partial h}{\partial T} \frac{\partial r}{\partial T} \frac{\partial^2 G}{\partial r \partial h}
\end{align*}
where the terms with second derivatives of $(r,h)$ wrt $(T,\mu_B)$ have been dropped, since the transformation between the two sets is linear. \\

\noindent
Derivatives of $G(R,\theta)$ wrt $(r,h)$:
\begin{align*}
\left( \frac{\partial^2 G}{\partial r^2} \right)_h &= \frac{\partial^2 G}{\partial R^2} \left( \frac{\partial R}{\partial r} \right)_h^2 + \frac{\partial G^2}{\partial \theta^2} \left( \frac{\partial \theta}{\partial r} \right)_h^2 + \frac{\partial G}{\partial R} \left( \frac{\partial^2 R}{\partial r^2} \right)_h + \frac{\partial G}{\partial \theta} \left( \frac{\partial^2 \theta}{\partial r^2} \right)_h + 2 \frac{\partial^2 G}{\partial R \partial \theta} \left( \frac{\partial R}{\partial r} \right)_h \left( \frac{\partial \theta}{\partial r} \right)_h \, \, , \\
\left( \frac{\partial^2 G}{\partial h^2} \right)_r &= \frac{\partial^2 G}{\partial R^2} \left( \frac{\partial R}{\partial h} \right)_r^2 + \frac{\partial G^2}{\partial \theta^2} \left( \frac{\partial \theta}{\partial h} \right)_r^2 + \frac{\partial G}{\partial R} \left( \frac{\partial^2 R}{\partial h^2} \right)_r + \frac{\partial G}{\partial \theta} \left( \frac{\partial^2 \theta}{\partial h^2} \right)_r + 2 \frac{\partial^2 G}{\partial R \partial \theta} \left( \frac{\partial R}{\partial h} \right)_r \left( \frac{\partial \theta}{\partial h} \right)_r \, \, , \\
\frac{\partial^2 G}{\partial r \partial h} &= \frac{\partial^2 G}{\partial R^2} \left( \frac{\partial R}{\partial h} \right)_r \left( \frac{\partial R}{\partial r} \right)_h + \frac{\partial G^2}{\partial \theta^2} \left( \frac{\partial \theta}{\partial h} \right)_r \left( \frac{\partial \theta}{\partial r} \right)_h + \frac{\partial G}{\partial R} \left( \frac{\partial^2 R}{\partial h^2} \right)_r + \\
& \quad + \frac{\partial^2 G}{\partial R \partial \theta} \left[ \left( \frac{\partial \theta}{\partial r} \right)_h \left( \frac{\partial R}{\partial h} \right)_r + \left( \frac{\partial R}{\partial r} \right)_h \left( \frac{\partial \theta}{\partial h} \right)_r \right] + \\
& \quad + \frac{\partial G}{\partial R} \left[ \left( \frac{\partial \theta}{\partial r} \right)_h \frac{\partial}{\partial \theta} + \left( \frac{\partial R}{\partial r} \right)_h \frac{\partial}{\partial R} \right] \left( \frac{\partial R}{\partial h} \right)_r + \frac{\partial G}{\partial \theta} \left[ \left( \frac{\partial \theta}{\partial r} \right)_h \frac{\partial}{\partial \theta} + \left( \frac{\partial R}{\partial r} \right)_h \frac{\partial}{\partial R} \right] \left( \frac{\partial \theta}{\partial h} \right)_r \, \, ,
\end{align*}

\noindent
Derivatives of $(R,\theta)$ wrt $(r,h)$:
\begin{align*}
\left( \frac{\partial^2 R}{\partial r^2} \right)_h &= - \left( \frac{\partial^2 r}{\partial R^2} \right)_h \left( \frac{\partial R}{\partial r} \right)^3 \, \, , & \left( \frac{\partial^2 R}{\partial h^2} \right)_r &= - \left( \frac{\partial^2 h}{\partial R^2} \right)_r \left( \frac{\partial R}{\partial h} \right)^3 \, \, , \\
\left( \frac{\partial^2 \theta}{\partial r^2} \right)_h &= - \left( \frac{\partial^2 r}{\partial \theta^2} \right)_h \left( \frac{\partial \theta}{\partial r} \right)^3 \, \, , & \left( \frac{\partial^2 \theta}{\partial h^2} \right)_r &= - \left( \frac{\partial^2 h}{\partial \theta^2} \right)_r \left( \frac{\partial \theta}{\partial h} \right)^3 \, \, ,
\end{align*}
where the expression for the derivatives of $(r,h)$ wrt $(R,\theta)$ are already quite long.

\section*{Appendix B}
In this appendix we present a few more results for the EoS, obtained with different parameter choices in the allowed ranges discussed in the main text. We mainly aim at exploring the effect of a different size for the critical region and a different location for the critical point.
\begin{figure}
\center
\includegraphics[width=.4\textwidth]{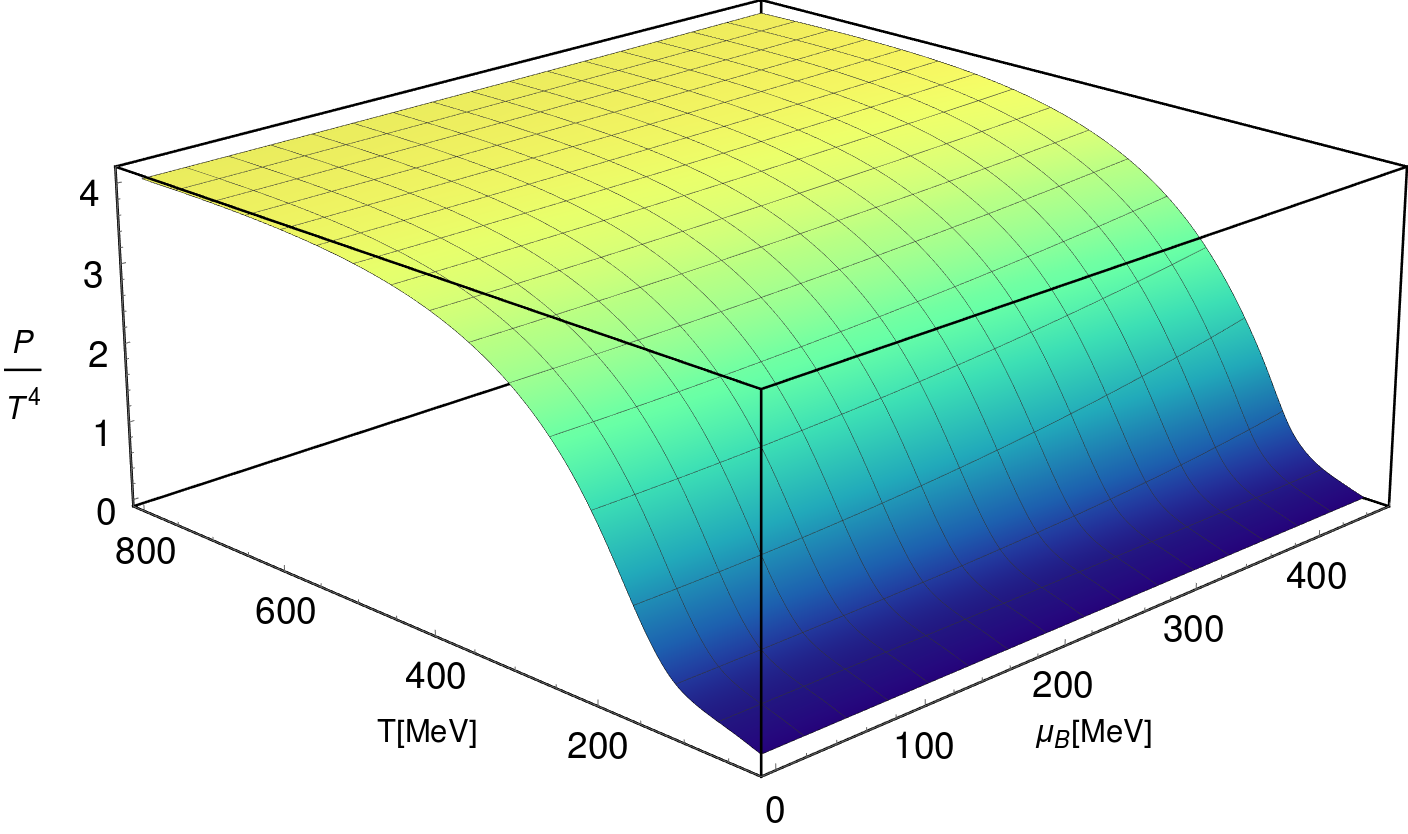}
\includegraphics[width=.4\textwidth]{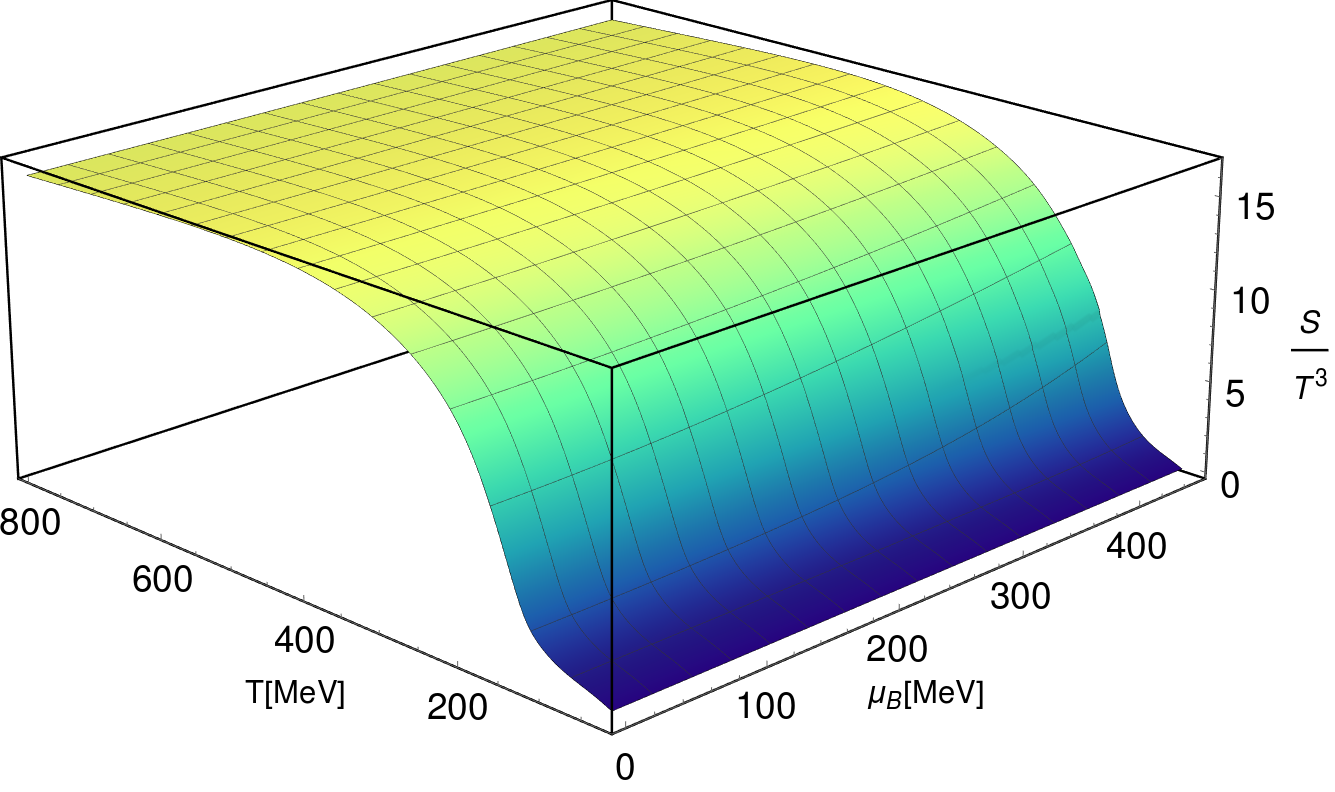}
\includegraphics[width=.4\textwidth]{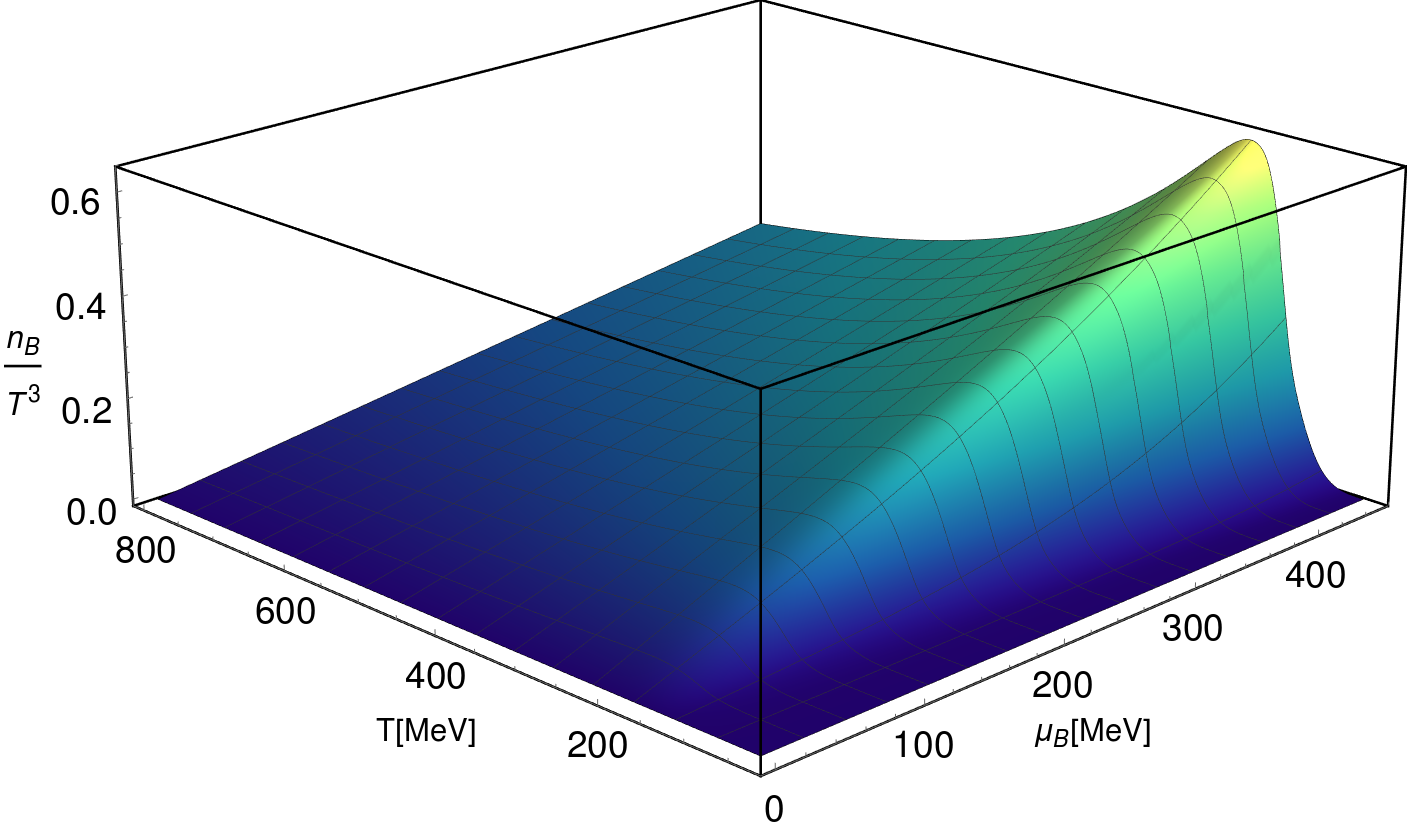}
\includegraphics[width=.4\textwidth]{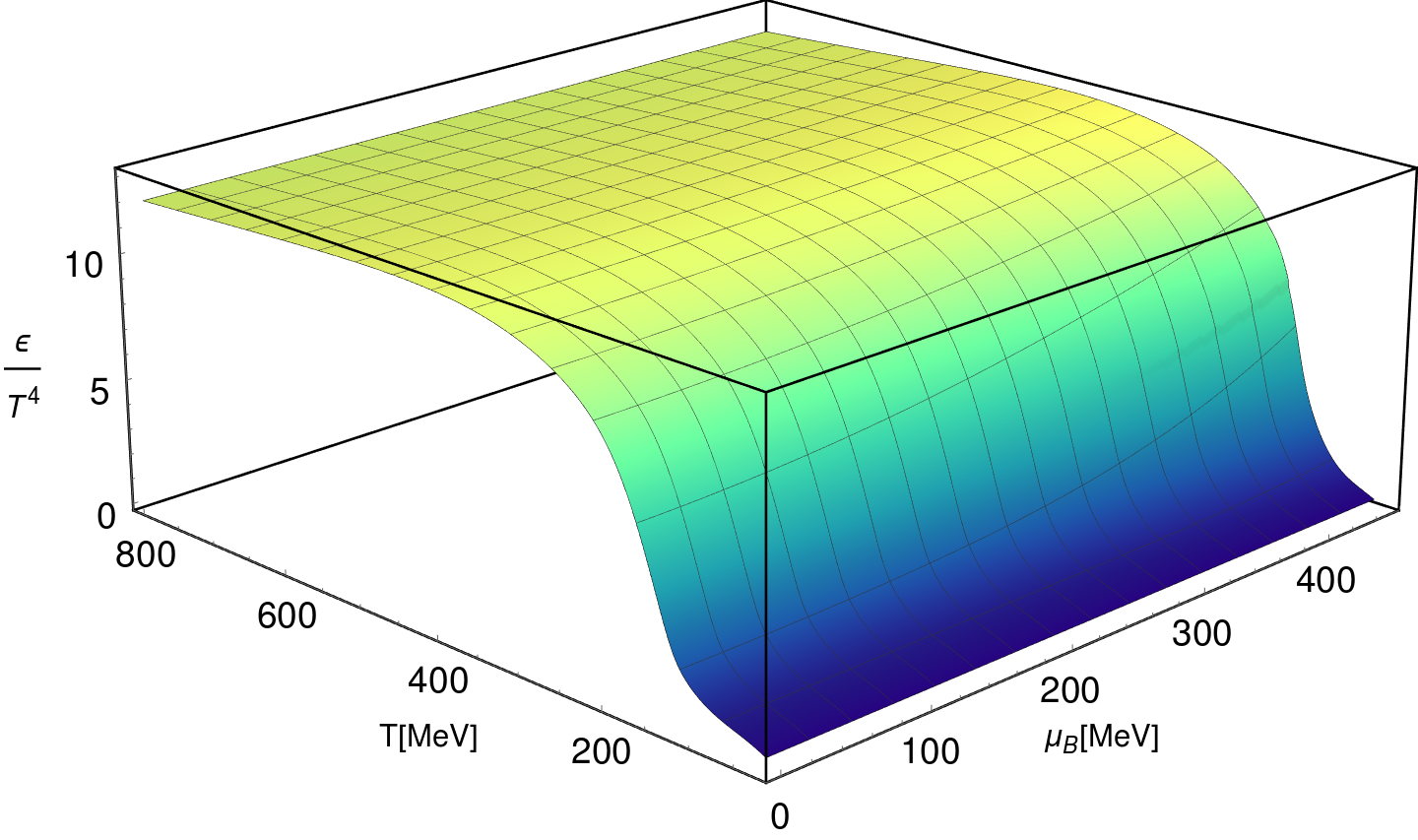}
\includegraphics[width=.4\textwidth]{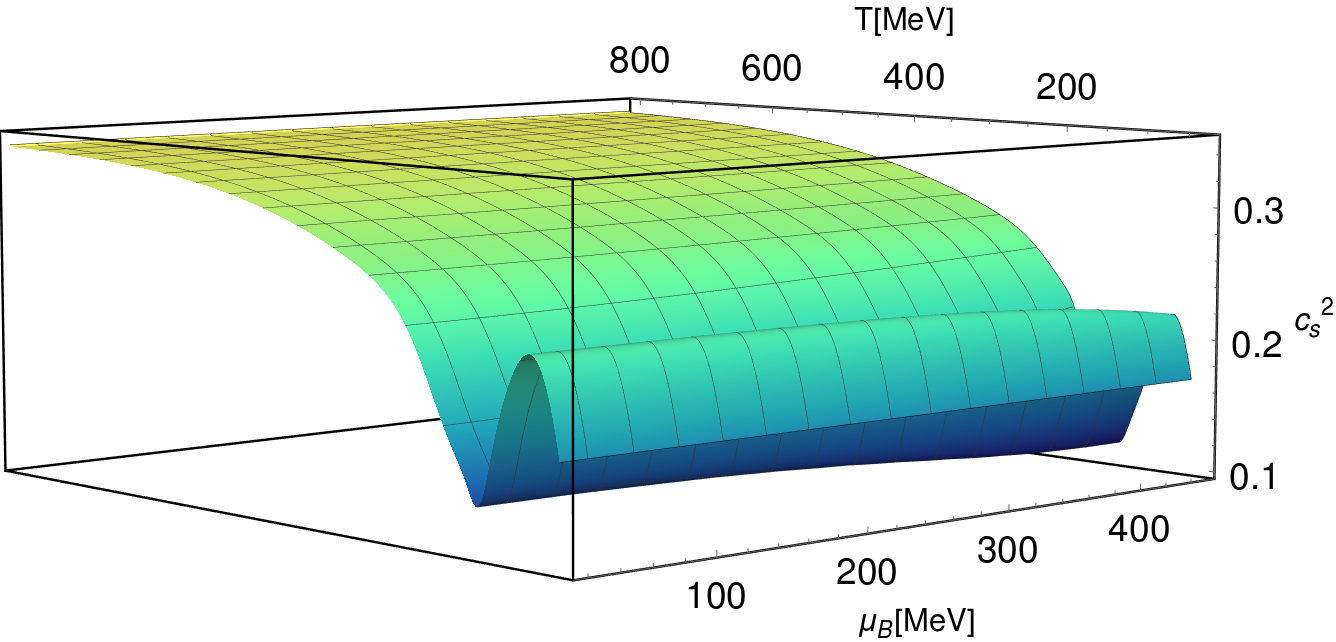}
\includegraphics[width=.4\textwidth]{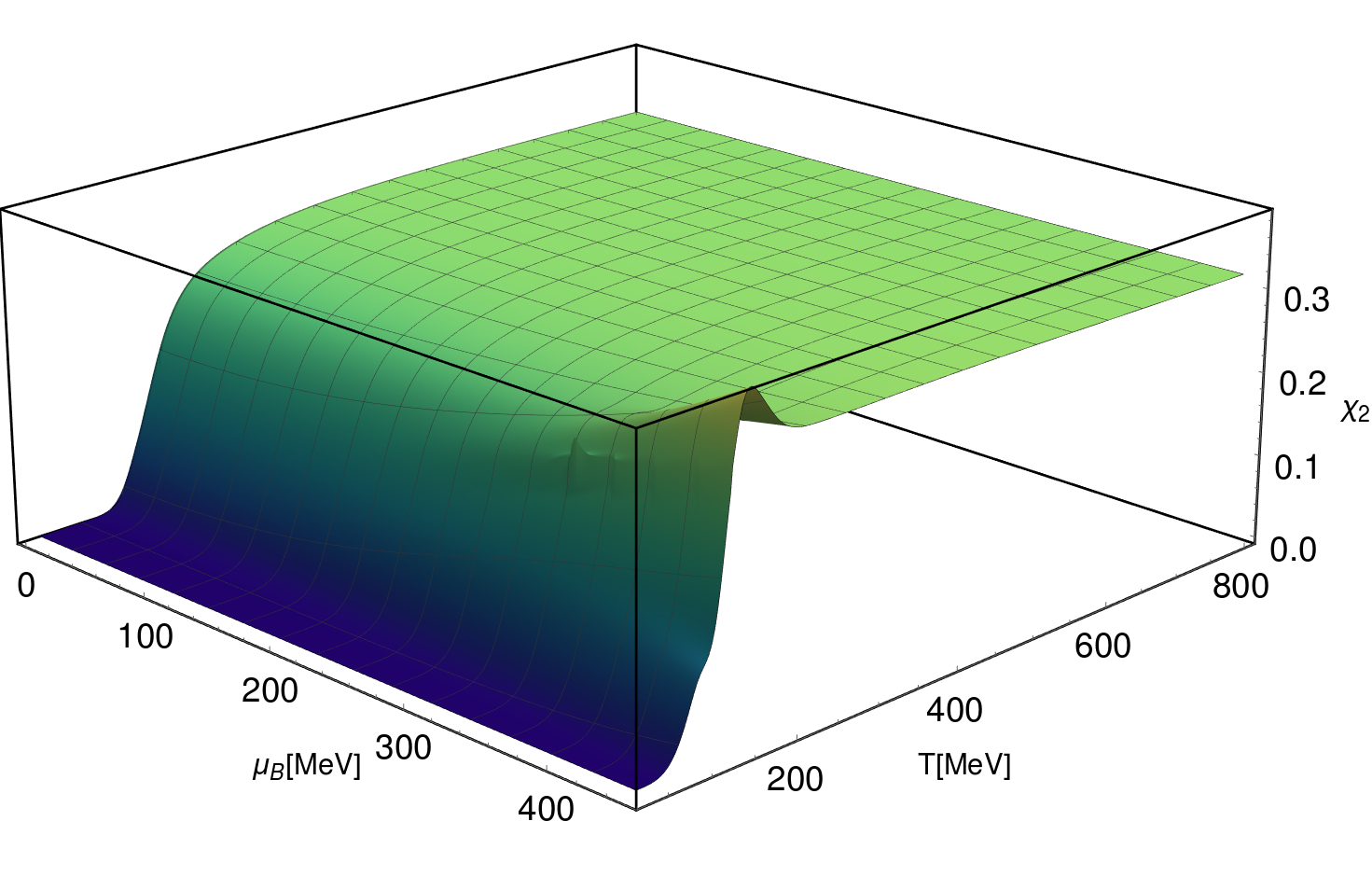}
\caption{Upper panels: pressure (left) and entropy density (right); middle panels: baryonic density (left) and energy density (right); lower panels: speed of sound (left) and $\chi_2$ (right) as functions of $T$ and $\mu_B$. All panels correspond to the same location of the CP and angles presented in the main text, but with $w=4$ and $\rho=1$.}
\label{FigAppB1}
\end{figure}
Figure \ref{FigAppB1} shows our EoS for the same location of the CP and angles presented in the main text, but for $w=4$ and $\rho=1$. This choice was made to show that the size of the critical region can be made arbitrarily small by increasing $w$. This is evident by the fact that the discontinuity in entropy, energy and baryonic density is much reduced in this case, as well as the dip in the speed of sound and peak in $\chi_2$.
\begin{figure}
\center
\includegraphics[width=.4\textwidth]{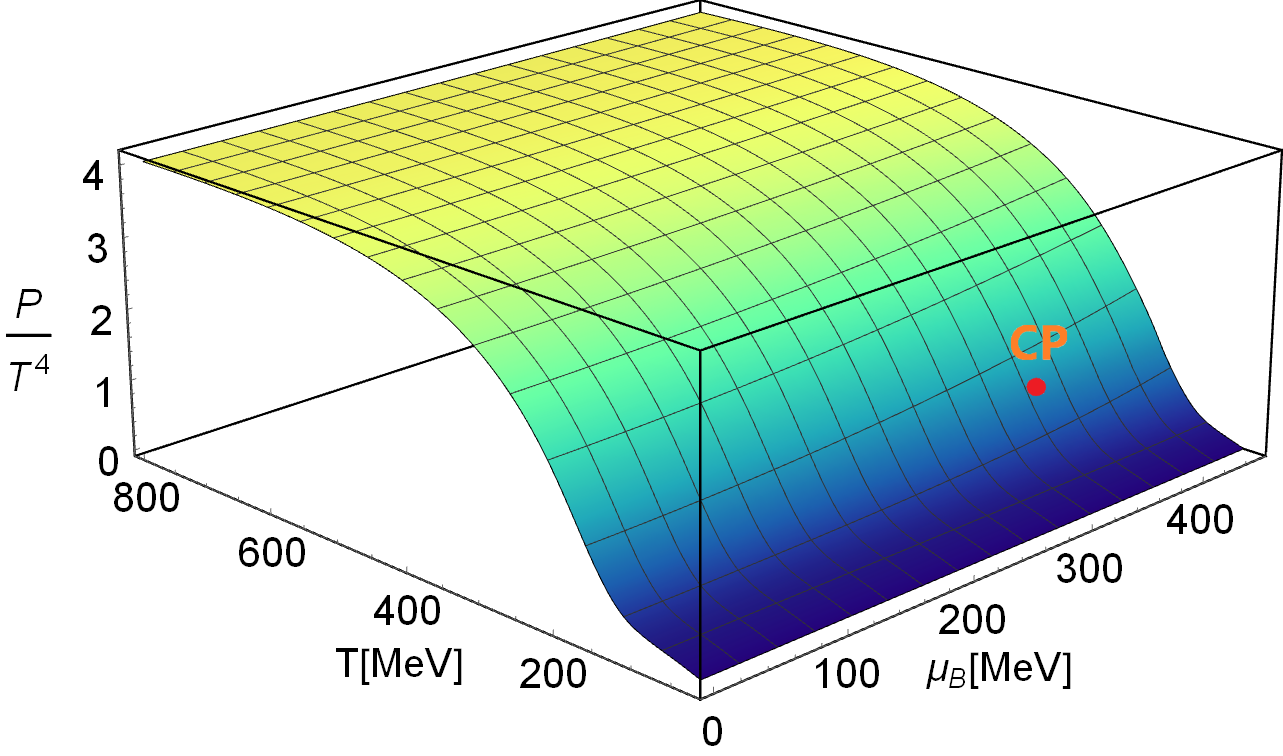}
\includegraphics[width=.4\textwidth]{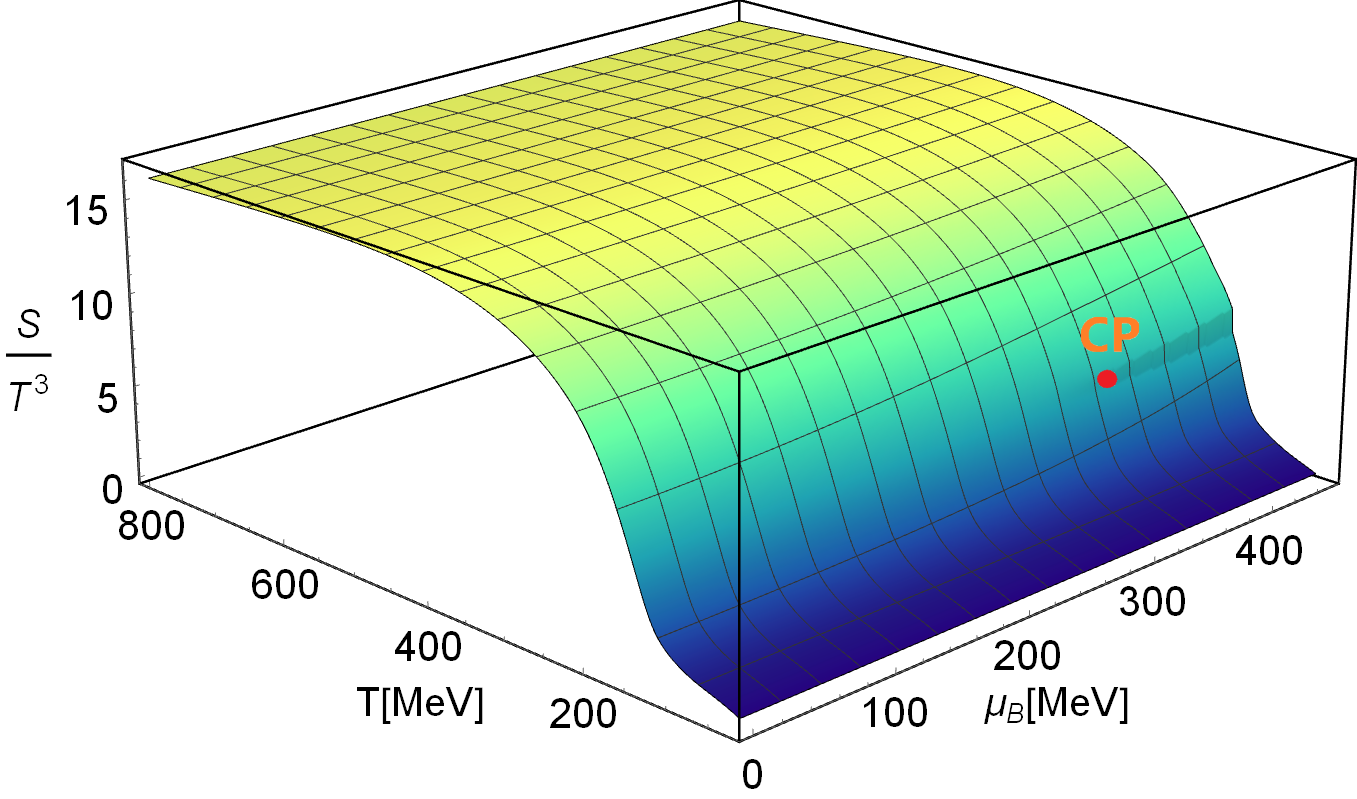}
\includegraphics[width=.4\textwidth]{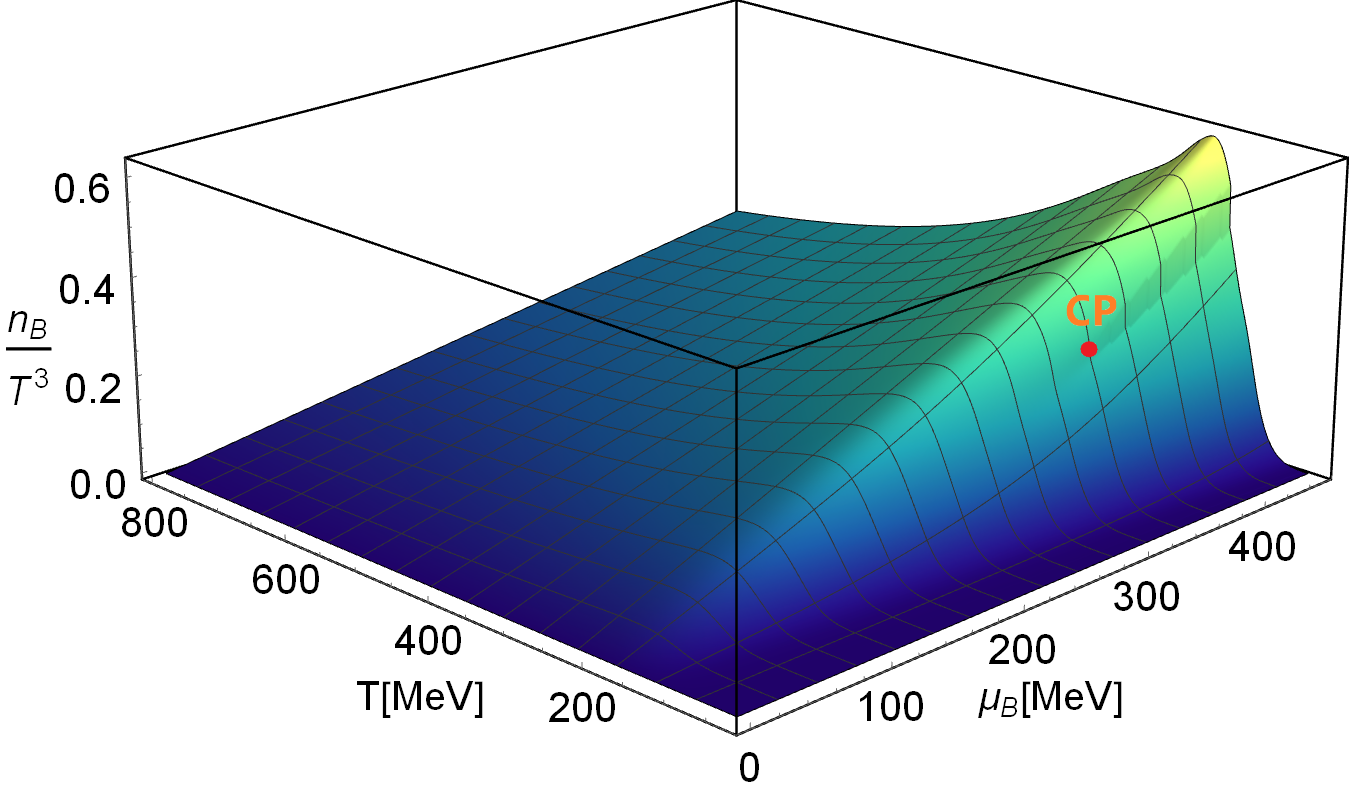}
\includegraphics[width=.4\textwidth]{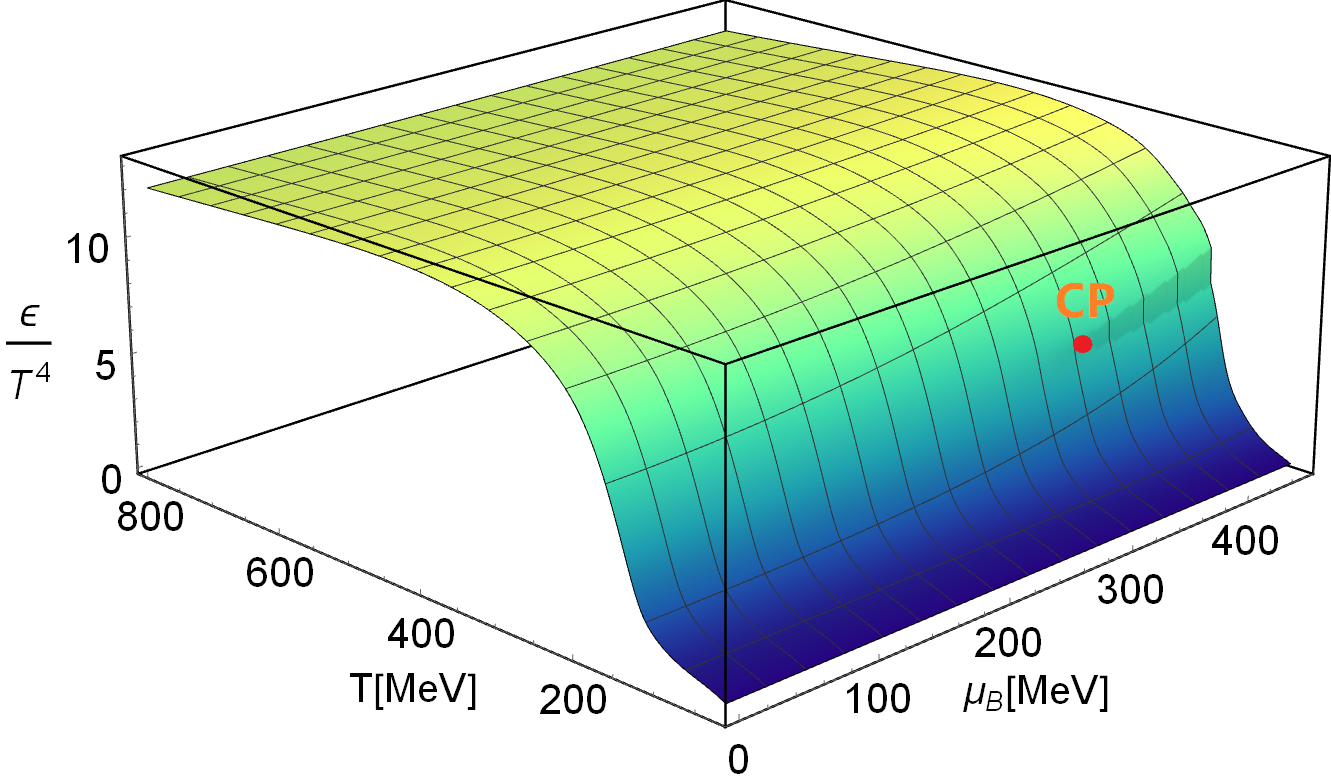}
\includegraphics[width=.4\textwidth]{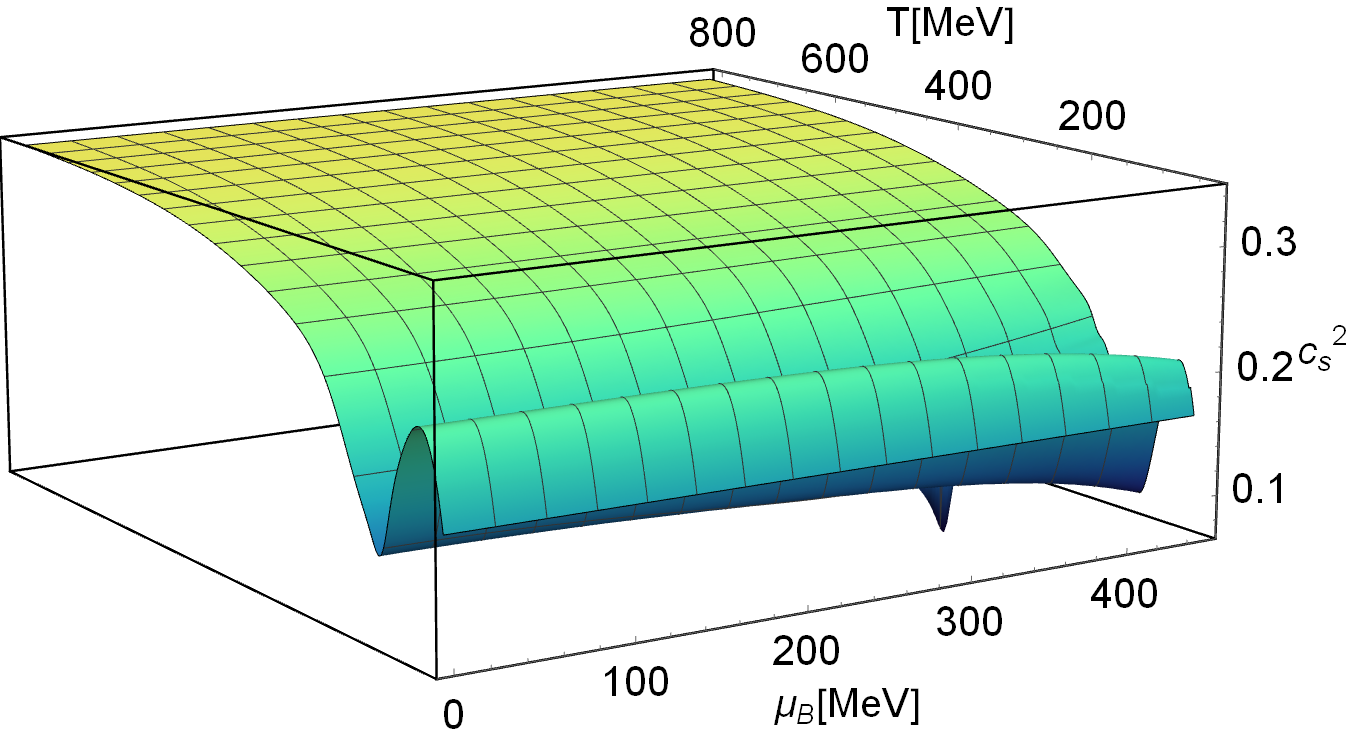}
\includegraphics[width=.4\textwidth]{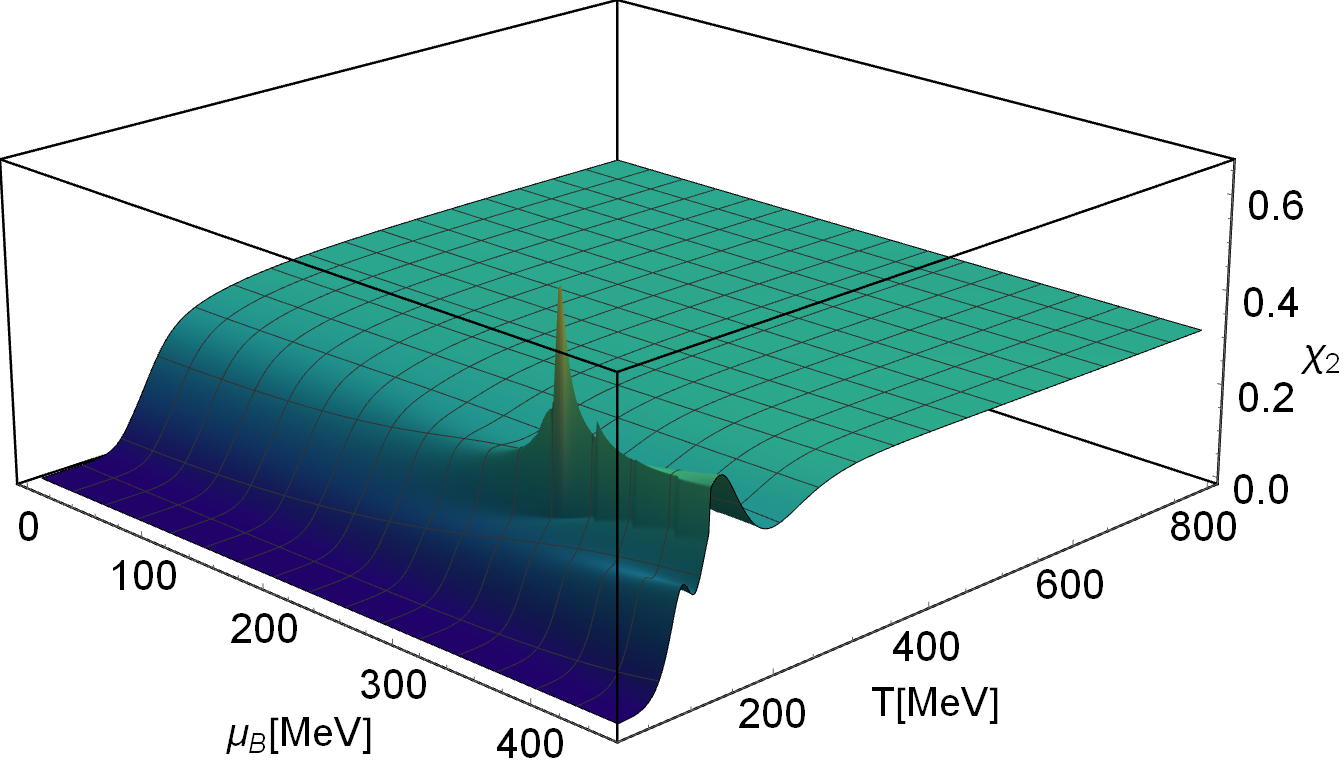}
\caption{Upper panels: pressure (left) and entropy density (right); middle panels: baryonic density (left) and energy density (right); lower panels: speed of sound (left) and $\chi_2$ (right) as functions of $T$ and $\mu_B$. All panels correspond to the same location of the CP and angles presented in the main text, but with $w=0.75$ and $\rho=2$.}
\label{FigAppB2}
\end{figure}

Figure \ref{FigAppB2} shows our EoS for the same location of the CP and angles presented in the main text, but for $w=0.75$ and $\rho=2$. With this choice of parameters, in particular with a smaller $w$, the size of the critical region and the effect of the critical point become larger. This is evident by the fact that the discontinuity in entropy, energy and baryonic density is larger in this case, as well as the dip in the speed of sound and peak in $\chi_2$.

\begin{figure}
\center
\includegraphics[width=.4\textwidth]{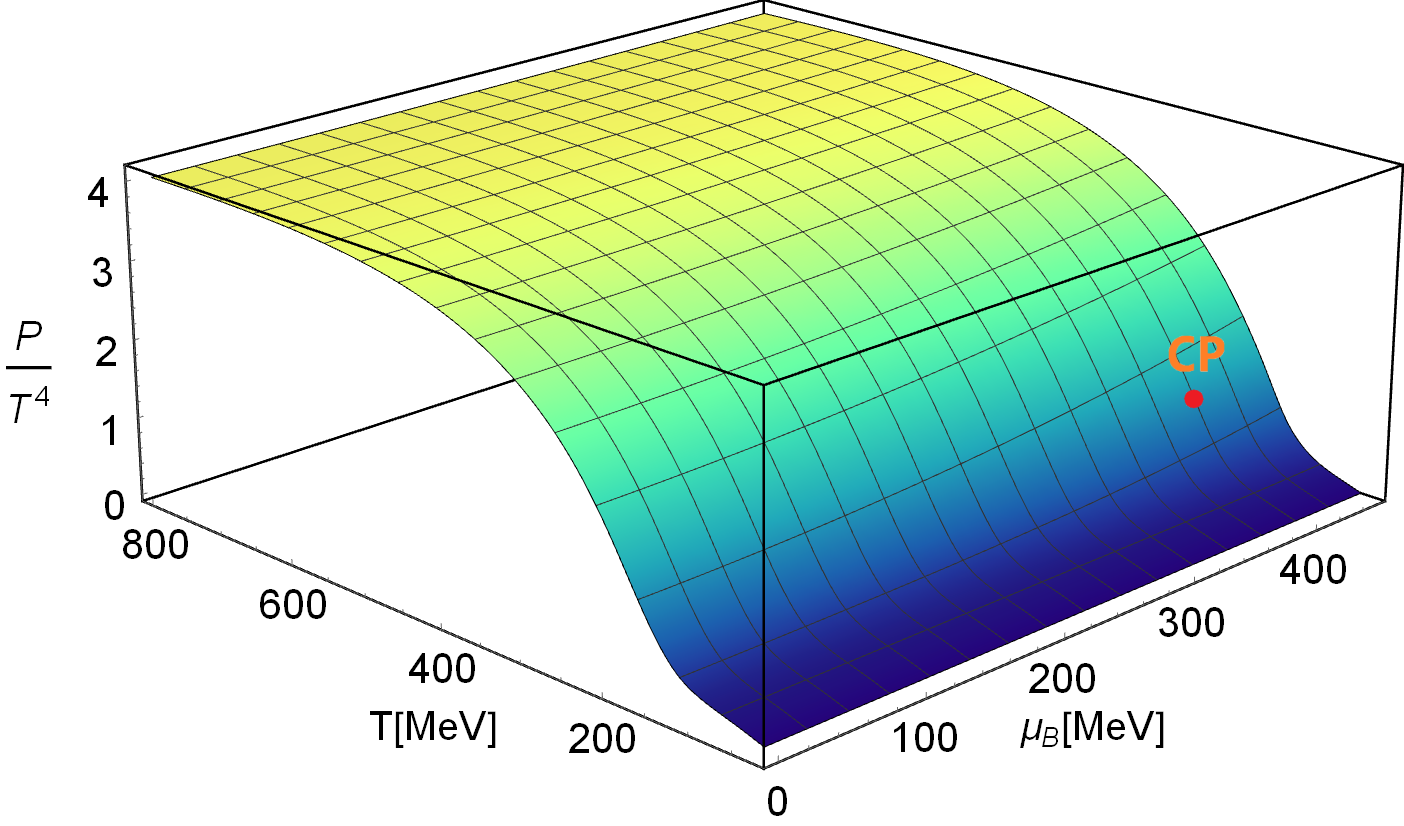}
\includegraphics[width=.4\textwidth]{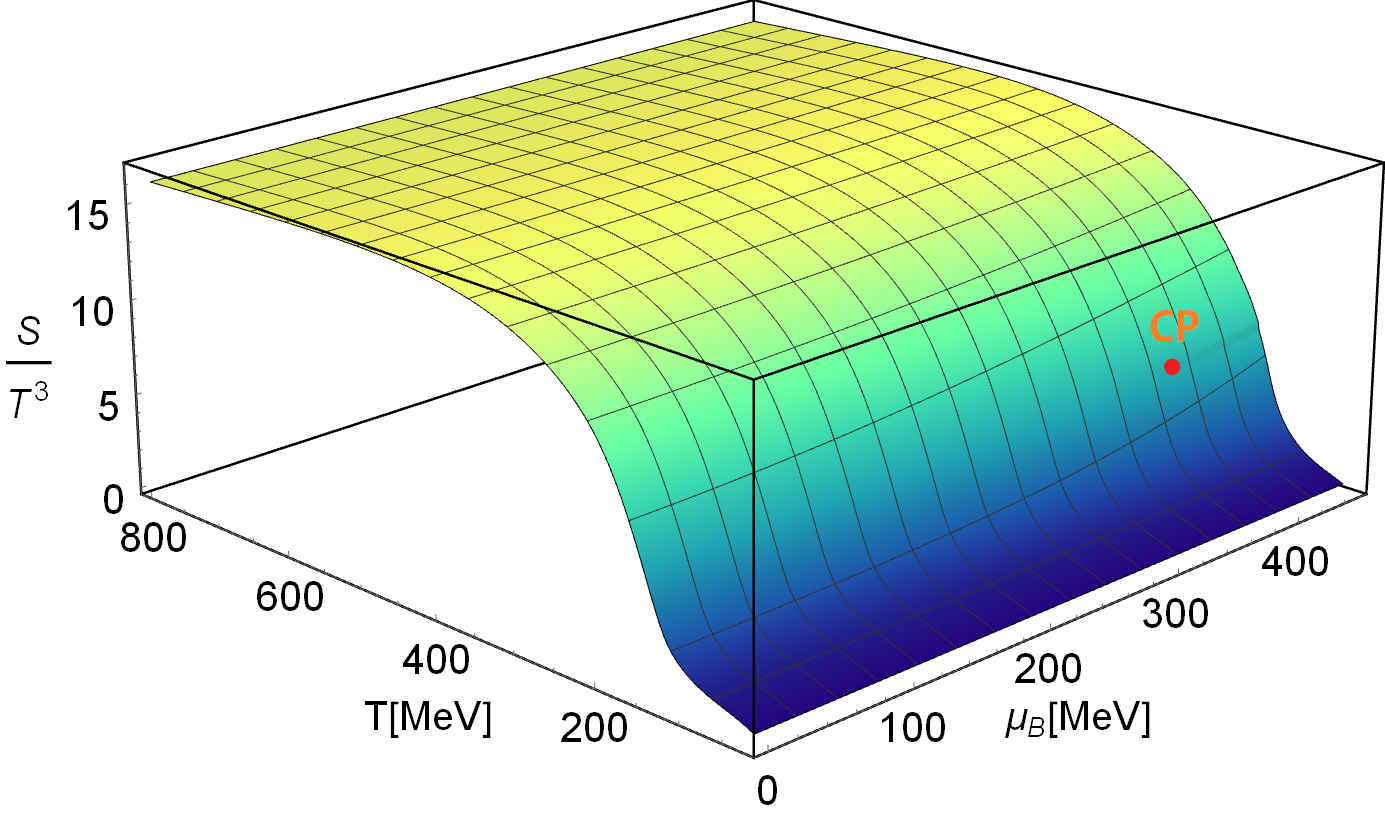}
\includegraphics[width=.4\textwidth]{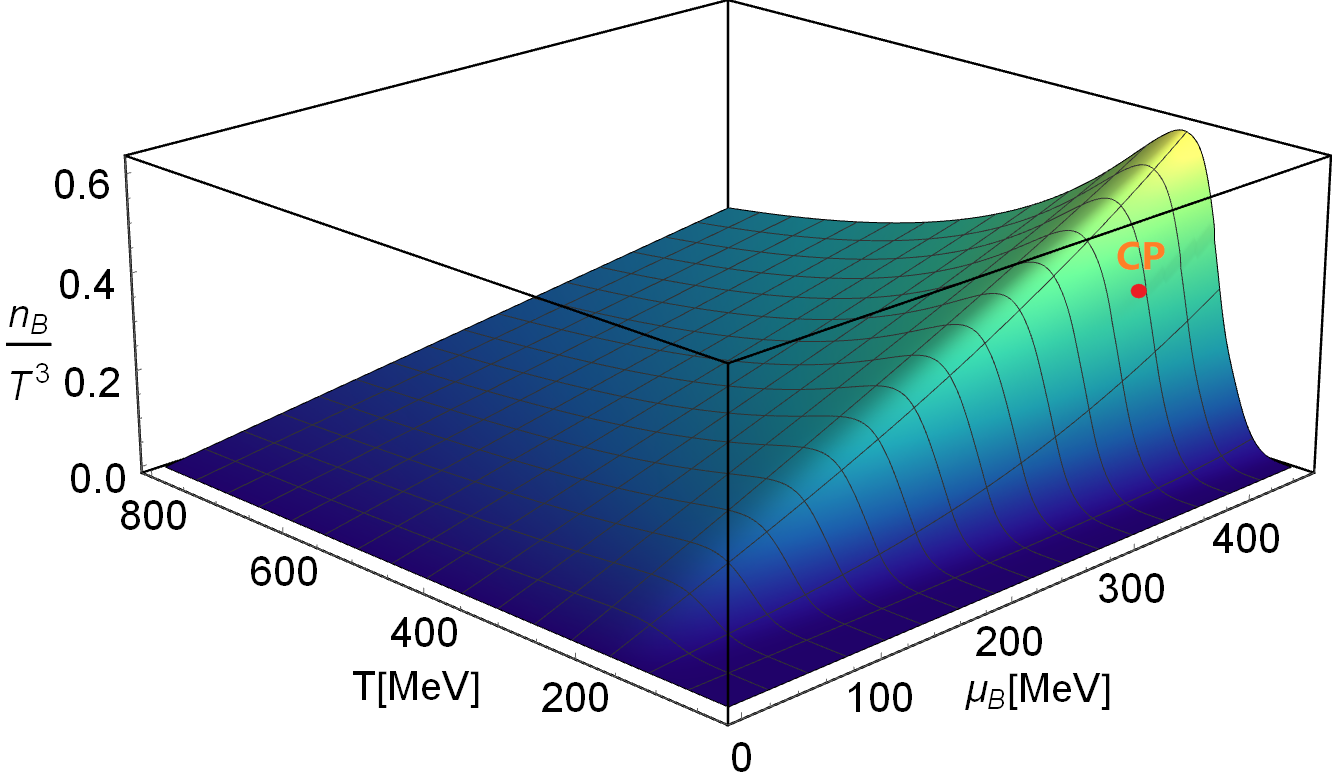}
\includegraphics[width=.4\textwidth]{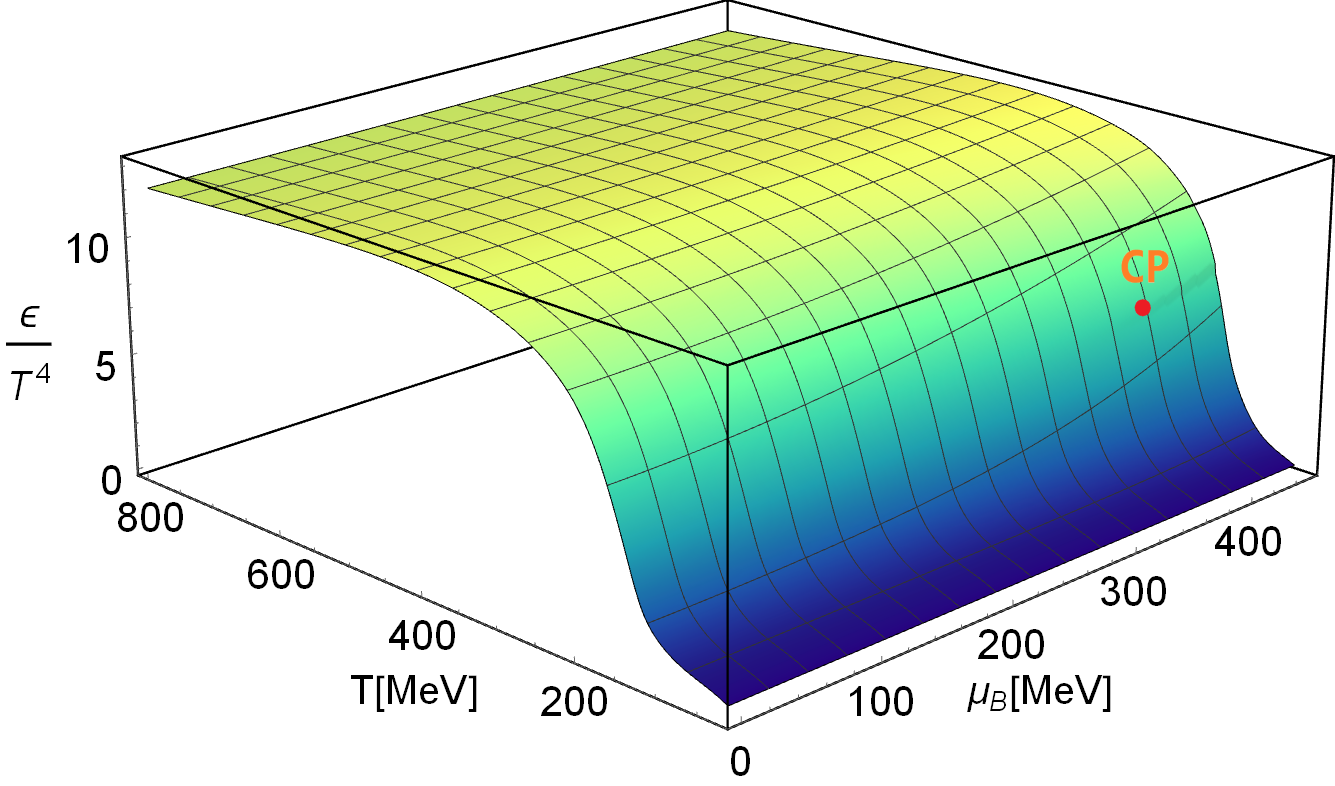}
\includegraphics[width=.4\textwidth]{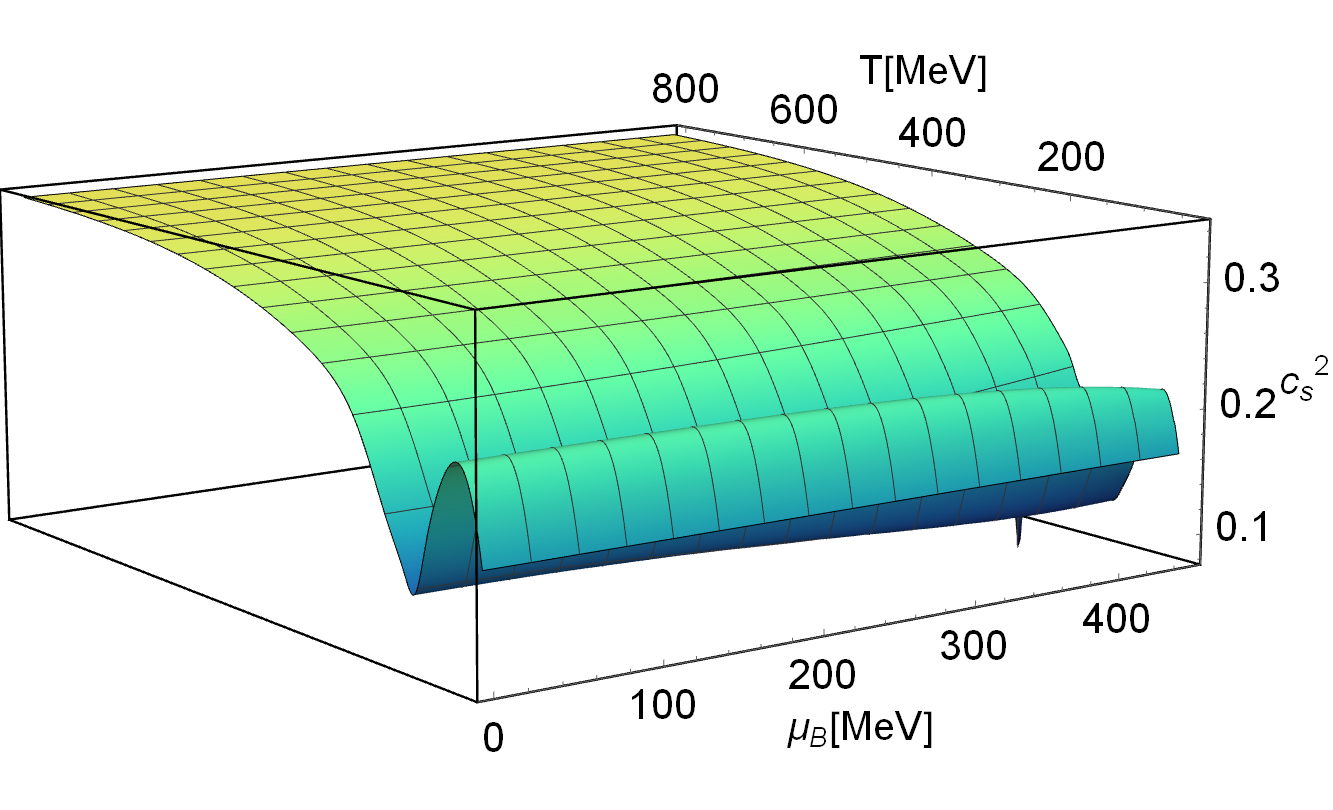}
\includegraphics[width=.4\textwidth]{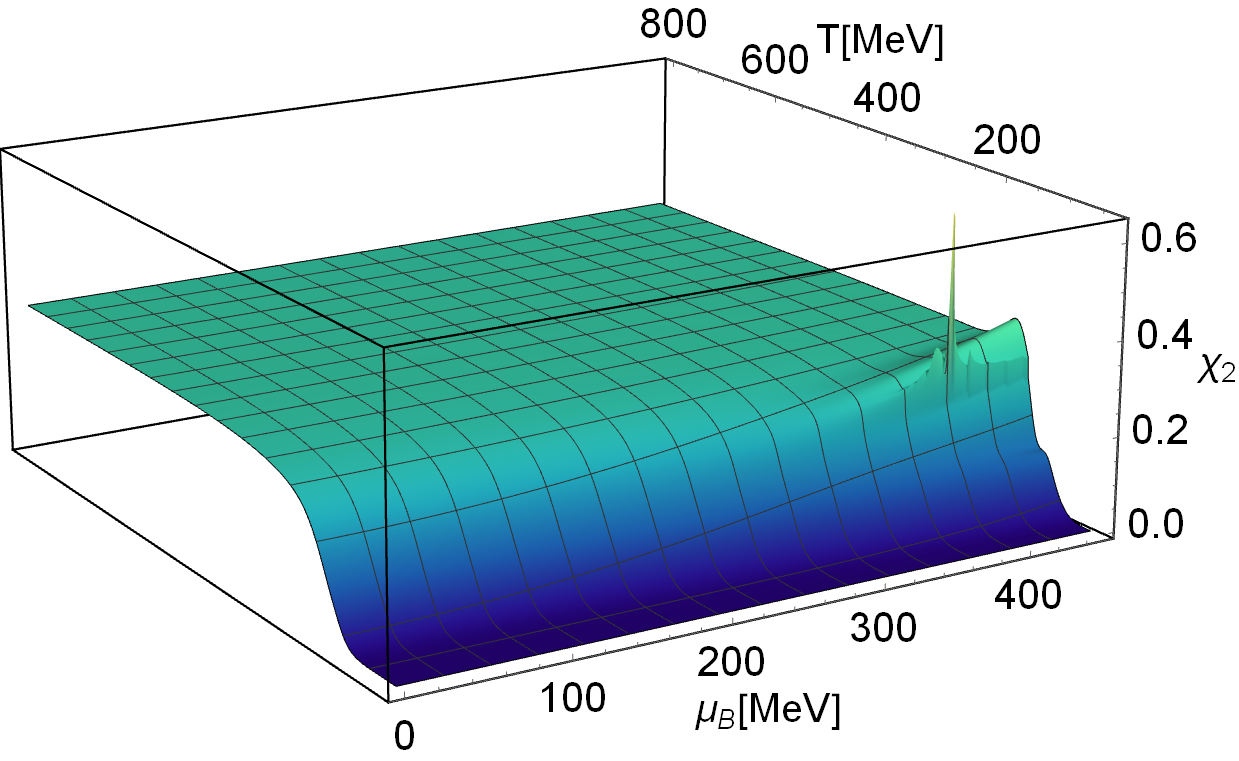}
\caption{Upper panels: pressure (left) and entropy density (right); middle panels: baryonic density (left) and energy density (right); lower panels: speed of sound (left) and $\chi_2$ (right) as functions of $T$ and $\mu_B$. All panels correspond to $\mu_{BC}=400$ MeV, $\alpha_1\simeq4.40^\circ$, $\alpha_2-\alpha_1=90^\circ$ and $w=2$, $\rho=2$.}
\label{FigAppB3}
\end{figure}

In Figure \ref{FigAppB3} we change the location of the CP to $\mu_{BC}$=400 MeV, the angles become $\alpha_1\simeq4.40^\circ$, $\alpha_2-\alpha_1=90^\circ$, and we set $w=2$ and $\rho=2$. 
When the data from the second beam energy scan at RHIC become available, it will be possible to perform a Bayesian analysis to systematically scan the parameter region (in a similar fashion as presented in this appendix, but for many more parameter sets) and test the results of hydrodynamic simulations with this family of EoSs as input against the data. This will hopefully help to constrain the size of the critical region, and the location of the critical point.
\section*{Appendix C}
\subsection{Treatment of the phase coexistence region}

The first order phase transition appearing at chemical potentials larger than the critical one, leads to a phase coexistence region in the plane of pressure vs net-baryon density. The knowledge of the EoS in such a region, even though it does not correspond to a stable state of the system, can be needed in hydrodynamic simulations \cite{Randrup:2009gp,Randrup:2010ax,Steinheimer:2012gc,Steinheimer:2013gla,Steinheimer:2013xxa}. 

\begin{figure}
    \centering
    \includegraphics[width=0.6\textwidth]{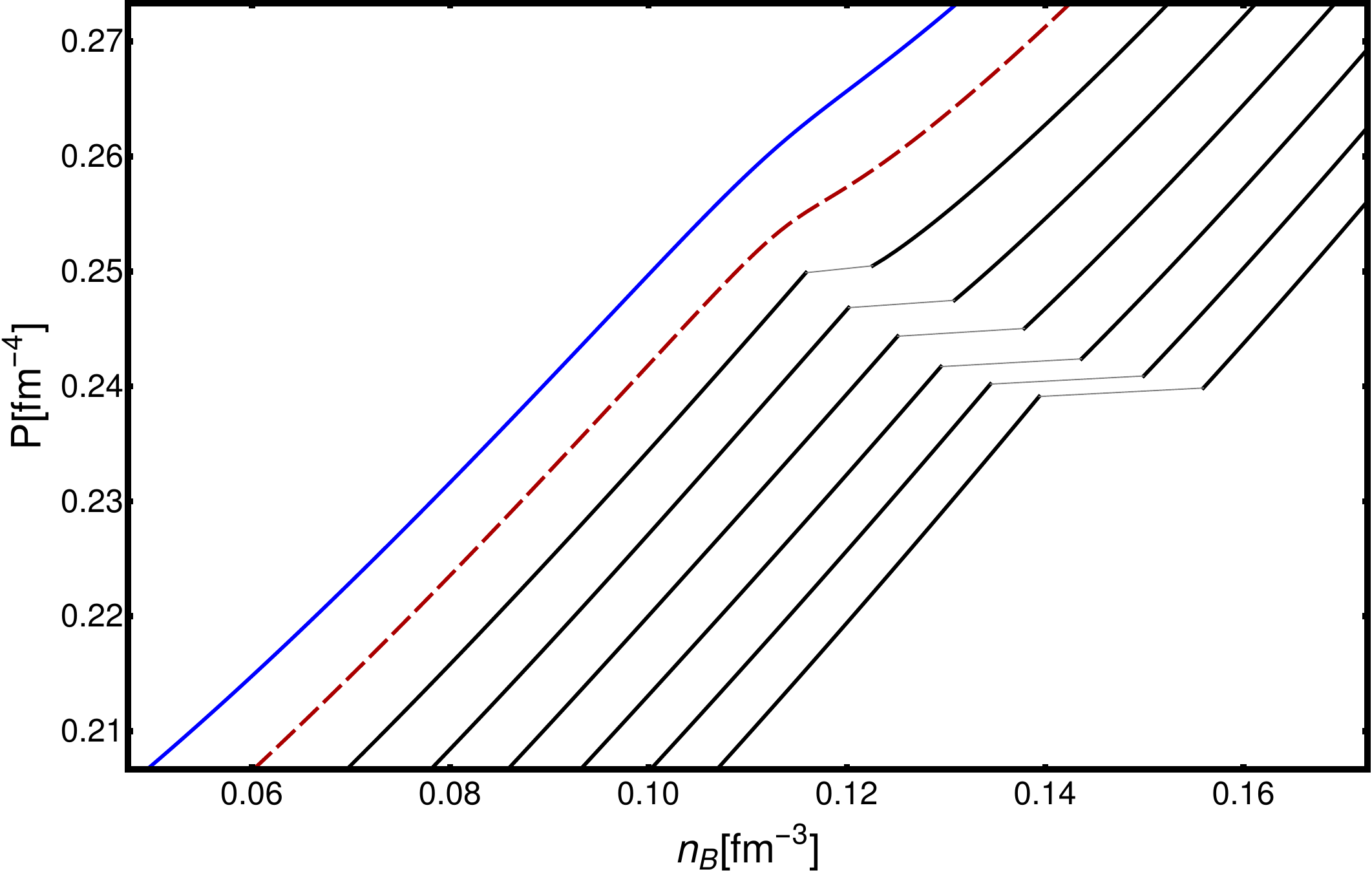}
    \caption{Isotherms in the plane of pressure vs baryon density for $T = 145~ \MeV$ (blue, solid line), $T = 144~ \MeV$ (red, dashed line) and $T = 143 - 138 ~\MeV$ (solid, black lines), for the choice of parameters in Section \ref{sec:paramRed}.}
    \label{fig:isotherms}
\end{figure}

Due to the fact that our treatment is carried out in the temperature vs chemical potential plane, our EoS does not contain information about the coexistence region. In order to use the standard Maxwell construction, we would need a thermodynamic potential defined even in the un-physical region, but this is unfortunately unavailable in our approach. Thus, we take a phenomenological approach to reconstruct the EoS in this region. In Fig. \ref{fig:isotherms} the isothermal curves in the pressure vs baryon density plane are shown, with the same choice of parameters as in Section \ref{sec:paramRed}: the solid blue line corresponds to a temperature $T>T_C$ and shows only a slight inflection, while the solid black lines correspond to temperatures $T<T_C$, and display a clear discontinuity; the dashed red line corresponds to $T \simeq T_C$, and shows the insurgence of the discontinuous behavior.

\begin{figure}
    \centering
    \includegraphics[width=\textwidth]{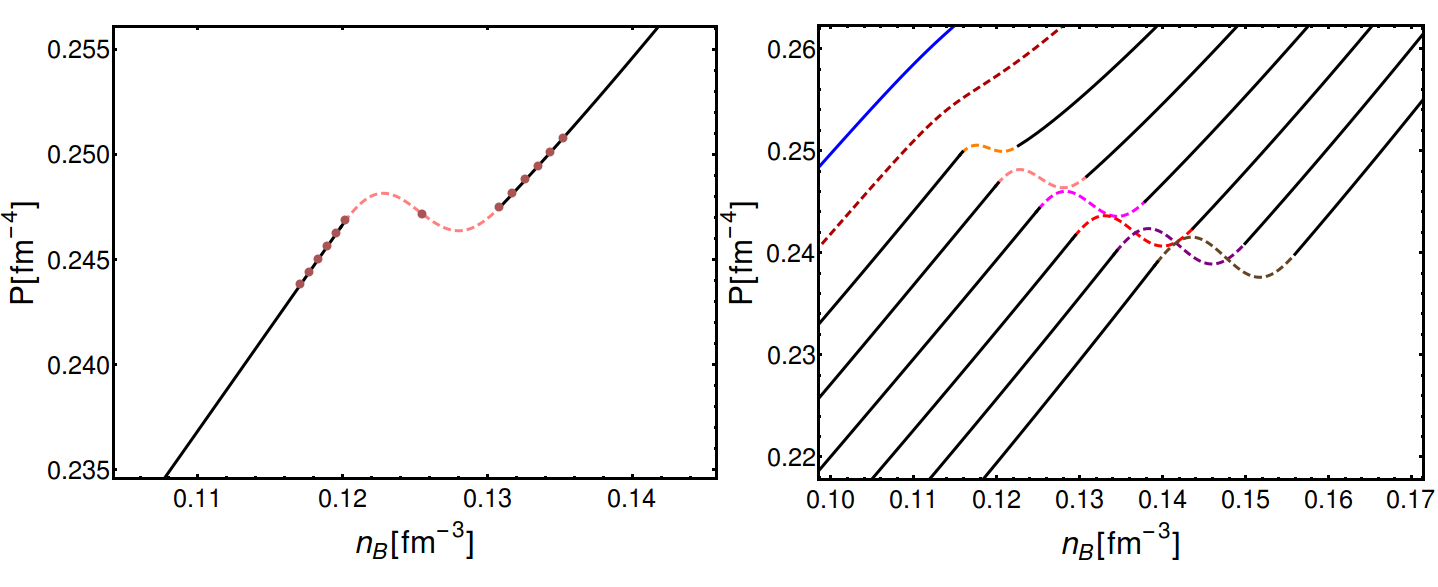}
    \caption{Left panel: the isotherm at $T = 142~ \MeV$ is shown (black solid) along with the points (dark pink) utilized to generate the curve in the coexistence region (pink dashed). Right panel: the same isotherms as in Fig. \ref{fig:isotherms}, with the modeled dependence of the pressure on the baryon density in the phase coexistence region.}
    \label{fig:isotherms_fit}
\end{figure}

Since the thermodynamic conditions for the coexistence of two phases (thermal, chemical and mechanical equilibrium) are all fulfilled by construction in our procedure, what is needed is a continuous curve connecting the two branches of each discontinuous isotherm. The phenomenological approach we follow here is to construct a form that is similar to many other analytic theories (e.g. real gases) and presents the typical double lobe structure. We perform a polynomial fit to the two branches of each isotherm and the middle point in the discontinuity. In the left panel of Fig. \ref{fig:isotherms_fit} the original, discontinuous isotherm at $T = 142 \MeV$ is shown (black solid), along with the points used to generate the curve in the coexistence region (dark pink) and the resulting function \textit{inside} the coexistence region (pink dashed).

The results of this treatment are shown in the right panel of Fig. \ref{fig:isotherms_fit}: the gap in the $(P,n_B)$ phase diagram corresponding to the coexistence region between two phases is now filled by a phenomenological dependence of $P(n_B)$ along the isotherms. This will allow one to use our EoS in hydrodynamic codes inside the phase coexistence region emerging from the presence of a first order chiral/deconfiment phase transition.
\clearpage
\bibliography{thermo}

\begin{thebibliography}{83}%
\makeatletter
\providecommand \@ifxundefined [1]{%
 \@ifx{#1\undefined}
}%
\providecommand \@ifnum [1]{%
 \ifnum #1\expandafter \@firstoftwo
 \else \expandafter \@secondoftwo
 \fi
}%
\providecommand \@ifx [1]{%
 \ifx #1\expandafter \@firstoftwo
 \else \expandafter \@secondoftwo
 \fi
}%
\providecommand \natexlab [1]{#1}%
\providecommand \enquote  [1]{``#1''}%
\providecommand \bibnamefont  [1]{#1}%
\providecommand \bibfnamefont [1]{#1}%
\providecommand \citenamefont [1]{#1}%
\providecommand \href@noop [0]{\@secondoftwo}%
\providecommand \href [0]{\begingroup \@sanitize@url \@href}%
\providecommand \@href[1]{\@@startlink{#1}\@@href}%
\providecommand \@@href[1]{\endgroup#1\@@endlink}%
\providecommand \@sanitize@url [0]{\catcode `\\12\catcode `\$12\catcode
  `\&12\catcode `\#12\catcode `\^12\catcode `\_12\catcode `\%12\relax}%
\providecommand \@@startlink[1]{}%
\providecommand \@@endlink[0]{}%
\providecommand \url  [0]{\begingroup\@sanitize@url \@url }%
\providecommand \@url [1]{\endgroup\@href {#1}{\urlprefix }}%
\providecommand \urlprefix  [0]{URL }%
\providecommand \Eprint [0]{\href }%
\providecommand \doibase [0]{http://dx.doi.org/}%
\providecommand \selectlanguage [0]{\@gobble}%
\providecommand \bibinfo  [0]{\@secondoftwo}%
\providecommand \bibfield  [0]{\@secondoftwo}%
\providecommand \translation [1]{[#1]}%
\providecommand \BibitemOpen [0]{}%
\providecommand \bibitemStop [0]{}%
\providecommand \bibitemNoStop [0]{.\EOS\space}%
\providecommand \EOS [0]{\spacefactor3000\relax}%
\providecommand \BibitemShut  [1]{\csname bibitem#1\endcsname}%
\let\auto@bib@innerbib\@empty
\bibitem [{\citenamefont {Aoki}\ \emph {et~al.}(2006)\citenamefont {Aoki},
  \citenamefont {Endrodi}, \citenamefont {Fodor}, \citenamefont {Katz},\ and\
  \citenamefont {Szabo}}]{Aoki:2006we}%
  \BibitemOpen
  \bibfield  {author} {\bibinfo {author} {\bibfnamefont {Y.}~\bibnamefont
  {Aoki}}, \bibinfo {author} {\bibfnamefont {G.}~\bibnamefont {Endrodi}},
  \bibinfo {author} {\bibfnamefont {Z.}~\bibnamefont {Fodor}}, \bibinfo
  {author} {\bibfnamefont {S.~D.}\ \bibnamefont {Katz}}, \ and\ \bibinfo
  {author} {\bibfnamefont {K.~K.}\ \bibnamefont {Szabo}},\ }\href {\doibase
  10.1038/nature05120} {\bibfield  {journal} {\bibinfo  {journal} {Nature}\
  }\textbf {\bibinfo {volume} {443}},\ \bibinfo {pages} {675} (\bibinfo {year}
  {2006})},\ \Eprint {http://arxiv.org/abs/hep-lat/0611014}
  {arXiv:hep-lat/0611014 [hep-lat]} \BibitemShut {NoStop}%
\bibitem [{\citenamefont {Luo}\ and\ \citenamefont {Xu}(2017)}]{Luo:2017faz}%
  \BibitemOpen
  \bibfield  {author} {\bibinfo {author} {\bibfnamefont {X.}~\bibnamefont
  {Luo}}\ and\ \bibinfo {author} {\bibfnamefont {N.}~\bibnamefont {Xu}},\
  }\href {\doibase 10.1007/s41365-017-0257-0} {\bibfield  {journal} {\bibinfo
  {journal} {Nucl. Sci. Tech.}\ }\textbf {\bibinfo {volume} {28}},\ \bibinfo
  {pages} {112} (\bibinfo {year} {2017})},\ \Eprint
  {http://arxiv.org/abs/1701.02105} {arXiv:1701.02105 [nucl-ex]} \BibitemShut
  {NoStop}%
\bibitem [{\citenamefont {Busza}\ \emph {et~al.}(2018)\citenamefont {Busza},
  \citenamefont {Rajagopal},\ and\ \citenamefont {van~der
  Schee}}]{Busza:2018rrf}%
  \BibitemOpen
  \bibfield  {author} {\bibinfo {author} {\bibfnamefont {W.}~\bibnamefont
  {Busza}}, \bibinfo {author} {\bibfnamefont {K.}~\bibnamefont {Rajagopal}}, \
  and\ \bibinfo {author} {\bibfnamefont {W.}~\bibnamefont {van~der Schee}},\
  }\href@noop {} {\  (\bibinfo {year} {2018})},\ \Eprint
  {http://arxiv.org/abs/1802.04801} {arXiv:1802.04801 [hep-ph]} \BibitemShut
  {NoStop}%
\bibitem [{\citenamefont {Stephanov}(2006)}]{Stephanov:2007fk}%
  \BibitemOpen
  \bibfield  {author} {\bibinfo {author} {\bibfnamefont {M.~A.}\ \bibnamefont
  {Stephanov}},\ }\bibfield  {booktitle} {\emph {\bibinfo {booktitle}
  {{Proceedings, 24th International Symposium on Lattice Field Theory (Lattice
  2006): Tucson, USA, July 23-28, 2006}}},\ }\href@noop {} {\bibfield
  {journal} {\bibinfo  {journal} {PoS}\ }\textbf {\bibinfo {volume}
  {LAT2006}},\ \bibinfo {pages} {024} (\bibinfo {year} {2006})},\ \Eprint
  {http://arxiv.org/abs/hep-lat/0701002} {arXiv:hep-lat/0701002 [hep-lat]}
  \BibitemShut {NoStop}%
\bibitem [{\citenamefont {Stephanov}\ \emph {et~al.}(1998)\citenamefont
  {Stephanov}, \citenamefont {Rajagopal},\ and\ \citenamefont
  {Shuryak}}]{Stephanov:1998dy}%
  \BibitemOpen
  \bibfield  {author} {\bibinfo {author} {\bibfnamefont {M.~A.}\ \bibnamefont
  {Stephanov}}, \bibinfo {author} {\bibfnamefont {K.}~\bibnamefont
  {Rajagopal}}, \ and\ \bibinfo {author} {\bibfnamefont {E.~V.}\ \bibnamefont
  {Shuryak}},\ }\href {\doibase 10.1103/PhysRevLett.81.4816} {\bibfield
  {journal} {\bibinfo  {journal} {Phys. Rev. Lett.}\ }\textbf {\bibinfo
  {volume} {81}},\ \bibinfo {pages} {4816} (\bibinfo {year} {1998})},\ \Eprint
  {http://arxiv.org/abs/hep-ph/9806219} {arXiv:hep-ph/9806219 [hep-ph]}
  \BibitemShut {NoStop}%
\bibitem [{\citenamefont {Stephanov}\ \emph {et~al.}(1999)\citenamefont
  {Stephanov}, \citenamefont {Rajagopal},\ and\ \citenamefont
  {Shuryak}}]{Stephanov:1999zu}%
  \BibitemOpen
  \bibfield  {author} {\bibinfo {author} {\bibfnamefont {M.~A.}\ \bibnamefont
  {Stephanov}}, \bibinfo {author} {\bibfnamefont {K.}~\bibnamefont
  {Rajagopal}}, \ and\ \bibinfo {author} {\bibfnamefont {E.~V.}\ \bibnamefont
  {Shuryak}},\ }\href {\doibase 10.1103/PhysRevD.60.114028} {\bibfield
  {journal} {\bibinfo  {journal} {Phys. Rev.}\ }\textbf {\bibinfo {volume}
  {D60}},\ \bibinfo {pages} {114028} (\bibinfo {year} {1999})},\ \Eprint
  {http://arxiv.org/abs/hep-ph/9903292} {arXiv:hep-ph/9903292 [hep-ph]}
  \BibitemShut {NoStop}%
\bibitem [{\citenamefont {Jeon}\ and\ \citenamefont
  {Heinz}(2015)}]{Jeon:2015dfa}%
  \BibitemOpen
  \bibfield  {author} {\bibinfo {author} {\bibfnamefont {S.}~\bibnamefont
  {Jeon}}\ and\ \bibinfo {author} {\bibfnamefont {U.}~\bibnamefont {Heinz}},\
  }\href {\doibase 10.1142/S0218301315300106} {\bibfield  {journal} {\bibinfo
  {journal} {Int. J. Mod. Phys.}\ }\textbf {\bibinfo {volume} {E24}},\ \bibinfo
  {pages} {1530010} (\bibinfo {year} {2015})},\ \Eprint
  {http://arxiv.org/abs/1503.03931} {arXiv:1503.03931 [hep-ph]} \BibitemShut
  {NoStop}%
\bibitem [{\citenamefont {Stephanov}(2010)}]{Stephanov:2009ra}%
  \BibitemOpen
  \bibfield  {author} {\bibinfo {author} {\bibfnamefont {M.~A.}\ \bibnamefont
  {Stephanov}},\ }\href {\doibase 10.1103/PhysRevD.81.054012} {\bibfield
  {journal} {\bibinfo  {journal} {Phys. Rev.}\ }\textbf {\bibinfo {volume}
  {D81}},\ \bibinfo {pages} {054012} (\bibinfo {year} {2010})},\ \Eprint
  {http://arxiv.org/abs/0911.1772} {arXiv:0911.1772 [hep-ph]} \BibitemShut
  {NoStop}%
\bibitem [{\citenamefont {Nahrgang}\ \emph {et~al.}(2011)\citenamefont
  {Nahrgang}, \citenamefont {Leupold}, \citenamefont {Herold},\ and\
  \citenamefont {Bleicher}}]{Nahrgang:2011mg}%
  \BibitemOpen
  \bibfield  {author} {\bibinfo {author} {\bibfnamefont {M.}~\bibnamefont
  {Nahrgang}}, \bibinfo {author} {\bibfnamefont {S.}~\bibnamefont {Leupold}},
  \bibinfo {author} {\bibfnamefont {C.}~\bibnamefont {Herold}}, \ and\ \bibinfo
  {author} {\bibfnamefont {M.}~\bibnamefont {Bleicher}},\ }\href {\doibase
  10.1103/PhysRevC.84.024912} {\bibfield  {journal} {\bibinfo  {journal} {Phys.
  Rev.}\ }\textbf {\bibinfo {volume} {C84}},\ \bibinfo {pages} {024912}
  (\bibinfo {year} {2011})},\ \Eprint {http://arxiv.org/abs/1105.0622}
  {arXiv:1105.0622 [nucl-th]} \BibitemShut {NoStop}%
\bibitem [{\citenamefont {Stephanov}\ and\ \citenamefont
  {Yin}(2017)}]{Stephanov:2017ghc}%
  \BibitemOpen
  \bibfield  {author} {\bibinfo {author} {\bibfnamefont {M.}~\bibnamefont
  {Stephanov}}\ and\ \bibinfo {author} {\bibfnamefont {Y.}~\bibnamefont
  {Yin}},\ }\href@noop {} {\  (\bibinfo {year} {2017})},\ \Eprint
  {http://arxiv.org/abs/1712.10305} {arXiv:1712.10305 [nucl-th]} \BibitemShut
  {NoStop}%
\bibitem [{\citenamefont {Nahrgang}\ \emph {et~al.}(2018)\citenamefont
  {Nahrgang}, \citenamefont {Bluhm}, \citenamefont {Schäfer},\ and\
  \citenamefont {Bass}}]{Nahrgang:2018afz}%
  \BibitemOpen
  \bibfield  {author} {\bibinfo {author} {\bibfnamefont {M.}~\bibnamefont
  {Nahrgang}}, \bibinfo {author} {\bibfnamefont {M.}~\bibnamefont {Bluhm}},
  \bibinfo {author} {\bibfnamefont {T.}~\bibnamefont {Schäfer}}, \ and\
  \bibinfo {author} {\bibfnamefont {S.~A.}\ \bibnamefont {Bass}},\ }\href@noop
  {} {\  (\bibinfo {year} {2018})},\ \Eprint {http://arxiv.org/abs/1804.05728}
  {arXiv:1804.05728 [nucl-th]} \BibitemShut {NoStop}%
\bibitem [{\citenamefont {Hama}\ \emph {et~al.}(2006)\citenamefont {Hama},
  \citenamefont {Andrade}, \citenamefont {Grassi}, \citenamefont {Socolowski},
  \citenamefont {Kodama}, \citenamefont {Tavares},\ and\ \citenamefont
  {Padula}}]{Hama:2005dz}%
  \BibitemOpen
  \bibfield  {author} {\bibinfo {author} {\bibfnamefont {Y.}~\bibnamefont
  {Hama}}, \bibinfo {author} {\bibfnamefont {R.~P.~G.}\ \bibnamefont
  {Andrade}}, \bibinfo {author} {\bibfnamefont {F.}~\bibnamefont {Grassi}},
  \bibinfo {author} {\bibfnamefont {O.}~\bibnamefont {Socolowski},
  \bibfnamefont {Jr.}}, \bibinfo {author} {\bibfnamefont {T.}~\bibnamefont
  {Kodama}}, \bibinfo {author} {\bibfnamefont {B.}~\bibnamefont {Tavares}}, \
  and\ \bibinfo {author} {\bibfnamefont {S.~S.}\ \bibnamefont {Padula}},\
  }\bibfield  {booktitle} {\emph {\bibinfo {booktitle} {{Proceedings, 18th
  International Conference on Ultra-Relativistic Nucleus-Nucleus Collisions
  (Quark Matter 2005): Budapest, Hungary, August 4-9, 2005}}},\ }\href
  {\doibase 10.1016/j.nuclphysa.2006.06.024} {\bibfield  {journal} {\bibinfo
  {journal} {Nucl. Phys.}\ }\textbf {\bibinfo {volume} {A774}},\ \bibinfo
  {pages} {169} (\bibinfo {year} {2006})},\ \Eprint
  {http://arxiv.org/abs/hep-ph/0510096} {arXiv:hep-ph/0510096 [hep-ph]}
  \BibitemShut {NoStop}%
\bibitem [{\citenamefont {Steinheimer}\ \emph {et~al.}(2011)\citenamefont
  {Steinheimer}, \citenamefont {Schramm},\ and\ \citenamefont
  {Stocker}}]{Steinheimer:2010ib}%
  \BibitemOpen
  \bibfield  {author} {\bibinfo {author} {\bibfnamefont {J.}~\bibnamefont
  {Steinheimer}}, \bibinfo {author} {\bibfnamefont {S.}~\bibnamefont
  {Schramm}}, \ and\ \bibinfo {author} {\bibfnamefont {H.}~\bibnamefont
  {Stocker}},\ }\href {\doibase 10.1088/0954-3899/38/3/035001} {\bibfield
  {journal} {\bibinfo  {journal} {J. Phys.}\ }\textbf {\bibinfo {volume}
  {G38}},\ \bibinfo {pages} {035001} (\bibinfo {year} {2011})},\ \Eprint
  {http://arxiv.org/abs/1009.5239} {arXiv:1009.5239 [hep-ph]} \BibitemShut
  {NoStop}%
\bibitem [{\citenamefont {Auvinen}\ \emph {et~al.}(2018)\citenamefont
  {Auvinen}, \citenamefont {Bernhard}, \citenamefont {Bass},\ and\
  \citenamefont {Karpenko}}]{Auvinen:2017fjw}%
  \BibitemOpen
  \bibfield  {author} {\bibinfo {author} {\bibfnamefont {J.}~\bibnamefont
  {Auvinen}}, \bibinfo {author} {\bibfnamefont {J.~E.}\ \bibnamefont
  {Bernhard}}, \bibinfo {author} {\bibfnamefont {S.~A.}\ \bibnamefont {Bass}},
  \ and\ \bibinfo {author} {\bibfnamefont {I.}~\bibnamefont {Karpenko}},\
  }\href {\doibase 10.1103/PhysRevC.97.044905} {\bibfield  {journal} {\bibinfo
  {journal} {Phys. Rev.}\ }\textbf {\bibinfo {volume} {C97}},\ \bibinfo {pages}
  {044905} (\bibinfo {year} {2018})},\ \Eprint
  {http://arxiv.org/abs/1706.03666} {arXiv:1706.03666 [hep-ph]} \BibitemShut
  {NoStop}%
\bibitem [{\citenamefont {Gardim}\ \emph {et~al.}(2018)\citenamefont {Gardim},
  \citenamefont {Grassi}, \citenamefont {Ishida}, \citenamefont {Luzum},
  \citenamefont {Magalhães},\ and\ \citenamefont
  {Noronha-Hostler}}]{Gardim:2017ruc}%
  \BibitemOpen
  \bibfield  {author} {\bibinfo {author} {\bibfnamefont {F.~G.}\ \bibnamefont
  {Gardim}}, \bibinfo {author} {\bibfnamefont {F.}~\bibnamefont {Grassi}},
  \bibinfo {author} {\bibfnamefont {P.}~\bibnamefont {Ishida}}, \bibinfo
  {author} {\bibfnamefont {M.}~\bibnamefont {Luzum}}, \bibinfo {author}
  {\bibfnamefont {P.~S.}\ \bibnamefont {Magalhães}}, \ and\ \bibinfo {author}
  {\bibfnamefont {J.}~\bibnamefont {Noronha-Hostler}},\ }\href {\doibase
  10.1103/PhysRevC.97.064919} {\bibfield  {journal} {\bibinfo  {journal} {Phys.
  Rev.}\ }\textbf {\bibinfo {volume} {C97}},\ \bibinfo {pages} {064919}
  (\bibinfo {year} {2018})},\ \Eprint {http://arxiv.org/abs/1712.03912}
  {arXiv:1712.03912 [nucl-th]} \BibitemShut {NoStop}%
\bibitem [{\citenamefont {Bellwied}\ \emph
  {et~al.}(2015{\natexlab{a}})\citenamefont {Bellwied}, \citenamefont
  {Borsanyi}, \citenamefont {Fodor}, \citenamefont {Günther}, \citenamefont
  {Katz}, \citenamefont {Ratti},\ and\ \citenamefont
  {Szabo}}]{Bellwied:2015rza}%
  \BibitemOpen
  \bibfield  {author} {\bibinfo {author} {\bibfnamefont {R.}~\bibnamefont
  {Bellwied}}, \bibinfo {author} {\bibfnamefont {S.}~\bibnamefont {Borsanyi}},
  \bibinfo {author} {\bibfnamefont {Z.}~\bibnamefont {Fodor}}, \bibinfo
  {author} {\bibfnamefont {J.}~\bibnamefont {Günther}}, \bibinfo {author}
  {\bibfnamefont {S.~D.}\ \bibnamefont {Katz}}, \bibinfo {author}
  {\bibfnamefont {C.}~\bibnamefont {Ratti}}, \ and\ \bibinfo {author}
  {\bibfnamefont {K.~K.}\ \bibnamefont {Szabo}},\ }\href {\doibase
  10.1016/j.physletb.2015.11.011} {\bibfield  {journal} {\bibinfo  {journal}
  {Phys. Lett.}\ }\textbf {\bibinfo {volume} {B751}},\ \bibinfo {pages} {559}
  (\bibinfo {year} {2015}{\natexlab{a}})},\ \Eprint
  {http://arxiv.org/abs/1507.07510} {arXiv:1507.07510 [hep-lat]} \BibitemShut
  {NoStop}%
\bibitem [{\citenamefont {Monnai}\ \emph {et~al.}(2017)\citenamefont {Monnai},
  \citenamefont {Mukherjee},\ and\ \citenamefont {Yin}}]{Monnai:2016kud}%
  \BibitemOpen
  \bibfield  {author} {\bibinfo {author} {\bibfnamefont {A.}~\bibnamefont
  {Monnai}}, \bibinfo {author} {\bibfnamefont {S.}~\bibnamefont {Mukherjee}}, \
  and\ \bibinfo {author} {\bibfnamefont {Y.}~\bibnamefont {Yin}},\ }\href
  {\doibase 10.1103/PhysRevC.95.034902} {\bibfield  {journal} {\bibinfo
  {journal} {Phys. Rev.}\ }\textbf {\bibinfo {volume} {C95}},\ \bibinfo {pages}
  {034902} (\bibinfo {year} {2017})},\ \Eprint
  {http://arxiv.org/abs/1606.00771} {arXiv:1606.00771 [nucl-th]} \BibitemShut
  {NoStop}%
\bibitem [{\citenamefont {Denicol}\ \emph {et~al.}(2018)\citenamefont
  {Denicol}, \citenamefont {Gale}, \citenamefont {Jeon}, \citenamefont
  {Monnai}, \citenamefont {Schenke},\ and\ \citenamefont
  {Shen}}]{Denicol:2018wdp}%
  \BibitemOpen
  \bibfield  {author} {\bibinfo {author} {\bibfnamefont {G.~S.}\ \bibnamefont
  {Denicol}}, \bibinfo {author} {\bibfnamefont {C.}~\bibnamefont {Gale}},
  \bibinfo {author} {\bibfnamefont {S.}~\bibnamefont {Jeon}}, \bibinfo {author}
  {\bibfnamefont {A.}~\bibnamefont {Monnai}}, \bibinfo {author} {\bibfnamefont
  {B.}~\bibnamefont {Schenke}}, \ and\ \bibinfo {author} {\bibfnamefont
  {C.}~\bibnamefont {Shen}},\ }\href {\doibase 10.1103/PhysRevC.98.034916}
  {\bibfield  {journal} {\bibinfo  {journal} {Phys. Rev.}\ }\textbf {\bibinfo
  {volume} {C98}},\ \bibinfo {pages} {034916} (\bibinfo {year} {2018})},\
  \Eprint {http://arxiv.org/abs/1804.10557} {arXiv:1804.10557 [nucl-th]}
  \BibitemShut {NoStop}%
\bibitem [{\citenamefont {Borsanyi}\ \emph
  {et~al.}(2010{\natexlab{a}})\citenamefont {Borsanyi}, \citenamefont
  {Endrodi}, \citenamefont {Fodor}, \citenamefont {Jakovac}, \citenamefont
  {Katz}, \citenamefont {Krieg}, \citenamefont {Ratti},\ and\ \citenamefont
  {Szabo}}]{Borsanyi:2010cj}%
  \BibitemOpen
  \bibfield  {author} {\bibinfo {author} {\bibfnamefont {S.}~\bibnamefont
  {Borsanyi}}, \bibinfo {author} {\bibfnamefont {G.}~\bibnamefont {Endrodi}},
  \bibinfo {author} {\bibfnamefont {Z.}~\bibnamefont {Fodor}}, \bibinfo
  {author} {\bibfnamefont {A.}~\bibnamefont {Jakovac}}, \bibinfo {author}
  {\bibfnamefont {S.~D.}\ \bibnamefont {Katz}}, \bibinfo {author}
  {\bibfnamefont {S.}~\bibnamefont {Krieg}}, \bibinfo {author} {\bibfnamefont
  {C.}~\bibnamefont {Ratti}}, \ and\ \bibinfo {author} {\bibfnamefont {K.~K.}\
  \bibnamefont {Szabo}},\ }\href {\doibase 10.1007/JHEP11(2010)077} {\bibfield
  {journal} {\bibinfo  {journal} {JHEP}\ }\textbf {\bibinfo {volume} {11}},\
  \bibinfo {pages} {077} (\bibinfo {year} {2010}{\natexlab{a}})},\ \Eprint
  {http://arxiv.org/abs/1007.2580} {arXiv:1007.2580 [hep-lat]} \BibitemShut
  {NoStop}%
\bibitem [{\citenamefont {Borsanyi}\ \emph {et~al.}(2014)\citenamefont
  {Borsanyi}, \citenamefont {Fodor}, \citenamefont {Hoelbling}, \citenamefont
  {Katz}, \citenamefont {Krieg},\ and\ \citenamefont
  {Szabo}}]{Borsanyi:2013bia}%
  \BibitemOpen
  \bibfield  {author} {\bibinfo {author} {\bibfnamefont {S.}~\bibnamefont
  {Borsanyi}}, \bibinfo {author} {\bibfnamefont {Z.}~\bibnamefont {Fodor}},
  \bibinfo {author} {\bibfnamefont {C.}~\bibnamefont {Hoelbling}}, \bibinfo
  {author} {\bibfnamefont {S.~D.}\ \bibnamefont {Katz}}, \bibinfo {author}
  {\bibfnamefont {S.}~\bibnamefont {Krieg}}, \ and\ \bibinfo {author}
  {\bibfnamefont {K.~K.}\ \bibnamefont {Szabo}},\ }\href {\doibase
  10.1016/j.physletb.2014.01.007} {\bibfield  {journal} {\bibinfo  {journal}
  {Phys. Lett.}\ }\textbf {\bibinfo {volume} {B730}},\ \bibinfo {pages} {99}
  (\bibinfo {year} {2014})},\ \Eprint {http://arxiv.org/abs/1309.5258}
  {arXiv:1309.5258 [hep-lat]} \BibitemShut {NoStop}%
\bibitem [{\citenamefont {Bazavov}\ \emph {et~al.}(2014)\citenamefont {Bazavov}
  \emph {et~al.}}]{Bazavov:2014pvz}%
  \BibitemOpen
  \bibfield  {author} {\bibinfo {author} {\bibfnamefont {A.}~\bibnamefont
  {Bazavov}} \emph {et~al.} (\bibinfo {collaboration} {HotQCD}),\ }\href
  {\doibase 10.1103/PhysRevD.90.094503} {\bibfield  {journal} {\bibinfo
  {journal} {Phys. Rev.}\ }\textbf {\bibinfo {volume} {D90}},\ \bibinfo {pages}
  {094503} (\bibinfo {year} {2014})},\ \Eprint {http://arxiv.org/abs/1407.6387}
  {arXiv:1407.6387 [hep-lat]} \BibitemShut {NoStop}%
\bibitem [{\citenamefont {Borsanyi}\ \emph {et~al.}(2016)\citenamefont
  {Borsanyi} \emph {et~al.}}]{Borsanyi:2016ksw}%
  \BibitemOpen
  \bibfield  {author} {\bibinfo {author} {\bibfnamefont {S.}~\bibnamefont
  {Borsanyi}} \emph {et~al.},\ }\href {\doibase 10.1038/nature20115} {\bibfield
   {journal} {\bibinfo  {journal} {Nature}\ }\textbf {\bibinfo {volume}
  {539}},\ \bibinfo {pages} {69} (\bibinfo {year} {2016})},\ \Eprint
  {http://arxiv.org/abs/1606.07494} {arXiv:1606.07494 [hep-lat]} \BibitemShut
  {NoStop}%
\bibitem [{\citenamefont {Allton}\ \emph {et~al.}(2002)\citenamefont {Allton},
  \citenamefont {Ejiri}, \citenamefont {Hands}, \citenamefont {Kaczmarek},
  \citenamefont {Karsch}, \citenamefont {Laermann}, \citenamefont {Schmidt},\
  and\ \citenamefont {Scorzato}}]{Allton:2002zi}%
  \BibitemOpen
  \bibfield  {author} {\bibinfo {author} {\bibfnamefont {C.~R.}\ \bibnamefont
  {Allton}}, \bibinfo {author} {\bibfnamefont {S.}~\bibnamefont {Ejiri}},
  \bibinfo {author} {\bibfnamefont {S.~J.}\ \bibnamefont {Hands}}, \bibinfo
  {author} {\bibfnamefont {O.}~\bibnamefont {Kaczmarek}}, \bibinfo {author}
  {\bibfnamefont {F.}~\bibnamefont {Karsch}}, \bibinfo {author} {\bibfnamefont
  {E.}~\bibnamefont {Laermann}}, \bibinfo {author} {\bibfnamefont
  {C.}~\bibnamefont {Schmidt}}, \ and\ \bibinfo {author} {\bibfnamefont
  {L.}~\bibnamefont {Scorzato}},\ }\href {\doibase 10.1103/PhysRevD.66.074507}
  {\bibfield  {journal} {\bibinfo  {journal} {Phys. Rev.}\ }\textbf {\bibinfo
  {volume} {D66}},\ \bibinfo {pages} {074507} (\bibinfo {year} {2002})},\
  \Eprint {http://arxiv.org/abs/hep-lat/0204010} {arXiv:hep-lat/0204010
  [hep-lat]} \BibitemShut {NoStop}%
\bibitem [{\citenamefont {Allton}\ \emph {et~al.}(2005)\citenamefont {Allton},
  \citenamefont {Doring}, \citenamefont {Ejiri}, \citenamefont {Hands},
  \citenamefont {Kaczmarek}, \citenamefont {Karsch}, \citenamefont {Laermann},\
  and\ \citenamefont {Redlich}}]{Allton:2005gk}%
  \BibitemOpen
  \bibfield  {author} {\bibinfo {author} {\bibfnamefont {C.~R.}\ \bibnamefont
  {Allton}}, \bibinfo {author} {\bibfnamefont {M.}~\bibnamefont {Doring}},
  \bibinfo {author} {\bibfnamefont {S.}~\bibnamefont {Ejiri}}, \bibinfo
  {author} {\bibfnamefont {S.~J.}\ \bibnamefont {Hands}}, \bibinfo {author}
  {\bibfnamefont {O.}~\bibnamefont {Kaczmarek}}, \bibinfo {author}
  {\bibfnamefont {F.}~\bibnamefont {Karsch}}, \bibinfo {author} {\bibfnamefont
  {E.}~\bibnamefont {Laermann}}, \ and\ \bibinfo {author} {\bibfnamefont
  {K.}~\bibnamefont {Redlich}},\ }\href {\doibase 10.1103/PhysRevD.71.054508}
  {\bibfield  {journal} {\bibinfo  {journal} {Phys. Rev.}\ }\textbf {\bibinfo
  {volume} {D71}},\ \bibinfo {pages} {054508} (\bibinfo {year} {2005})},\
  \Eprint {http://arxiv.org/abs/hep-lat/0501030} {arXiv:hep-lat/0501030
  [hep-lat]} \BibitemShut {NoStop}%
\bibitem [{\citenamefont {Gavai}\ and\ \citenamefont
  {Gupta}(2008)}]{Gavai:2008zr}%
  \BibitemOpen
  \bibfield  {author} {\bibinfo {author} {\bibfnamefont {R.~V.}\ \bibnamefont
  {Gavai}}\ and\ \bibinfo {author} {\bibfnamefont {S.}~\bibnamefont {Gupta}},\
  }\href {\doibase 10.1103/PhysRevD.78.114503} {\bibfield  {journal} {\bibinfo
  {journal} {Phys. Rev.}\ }\textbf {\bibinfo {volume} {D78}},\ \bibinfo {pages}
  {114503} (\bibinfo {year} {2008})},\ \Eprint {http://arxiv.org/abs/0806.2233}
  {arXiv:0806.2233 [hep-lat]} \BibitemShut {NoStop}%
\bibitem [{\citenamefont {Basak}\ \emph {et~al.}(2008)\citenamefont {Basak}
  \emph {et~al.}}]{Basak:2009uv}%
  \BibitemOpen
  \bibfield  {author} {\bibinfo {author} {\bibfnamefont {S.}~\bibnamefont
  {Basak}} \emph {et~al.} (\bibinfo {collaboration} {MILC}),\ }\bibfield
  {booktitle} {\emph {\bibinfo {booktitle} {{Proceedings, 26th International
  Symposium on Lattice field theory (Lattice 2008): Williamsburg, USA, July
  14-19, 2008}}},\ }\href@noop {} {\bibfield  {journal} {\bibinfo  {journal}
  {PoS}\ }\textbf {\bibinfo {volume} {LATTICE2008}},\ \bibinfo {pages} {171}
  (\bibinfo {year} {2008})},\ \Eprint {http://arxiv.org/abs/0910.0276}
  {arXiv:0910.0276 [hep-lat]} \BibitemShut {NoStop}%
\bibitem [{\citenamefont {Kaczmarek}\ \emph {et~al.}(2011)\citenamefont
  {Kaczmarek}, \citenamefont {Karsch}, \citenamefont {Laermann}, \citenamefont
  {Miao}, \citenamefont {Mukherjee}, \citenamefont {Petreczky}, \citenamefont
  {Schmidt}, \citenamefont {Soeldner},\ and\ \citenamefont
  {Unger}}]{Kaczmarek:2011zz}%
  \BibitemOpen
  \bibfield  {author} {\bibinfo {author} {\bibfnamefont {O.}~\bibnamefont
  {Kaczmarek}}, \bibinfo {author} {\bibfnamefont {F.}~\bibnamefont {Karsch}},
  \bibinfo {author} {\bibfnamefont {E.}~\bibnamefont {Laermann}}, \bibinfo
  {author} {\bibfnamefont {C.}~\bibnamefont {Miao}}, \bibinfo {author}
  {\bibfnamefont {S.}~\bibnamefont {Mukherjee}}, \bibinfo {author}
  {\bibfnamefont {P.}~\bibnamefont {Petreczky}}, \bibinfo {author}
  {\bibfnamefont {C.}~\bibnamefont {Schmidt}}, \bibinfo {author} {\bibfnamefont
  {W.}~\bibnamefont {Soeldner}}, \ and\ \bibinfo {author} {\bibfnamefont
  {W.}~\bibnamefont {Unger}},\ }\href {\doibase 10.1103/PhysRevD.83.014504}
  {\bibfield  {journal} {\bibinfo  {journal} {Phys. Rev.}\ }\textbf {\bibinfo
  {volume} {D83}},\ \bibinfo {pages} {014504} (\bibinfo {year} {2011})},\
  \Eprint {http://arxiv.org/abs/1011.3130} {arXiv:1011.3130 [hep-lat]}
  \BibitemShut {NoStop}%
\bibitem [{\citenamefont {de~Forcrand}\ and\ \citenamefont
  {Philipsen}(2002)}]{deForcrand:2002hgr}%
  \BibitemOpen
  \bibfield  {author} {\bibinfo {author} {\bibfnamefont {P.}~\bibnamefont
  {de~Forcrand}}\ and\ \bibinfo {author} {\bibfnamefont {O.}~\bibnamefont
  {Philipsen}},\ }\href {\doibase 10.1016/S0550-3213(02)00626-0} {\bibfield
  {journal} {\bibinfo  {journal} {Nucl. Phys.}\ }\textbf {\bibinfo {volume}
  {B642}},\ \bibinfo {pages} {290} (\bibinfo {year} {2002})},\ \Eprint
  {http://arxiv.org/abs/hep-lat/0205016} {arXiv:hep-lat/0205016 [hep-lat]}
  \BibitemShut {NoStop}%
\bibitem [{\citenamefont {D'Elia}\ and\ \citenamefont
  {Lombardo}(2003)}]{DElia:2002tig}%
  \BibitemOpen
  \bibfield  {author} {\bibinfo {author} {\bibfnamefont {M.}~\bibnamefont
  {D'Elia}}\ and\ \bibinfo {author} {\bibfnamefont {M.-P.}\ \bibnamefont
  {Lombardo}},\ }\href {\doibase 10.1103/PhysRevD.67.014505} {\bibfield
  {journal} {\bibinfo  {journal} {Phys. Rev.}\ }\textbf {\bibinfo {volume}
  {D67}},\ \bibinfo {pages} {014505} (\bibinfo {year} {2003})},\ \Eprint
  {http://arxiv.org/abs/hep-lat/0209146} {arXiv:hep-lat/0209146 [hep-lat]}
  \BibitemShut {NoStop}%
\bibitem [{\citenamefont {Wu}\ \emph {et~al.}(2007)\citenamefont {Wu},
  \citenamefont {Luo},\ and\ \citenamefont {Chen}}]{Wu:2006su}%
  \BibitemOpen
  \bibfield  {author} {\bibinfo {author} {\bibfnamefont {L.-K.}\ \bibnamefont
  {Wu}}, \bibinfo {author} {\bibfnamefont {X.-Q.}\ \bibnamefont {Luo}}, \ and\
  \bibinfo {author} {\bibfnamefont {H.-S.}\ \bibnamefont {Chen}},\ }\href
  {\doibase 10.1103/PhysRevD.76.034505} {\bibfield  {journal} {\bibinfo
  {journal} {Phys. Rev.}\ }\textbf {\bibinfo {volume} {D76}},\ \bibinfo {pages}
  {034505} (\bibinfo {year} {2007})},\ \Eprint
  {http://arxiv.org/abs/hep-lat/0611035} {arXiv:hep-lat/0611035 [hep-lat]}
  \BibitemShut {NoStop}%
\bibitem [{\citenamefont {D'Elia}\ \emph {et~al.}(2007)\citenamefont {D'Elia},
  \citenamefont {Di~Renzo},\ and\ \citenamefont {Lombardo}}]{DElia:2007bkz}%
  \BibitemOpen
  \bibfield  {author} {\bibinfo {author} {\bibfnamefont {M.}~\bibnamefont
  {D'Elia}}, \bibinfo {author} {\bibfnamefont {F.}~\bibnamefont {Di~Renzo}}, \
  and\ \bibinfo {author} {\bibfnamefont {M.~P.}\ \bibnamefont {Lombardo}},\
  }\href {\doibase 10.1103/PhysRevD.76.114509} {\bibfield  {journal} {\bibinfo
  {journal} {Phys. Rev.}\ }\textbf {\bibinfo {volume} {D76}},\ \bibinfo {pages}
  {114509} (\bibinfo {year} {2007})},\ \Eprint {http://arxiv.org/abs/0705.3814}
  {arXiv:0705.3814 [hep-lat]} \BibitemShut {NoStop}%
\bibitem [{\citenamefont {Conradi}\ and\ \citenamefont
  {D'Elia}(2007)}]{Conradi:2007be}%
  \BibitemOpen
  \bibfield  {author} {\bibinfo {author} {\bibfnamefont {S.}~\bibnamefont
  {Conradi}}\ and\ \bibinfo {author} {\bibfnamefont {M.}~\bibnamefont
  {D'Elia}},\ }\href {\doibase 10.1103/PhysRevD.76.074501} {\bibfield
  {journal} {\bibinfo  {journal} {Phys. Rev.}\ }\textbf {\bibinfo {volume}
  {D76}},\ \bibinfo {pages} {074501} (\bibinfo {year} {2007})},\ \Eprint
  {http://arxiv.org/abs/0707.1987} {arXiv:0707.1987 [hep-lat]} \BibitemShut
  {NoStop}%
\bibitem [{\citenamefont {de~Forcrand}\ and\ \citenamefont
  {Philipsen}(2008)}]{deForcrand:2008vr}%
  \BibitemOpen
  \bibfield  {author} {\bibinfo {author} {\bibfnamefont {P.}~\bibnamefont
  {de~Forcrand}}\ and\ \bibinfo {author} {\bibfnamefont {O.}~\bibnamefont
  {Philipsen}},\ }\href {\doibase 10.1088/1126-6708/2008/11/012} {\bibfield
  {journal} {\bibinfo  {journal} {JHEP}\ }\textbf {\bibinfo {volume} {11}},\
  \bibinfo {pages} {012} (\bibinfo {year} {2008})},\ \Eprint
  {http://arxiv.org/abs/0808.1096} {arXiv:0808.1096 [hep-lat]} \BibitemShut
  {NoStop}%
\bibitem [{\citenamefont {D'Elia}\ and\ \citenamefont
  {Sanfilippo}(2009)}]{DElia:2009pdy}%
  \BibitemOpen
  \bibfield  {author} {\bibinfo {author} {\bibfnamefont {M.}~\bibnamefont
  {D'Elia}}\ and\ \bibinfo {author} {\bibfnamefont {F.}~\bibnamefont
  {Sanfilippo}},\ }\href {\doibase 10.1103/PhysRevD.80.014502} {\bibfield
  {journal} {\bibinfo  {journal} {Phys. Rev.}\ }\textbf {\bibinfo {volume}
  {D80}},\ \bibinfo {pages} {014502} (\bibinfo {year} {2009})},\ \Eprint
  {http://arxiv.org/abs/0904.1400} {arXiv:0904.1400 [hep-lat]} \BibitemShut
  {NoStop}%
\bibitem [{\citenamefont {Moscicki}\ \emph {et~al.}(2010)\citenamefont
  {Moscicki}, \citenamefont {Wos}, \citenamefont {Lamanna}, \citenamefont
  {de~Forcrand},\ and\ \citenamefont {Philipsen}}]{Moscicki:2009id}%
  \BibitemOpen
  \bibfield  {author} {\bibinfo {author} {\bibfnamefont {J.~T.}\ \bibnamefont
  {Moscicki}}, \bibinfo {author} {\bibfnamefont {M.}~\bibnamefont {Wos}},
  \bibinfo {author} {\bibfnamefont {M.}~\bibnamefont {Lamanna}}, \bibinfo
  {author} {\bibfnamefont {P.}~\bibnamefont {de~Forcrand}}, \ and\ \bibinfo
  {author} {\bibfnamefont {O.}~\bibnamefont {Philipsen}},\ }\href {\doibase
  10.1016/j.cpc.2010.06.027} {\bibfield  {journal} {\bibinfo  {journal}
  {Comput. Phys. Commun.}\ }\textbf {\bibinfo {volume} {181}},\ \bibinfo
  {pages} {1715} (\bibinfo {year} {2010})},\ \Eprint
  {http://arxiv.org/abs/0911.5682} {arXiv:0911.5682 [cs.DC]} \BibitemShut
  {NoStop}%
\bibitem [{\citenamefont {Borsanyi}\ \emph
  {et~al.}(2012{\natexlab{a}})\citenamefont {Borsanyi}, \citenamefont
  {Endrodi}, \citenamefont {Fodor}, \citenamefont {Katz}, \citenamefont
  {Krieg}, \citenamefont {Ratti},\ and\ \citenamefont
  {Szabo}}]{Borsanyi:2012cr}%
  \BibitemOpen
  \bibfield  {author} {\bibinfo {author} {\bibfnamefont {S.}~\bibnamefont
  {Borsanyi}}, \bibinfo {author} {\bibfnamefont {G.}~\bibnamefont {Endrodi}},
  \bibinfo {author} {\bibfnamefont {Z.}~\bibnamefont {Fodor}}, \bibinfo
  {author} {\bibfnamefont {S.~D.}\ \bibnamefont {Katz}}, \bibinfo {author}
  {\bibfnamefont {S.}~\bibnamefont {Krieg}}, \bibinfo {author} {\bibfnamefont
  {C.}~\bibnamefont {Ratti}}, \ and\ \bibinfo {author} {\bibfnamefont {K.~K.}\
  \bibnamefont {Szabo}},\ }\href {\doibase 10.1007/JHEP08(2012)053} {\bibfield
  {journal} {\bibinfo  {journal} {JHEP}\ }\textbf {\bibinfo {volume} {08}},\
  \bibinfo {pages} {053} (\bibinfo {year} {2012}{\natexlab{a}})},\ \Eprint
  {http://arxiv.org/abs/1204.6710} {arXiv:1204.6710 [hep-lat]} \BibitemShut
  {NoStop}%
\bibitem [{\citenamefont {Hegde}(2014)}]{Hegde:2014sta}%
  \BibitemOpen
  \bibfield  {author} {\bibinfo {author} {\bibfnamefont {P.}~\bibnamefont
  {Hegde}} (\bibinfo {collaboration} {BNL-Bielefeld-CCNU}),\ }\bibfield
  {booktitle} {\emph {\bibinfo {booktitle} {{Proceedings, 24th International
  Conference on Ultra-Relativistic Nucleus-Nucleus Collisions (Quark Matter
  2014): Darmstadt, Germany, May 19-24, 2014}}},\ }\href {\doibase
  10.1016/j.nuclphysa.2014.08.089} {\bibfield  {journal} {\bibinfo  {journal}
  {Nucl. Phys.}\ }\textbf {\bibinfo {volume} {A931}},\ \bibinfo {pages} {851}
  (\bibinfo {year} {2014})},\ \Eprint {http://arxiv.org/abs/1408.6305}
  {arXiv:1408.6305 [hep-lat]} \BibitemShut {NoStop}%
\bibitem [{\citenamefont {Gunther}\ \emph {et~al.}(2017)\citenamefont
  {Gunther}, \citenamefont {Bellwied}, \citenamefont {Borsanyi}, \citenamefont
  {Fodor}, \citenamefont {Katz}, \citenamefont {Pasztor},\ and\ \citenamefont
  {Ratti}}]{Gunther:2016vcp}%
  \BibitemOpen
  \bibfield  {author} {\bibinfo {author} {\bibfnamefont {J.}~\bibnamefont
  {Gunther}}, \bibinfo {author} {\bibfnamefont {R.}~\bibnamefont {Bellwied}},
  \bibinfo {author} {\bibfnamefont {S.}~\bibnamefont {Borsanyi}}, \bibinfo
  {author} {\bibfnamefont {Z.}~\bibnamefont {Fodor}}, \bibinfo {author}
  {\bibfnamefont {S.~D.}\ \bibnamefont {Katz}}, \bibinfo {author}
  {\bibfnamefont {A.}~\bibnamefont {Pasztor}}, \ and\ \bibinfo {author}
  {\bibfnamefont {C.}~\bibnamefont {Ratti}},\ }\bibfield  {booktitle} {\emph
  {\bibinfo {booktitle} {{Proceedings, 12th Conference on Quark Confinement and
  the Hadron Spectrum (Confinement XII): Thessaloniki, Greece}}},\ }\href
  {\doibase 10.1051/epjconf/201713707008} {\bibfield  {journal} {\bibinfo
  {journal} {EPJ Web Conf.}\ }\textbf {\bibinfo {volume} {137}},\ \bibinfo
  {pages} {07008} (\bibinfo {year} {2017})},\ \Eprint
  {http://arxiv.org/abs/1607.02493} {arXiv:1607.02493 [hep-lat]} \BibitemShut
  {NoStop}%
\bibitem [{\citenamefont {Bazavov}\ \emph {et~al.}(2017)\citenamefont {Bazavov}
  \emph {et~al.}}]{Bazavov:2017dus}%
  \BibitemOpen
  \bibfield  {author} {\bibinfo {author} {\bibfnamefont {A.}~\bibnamefont
  {Bazavov}} \emph {et~al.},\ }\href {\doibase 10.1103/PhysRevD.95.054504}
  {\bibfield  {journal} {\bibinfo  {journal} {Phys. Rev.}\ }\textbf {\bibinfo
  {volume} {D95}},\ \bibinfo {pages} {054504} (\bibinfo {year} {2017})},\
  \Eprint {http://arxiv.org/abs/1701.04325} {arXiv:1701.04325 [hep-lat]}
  \BibitemShut {NoStop}%
\bibitem [{\citenamefont {D'Elia}\ \emph {et~al.}(2017)\citenamefont {D'Elia},
  \citenamefont {Gagliardi},\ and\ \citenamefont {Sanfilippo}}]{DElia:2016jqh}%
  \BibitemOpen
  \bibfield  {author} {\bibinfo {author} {\bibfnamefont {M.}~\bibnamefont
  {D'Elia}}, \bibinfo {author} {\bibfnamefont {G.}~\bibnamefont {Gagliardi}}, \
  and\ \bibinfo {author} {\bibfnamefont {F.}~\bibnamefont {Sanfilippo}},\
  }\href {\doibase 10.1103/PhysRevD.95.094503} {\bibfield  {journal} {\bibinfo
  {journal} {Phys. Rev.}\ }\textbf {\bibinfo {volume} {D95}},\ \bibinfo {pages}
  {094503} (\bibinfo {year} {2017})},\ \Eprint
  {http://arxiv.org/abs/1611.08285} {arXiv:1611.08285 [hep-lat]} \BibitemShut
  {NoStop}%
\bibitem [{\citenamefont {Borsanyi}\ \emph {et~al.}(2018)\citenamefont
  {Borsanyi}, \citenamefont {Fodor}, \citenamefont {Guenther}, \citenamefont
  {Katz}, \citenamefont {Szabo}, \citenamefont {Pasztor}, \citenamefont
  {Portillo},\ and\ \citenamefont {Ratti}}]{Borsanyi:2018grb}%
  \BibitemOpen
  \bibfield  {author} {\bibinfo {author} {\bibfnamefont {S.}~\bibnamefont
  {Borsanyi}}, \bibinfo {author} {\bibfnamefont {Z.}~\bibnamefont {Fodor}},
  \bibinfo {author} {\bibfnamefont {J.~N.}\ \bibnamefont {Guenther}}, \bibinfo
  {author} {\bibfnamefont {S.~K.}\ \bibnamefont {Katz}}, \bibinfo {author}
  {\bibfnamefont {K.~K.}\ \bibnamefont {Szabo}}, \bibinfo {author}
  {\bibfnamefont {A.}~\bibnamefont {Pasztor}}, \bibinfo {author} {\bibfnamefont
  {I.}~\bibnamefont {Portillo}}, \ and\ \bibinfo {author} {\bibfnamefont
  {C.}~\bibnamefont {Ratti}},\ }\href {\doibase 10.1007/JHEP10(2018)205}
  {\bibfield  {journal} {\bibinfo  {journal} {JHEP}\ }\textbf {\bibinfo
  {volume} {10}},\ \bibinfo {pages} {205} (\bibinfo {year} {2018})},\ \Eprint
  {http://arxiv.org/abs/1805.04445} {arXiv:1805.04445 [hep-lat]} \BibitemShut
  {NoStop}%
\bibitem [{\citenamefont {Ratti}(2018)}]{Ratti:2018ksb}%
  \BibitemOpen
  \bibfield  {author} {\bibinfo {author} {\bibfnamefont {C.}~\bibnamefont
  {Ratti}},\ }\href@noop {} {\  (\bibinfo {year} {2018})},\ \Eprint
  {http://arxiv.org/abs/1804.07810} {arXiv:1804.07810 [hep-lat]} \BibitemShut
  {NoStop}%
\bibitem [{\citenamefont {Rajagopal}\ and\ \citenamefont
  {Wilczek}(1993)}]{Rajagopal:1992qz}%
  \BibitemOpen
  \bibfield  {author} {\bibinfo {author} {\bibfnamefont {K.}~\bibnamefont
  {Rajagopal}}\ and\ \bibinfo {author} {\bibfnamefont {F.}~\bibnamefont
  {Wilczek}},\ }\href {\doibase 10.1016/0550-3213(93)90502-G} {\bibfield
  {journal} {\bibinfo  {journal} {Nucl. Phys.}\ }\textbf {\bibinfo {volume}
  {B399}},\ \bibinfo {pages} {395} (\bibinfo {year} {1993})},\ \Eprint
  {http://arxiv.org/abs/hep-ph/9210253} {arXiv:hep-ph/9210253 [hep-ph]}
  \BibitemShut {NoStop}%
\bibitem [{\citenamefont {Berges}\ and\ \citenamefont
  {Rajagopal}(1999)}]{Berges:1998rc}%
  \BibitemOpen
  \bibfield  {author} {\bibinfo {author} {\bibfnamefont {J.}~\bibnamefont
  {Berges}}\ and\ \bibinfo {author} {\bibfnamefont {K.}~\bibnamefont
  {Rajagopal}},\ }\href {\doibase 10.1016/S0550-3213(98)00620-8} {\bibfield
  {journal} {\bibinfo  {journal} {Nucl. Phys.}\ }\textbf {\bibinfo {volume}
  {B538}},\ \bibinfo {pages} {215} (\bibinfo {year} {1999})},\ \Eprint
  {http://arxiv.org/abs/hep-ph/9804233} {arXiv:hep-ph/9804233 [hep-ph]}
  \BibitemShut {NoStop}%
\bibitem [{\citenamefont {Halasz}\ \emph {et~al.}(1998)\citenamefont {Halasz},
  \citenamefont {Jackson}, \citenamefont {Shrock}, \citenamefont {Stephanov},\
  and\ \citenamefont {Verbaarschot}}]{Halasz:1998qr}%
  \BibitemOpen
  \bibfield  {author} {\bibinfo {author} {\bibfnamefont {A.~M.}\ \bibnamefont
  {Halasz}}, \bibinfo {author} {\bibfnamefont {A.~D.}\ \bibnamefont {Jackson}},
  \bibinfo {author} {\bibfnamefont {R.~E.}\ \bibnamefont {Shrock}}, \bibinfo
  {author} {\bibfnamefont {M.~A.}\ \bibnamefont {Stephanov}}, \ and\ \bibinfo
  {author} {\bibfnamefont {J.~J.~M.}\ \bibnamefont {Verbaarschot}},\ }\href
  {\doibase 10.1103/PhysRevD.58.096007} {\bibfield  {journal} {\bibinfo
  {journal} {Phys. Rev.}\ }\textbf {\bibinfo {volume} {D58}},\ \bibinfo {pages}
  {096007} (\bibinfo {year} {1998})},\ \Eprint
  {http://arxiv.org/abs/hep-ph/9804290} {arXiv:hep-ph/9804290 [hep-ph]}
  \BibitemShut {NoStop}%
\bibitem [{\citenamefont {Karsch}\ \emph {et~al.}(2001)\citenamefont {Karsch},
  \citenamefont {Laermann},\ and\ \citenamefont {Schmidt}}]{Karsch:2001nf}%
  \BibitemOpen
  \bibfield  {author} {\bibinfo {author} {\bibfnamefont {F.}~\bibnamefont
  {Karsch}}, \bibinfo {author} {\bibfnamefont {E.}~\bibnamefont {Laermann}}, \
  and\ \bibinfo {author} {\bibfnamefont {C.}~\bibnamefont {Schmidt}},\ }\href
  {\doibase 10.1016/S0370-2693(01)01114-5} {\bibfield  {journal} {\bibinfo
  {journal} {Phys. Lett.}\ }\textbf {\bibinfo {volume} {B520}},\ \bibinfo
  {pages} {41} (\bibinfo {year} {2001})},\ \Eprint
  {http://arxiv.org/abs/hep-lat/0107020} {arXiv:hep-lat/0107020 [hep-lat]}
  \BibitemShut {NoStop}%
\bibitem [{\citenamefont {de~Forcrand}\ and\ \citenamefont
  {Philipsen}(2003)}]{deForcrand:2003vyj}%
  \BibitemOpen
  \bibfield  {author} {\bibinfo {author} {\bibfnamefont {P.}~\bibnamefont
  {de~Forcrand}}\ and\ \bibinfo {author} {\bibfnamefont {O.}~\bibnamefont
  {Philipsen}},\ }\href {\doibase 10.1016/j.nuclphysb.2003.09.005} {\bibfield
  {journal} {\bibinfo  {journal} {Nucl. Phys.}\ }\textbf {\bibinfo {volume}
  {B673}},\ \bibinfo {pages} {170} (\bibinfo {year} {2003})},\ \Eprint
  {http://arxiv.org/abs/hep-lat/0307020} {arXiv:hep-lat/0307020 [hep-lat]}
  \BibitemShut {NoStop}%
\bibitem [{\citenamefont {Nonaka}\ and\ \citenamefont
  {Asakawa}(2005)}]{Nonaka:2004pg}%
  \BibitemOpen
  \bibfield  {author} {\bibinfo {author} {\bibfnamefont {C.}~\bibnamefont
  {Nonaka}}\ and\ \bibinfo {author} {\bibfnamefont {M.}~\bibnamefont
  {Asakawa}},\ }\href {\doibase 10.1103/PhysRevC.71.044904} {\bibfield
  {journal} {\bibinfo  {journal} {Phys. Rev.}\ }\textbf {\bibinfo {volume}
  {C71}},\ \bibinfo {pages} {044904} (\bibinfo {year} {2005})},\ \Eprint
  {http://arxiv.org/abs/nucl-th/0410078} {arXiv:nucl-th/0410078 [nucl-th]}
  \BibitemShut {NoStop}%
\bibitem [{\citenamefont {Bluhm}\ and\ \citenamefont
  {Kampfer}(2006)}]{Bluhm:2006av}%
  \BibitemOpen
  \bibfield  {author} {\bibinfo {author} {\bibfnamefont {M.}~\bibnamefont
  {Bluhm}}\ and\ \bibinfo {author} {\bibfnamefont {B.}~\bibnamefont
  {Kampfer}},\ }\bibfield  {booktitle} {\emph {\bibinfo {booktitle} {{Critical
  point and onset of deconfinement. Proceedings, 3rd Conference, CPOD2006,
  Florence, Itlay, July 3-6, 2006}}},\ }\href@noop {} {\bibfield  {journal}
  {\bibinfo  {journal} {PoS}\ }\textbf {\bibinfo {volume} {CPOD2006}},\
  \bibinfo {pages} {004} (\bibinfo {year} {2006})},\ \Eprint
  {http://arxiv.org/abs/hep-ph/0611083} {arXiv:hep-ph/0611083 [hep-ph]}
  \BibitemShut {NoStop}%
\bibitem [{\citenamefont {{The code on which this work is based can be
  downloaded at the following link}}()}]{code:2018}%
  \BibitemOpen
  \bibfield  {author} {\bibinfo {author} {\bibnamefont {{The code on which this
  work is based can be downloaded at the following link}}},\ }\href@noop {}
  {}\bibinfo {howpublished}
  {\url{https://www.bnl.gov/physics/best/resources.php}}\BibitemShut {NoStop}%
\bibitem [{\citenamefont {Hatta}\ and\ \citenamefont
  {Stephanov}(2003)}]{Hatta:2003wn}%
  \BibitemOpen
  \bibfield  {author} {\bibinfo {author} {\bibfnamefont {Y.}~\bibnamefont
  {Hatta}}\ and\ \bibinfo {author} {\bibfnamefont {M.~A.}\ \bibnamefont
  {Stephanov}},\ }\href {\doibase 10.1103/PhysRevLett.91.102003,
  10.1103/PhysRevLett.91.129901} {\bibfield  {journal} {\bibinfo  {journal}
  {Phys. Rev. Lett.}\ }\textbf {\bibinfo {volume} {91}},\ \bibinfo {pages}
  {102003} (\bibinfo {year} {2003})},\ \bibinfo {note} {[Erratum: Phys. Rev.
  Lett.91,129901(2003)]},\ \Eprint {http://arxiv.org/abs/hep-ph/0302002}
  {arXiv:hep-ph/0302002 [hep-ph]} \BibitemShut {NoStop}%
\bibitem [{\citenamefont {Berdnikov}\ and\ \citenamefont
  {Rajagopal}(2000)}]{Berdnikov:1999ph}%
  \BibitemOpen
  \bibfield  {author} {\bibinfo {author} {\bibfnamefont {B.}~\bibnamefont
  {Berdnikov}}\ and\ \bibinfo {author} {\bibfnamefont {K.}~\bibnamefont
  {Rajagopal}},\ }\href {\doibase 10.1103/PhysRevD.61.105017} {\bibfield
  {journal} {\bibinfo  {journal} {Phys. Rev.}\ }\textbf {\bibinfo {volume}
  {D61}},\ \bibinfo {pages} {105017} (\bibinfo {year} {2000})},\ \Eprint
  {http://arxiv.org/abs/hep-ph/9912274} {arXiv:hep-ph/9912274 [hep-ph]}
  \BibitemShut {NoStop}%
\bibitem [{\citenamefont {Mukherjee}\ \emph {et~al.}(2015)\citenamefont
  {Mukherjee}, \citenamefont {Venugopalan},\ and\ \citenamefont
  {Yin}}]{Mukherjee:2015swa}%
  \BibitemOpen
  \bibfield  {author} {\bibinfo {author} {\bibfnamefont {S.}~\bibnamefont
  {Mukherjee}}, \bibinfo {author} {\bibfnamefont {R.}~\bibnamefont
  {Venugopalan}}, \ and\ \bibinfo {author} {\bibfnamefont {Y.}~\bibnamefont
  {Yin}},\ }\href {\doibase 10.1103/PhysRevC.92.034912} {\bibfield  {journal}
  {\bibinfo  {journal} {Phys. Rev.}\ }\textbf {\bibinfo {volume} {C92}},\
  \bibinfo {pages} {034912} (\bibinfo {year} {2015})},\ \Eprint
  {http://arxiv.org/abs/1506.00645} {arXiv:1506.00645 [hep-ph]} \BibitemShut
  {NoStop}%
\bibitem [{\citenamefont {Mukherjee}\ \emph {et~al.}(2016)\citenamefont
  {Mukherjee}, \citenamefont {Venugopalan},\ and\ \citenamefont
  {Yin}}]{Mukherjee:2016kyu}%
  \BibitemOpen
  \bibfield  {author} {\bibinfo {author} {\bibfnamefont {S.}~\bibnamefont
  {Mukherjee}}, \bibinfo {author} {\bibfnamefont {R.}~\bibnamefont
  {Venugopalan}}, \ and\ \bibinfo {author} {\bibfnamefont {Y.}~\bibnamefont
  {Yin}},\ }\href {\doibase 10.1103/PhysRevLett.117.222301} {\bibfield
  {journal} {\bibinfo  {journal} {Phys. Rev. Lett.}\ }\textbf {\bibinfo
  {volume} {117}},\ \bibinfo {pages} {222301} (\bibinfo {year} {2016})},\
  \Eprint {http://arxiv.org/abs/1605.09341} {arXiv:1605.09341 [hep-ph]}
  \BibitemShut {NoStop}%
\bibitem [{\citenamefont {Guida}\ and\ \citenamefont
  {Zinn-Justin}(1997)}]{Guida:1996ep}%
  \BibitemOpen
  \bibfield  {author} {\bibinfo {author} {\bibfnamefont {R.}~\bibnamefont
  {Guida}}\ and\ \bibinfo {author} {\bibfnamefont {J.}~\bibnamefont
  {Zinn-Justin}},\ }\href {\doibase 10.1016/S0550-3213(96)00704-3} {\bibfield
  {journal} {\bibinfo  {journal} {Nucl. Phys.}\ }\textbf {\bibinfo {volume}
  {B489}},\ \bibinfo {pages} {626} (\bibinfo {year} {1997})},\ \Eprint
  {http://arxiv.org/abs/hep-th/9610223} {arXiv:hep-th/9610223 [hep-th]}
  \BibitemShut {NoStop}%
\bibitem [{\citenamefont {Schofield}\ \emph {et~al.}(1969)\citenamefont
  {Schofield}, \citenamefont {Litster},\ and\ \citenamefont
  {Ho}}]{Schofield:1969zz}%
  \BibitemOpen
  \bibfield  {author} {\bibinfo {author} {\bibfnamefont {P.}~\bibnamefont
  {Schofield}}, \bibinfo {author} {\bibfnamefont {J.~D.}\ \bibnamefont
  {Litster}}, \ and\ \bibinfo {author} {\bibfnamefont {J.~T.}\ \bibnamefont
  {Ho}},\ }\href {\doibase 10.1103/PhysRevLett.23.1098} {\bibfield  {journal}
  {\bibinfo  {journal} {Phys. Rev. Lett.}\ }\textbf {\bibinfo {volume} {23}},\
  \bibinfo {pages} {1098} (\bibinfo {year} {1969})}\BibitemShut {NoStop}%
\bibitem [{\citenamefont {Rehr}\ and\ \citenamefont
  {Mermin}(1973)}]{Rehr:1973zz}%
  \BibitemOpen
  \bibfield  {author} {\bibinfo {author} {\bibfnamefont {J.~J.}\ \bibnamefont
  {Rehr}}\ and\ \bibinfo {author} {\bibfnamefont {N.~D.}\ \bibnamefont
  {Mermin}},\ }\href {\doibase 10.1103/PhysRevA.8.472} {\bibfield  {journal}
  {\bibinfo  {journal} {Phys. Rev.}\ }\textbf {\bibinfo {volume} {A8}},\
  \bibinfo {pages} {472} (\bibinfo {year} {1973})}\BibitemShut {NoStop}%
\bibitem [{\citenamefont {Aoki}\ \emph {et~al.}(2009)\citenamefont {Aoki},
  \citenamefont {Borsanyi}, \citenamefont {Durr}, \citenamefont {Fodor},
  \citenamefont {Katz}, \citenamefont {Krieg},\ and\ \citenamefont
  {Szabo}}]{Aoki:2009sc}%
  \BibitemOpen
  \bibfield  {author} {\bibinfo {author} {\bibfnamefont {Y.}~\bibnamefont
  {Aoki}}, \bibinfo {author} {\bibfnamefont {S.}~\bibnamefont {Borsanyi}},
  \bibinfo {author} {\bibfnamefont {S.}~\bibnamefont {Durr}}, \bibinfo {author}
  {\bibfnamefont {Z.}~\bibnamefont {Fodor}}, \bibinfo {author} {\bibfnamefont
  {S.~D.}\ \bibnamefont {Katz}}, \bibinfo {author} {\bibfnamefont
  {S.}~\bibnamefont {Krieg}}, \ and\ \bibinfo {author} {\bibfnamefont {K.~K.}\
  \bibnamefont {Szabo}},\ }\href {\doibase 10.1088/1126-6708/2009/06/088}
  {\bibfield  {journal} {\bibinfo  {journal} {JHEP}\ }\textbf {\bibinfo
  {volume} {06}},\ \bibinfo {pages} {088} (\bibinfo {year} {2009})},\ \Eprint
  {http://arxiv.org/abs/0903.4155} {arXiv:0903.4155 [hep-lat]} \BibitemShut
  {NoStop}%
\bibitem [{\citenamefont {Borsanyi}\ \emph
  {et~al.}(2010{\natexlab{b}})\citenamefont {Borsanyi}, \citenamefont {Fodor},
  \citenamefont {Hoelbling}, \citenamefont {Katz}, \citenamefont {Krieg},
  \citenamefont {Ratti},\ and\ \citenamefont {Szabo}}]{Borsanyi:2010bp}%
  \BibitemOpen
  \bibfield  {author} {\bibinfo {author} {\bibfnamefont {S.}~\bibnamefont
  {Borsanyi}}, \bibinfo {author} {\bibfnamefont {Z.}~\bibnamefont {Fodor}},
  \bibinfo {author} {\bibfnamefont {C.}~\bibnamefont {Hoelbling}}, \bibinfo
  {author} {\bibfnamefont {S.~D.}\ \bibnamefont {Katz}}, \bibinfo {author}
  {\bibfnamefont {S.}~\bibnamefont {Krieg}}, \bibinfo {author} {\bibfnamefont
  {C.}~\bibnamefont {Ratti}}, \ and\ \bibinfo {author} {\bibfnamefont {K.~K.}\
  \bibnamefont {Szabo}} (\bibinfo {collaboration} {Wuppertal-Budapest}),\
  }\href {\doibase 10.1007/JHEP09(2010)073} {\bibfield  {journal} {\bibinfo
  {journal} {JHEP}\ }\textbf {\bibinfo {volume} {09}},\ \bibinfo {pages} {073}
  (\bibinfo {year} {2010}{\natexlab{b}})},\ \Eprint
  {http://arxiv.org/abs/1005.3508} {arXiv:1005.3508 [hep-lat]} \BibitemShut
  {NoStop}%
\bibitem [{\citenamefont {Bhattacharya}\ \emph {et~al.}(2014)\citenamefont
  {Bhattacharya} \emph {et~al.}}]{Bhattacharya:2014ara}%
  \BibitemOpen
  \bibfield  {author} {\bibinfo {author} {\bibfnamefont {T.}~\bibnamefont
  {Bhattacharya}} \emph {et~al.},\ }\href {\doibase
  10.1103/PhysRevLett.113.082001} {\bibfield  {journal} {\bibinfo  {journal}
  {Phys. Rev. Lett.}\ }\textbf {\bibinfo {volume} {113}},\ \bibinfo {pages}
  {082001} (\bibinfo {year} {2014})},\ \Eprint {http://arxiv.org/abs/1402.5175}
  {arXiv:1402.5175 [hep-lat]} \BibitemShut {NoStop}%
\bibitem [{\citenamefont {Bazavov}\ \emph {et~al.}(2012)\citenamefont {Bazavov}
  \emph {et~al.}}]{Bazavov:2011nk}%
  \BibitemOpen
  \bibfield  {author} {\bibinfo {author} {\bibfnamefont {A.}~\bibnamefont
  {Bazavov}} \emph {et~al.},\ }\href {\doibase 10.1103/PhysRevD.85.054503}
  {\bibfield  {journal} {\bibinfo  {journal} {Phys. Rev.}\ }\textbf {\bibinfo
  {volume} {D85}},\ \bibinfo {pages} {054503} (\bibinfo {year} {2012})},\
  \Eprint {http://arxiv.org/abs/1111.1710} {arXiv:1111.1710 [hep-lat]}
  \BibitemShut {NoStop}%
\bibitem [{\citenamefont {Cea}\ \emph {et~al.}(2016)\citenamefont {Cea},
  \citenamefont {Cosmai},\ and\ \citenamefont {Papa}}]{Cea:2015cya}%
  \BibitemOpen
  \bibfield  {author} {\bibinfo {author} {\bibfnamefont {P.}~\bibnamefont
  {Cea}}, \bibinfo {author} {\bibfnamefont {L.}~\bibnamefont {Cosmai}}, \ and\
  \bibinfo {author} {\bibfnamefont {A.}~\bibnamefont {Papa}},\ }\href {\doibase
  10.1103/PhysRevD.93.014507} {\bibfield  {journal} {\bibinfo  {journal} {Phys.
  Rev.}\ }\textbf {\bibinfo {volume} {D93}},\ \bibinfo {pages} {014507}
  (\bibinfo {year} {2016})},\ \Eprint {http://arxiv.org/abs/1508.07599}
  {arXiv:1508.07599 [hep-lat]} \BibitemShut {NoStop}%
\bibitem [{\citenamefont {Bonati}\ \emph {et~al.}(2015)\citenamefont {Bonati},
  \citenamefont {D'Elia}, \citenamefont {Mariti}, \citenamefont {Mesiti},
  \citenamefont {Negro},\ and\ \citenamefont {Sanfilippo}}]{Bonati:2015bha}%
  \BibitemOpen
  \bibfield  {author} {\bibinfo {author} {\bibfnamefont {C.}~\bibnamefont
  {Bonati}}, \bibinfo {author} {\bibfnamefont {M.}~\bibnamefont {D'Elia}},
  \bibinfo {author} {\bibfnamefont {M.}~\bibnamefont {Mariti}}, \bibinfo
  {author} {\bibfnamefont {M.}~\bibnamefont {Mesiti}}, \bibinfo {author}
  {\bibfnamefont {F.}~\bibnamefont {Negro}}, \ and\ \bibinfo {author}
  {\bibfnamefont {F.}~\bibnamefont {Sanfilippo}},\ }\href {\doibase
  10.1103/PhysRevD.92.054503} {\bibfield  {journal} {\bibinfo  {journal} {Phys.
  Rev.}\ }\textbf {\bibinfo {volume} {D92}},\ \bibinfo {pages} {054503}
  (\bibinfo {year} {2015})},\ \Eprint {http://arxiv.org/abs/1507.03571}
  {arXiv:1507.03571 [hep-lat]} \BibitemShut {NoStop}%
\bibitem [{\citenamefont {Bellwied}\ \emph
  {et~al.}(2015{\natexlab{b}})\citenamefont {Bellwied}, \citenamefont
  {Borsanyi}, \citenamefont {Fodor}, \citenamefont {Katz}, \citenamefont
  {Pasztor}, \citenamefont {Ratti},\ and\ \citenamefont
  {Szabo}}]{Bellwied:2015lba}%
  \BibitemOpen
  \bibfield  {author} {\bibinfo {author} {\bibfnamefont {R.}~\bibnamefont
  {Bellwied}}, \bibinfo {author} {\bibfnamefont {S.}~\bibnamefont {Borsanyi}},
  \bibinfo {author} {\bibfnamefont {Z.}~\bibnamefont {Fodor}}, \bibinfo
  {author} {\bibfnamefont {S.~D.}\ \bibnamefont {Katz}}, \bibinfo {author}
  {\bibfnamefont {A.}~\bibnamefont {Pasztor}}, \bibinfo {author} {\bibfnamefont
  {C.}~\bibnamefont {Ratti}}, \ and\ \bibinfo {author} {\bibfnamefont {K.~K.}\
  \bibnamefont {Szabo}},\ }\href {\doibase 10.1103/PhysRevD.92.114505}
  {\bibfield  {journal} {\bibinfo  {journal} {Phys. Rev.}\ }\textbf {\bibinfo
  {volume} {D92}},\ \bibinfo {pages} {114505} (\bibinfo {year}
  {2015}{\natexlab{b}})},\ \Eprint {http://arxiv.org/abs/1507.04627}
  {arXiv:1507.04627 [hep-lat]} \BibitemShut {NoStop}%
\bibitem [{\citenamefont {Borsanyi}\ \emph
  {et~al.}(2012{\natexlab{b}})\citenamefont {Borsanyi}, \citenamefont {Fodor},
  \citenamefont {Katz}, \citenamefont {Krieg}, \citenamefont {Ratti},\ and\
  \citenamefont {Szabo}}]{Borsanyi:2011sw}%
  \BibitemOpen
  \bibfield  {author} {\bibinfo {author} {\bibfnamefont {S.}~\bibnamefont
  {Borsanyi}}, \bibinfo {author} {\bibfnamefont {Z.}~\bibnamefont {Fodor}},
  \bibinfo {author} {\bibfnamefont {S.~D.}\ \bibnamefont {Katz}}, \bibinfo
  {author} {\bibfnamefont {S.}~\bibnamefont {Krieg}}, \bibinfo {author}
  {\bibfnamefont {C.}~\bibnamefont {Ratti}}, \ and\ \bibinfo {author}
  {\bibfnamefont {K.}~\bibnamefont {Szabo}},\ }\href {\doibase
  10.1007/JHEP01(2012)138} {\bibfield  {journal} {\bibinfo  {journal} {JHEP}\
  }\textbf {\bibinfo {volume} {01}},\ \bibinfo {pages} {138} (\bibinfo {year}
  {2012}{\natexlab{b}})},\ \Eprint {http://arxiv.org/abs/1112.4416}
  {arXiv:1112.4416 [hep-lat]} \BibitemShut {NoStop}%
\bibitem [{\citenamefont {Patrignani}\ \emph {et~al.}(2016)\citenamefont
  {Patrignani} \emph {et~al.}}]{Patrignani:2016xqp}%
  \BibitemOpen
  \bibfield  {author} {\bibinfo {author} {\bibfnamefont {C.}~\bibnamefont
  {Patrignani}} \emph {et~al.} (\bibinfo {collaboration} {Particle Data
  Group}),\ }\href {\doibase 10.1088/1674-1137/40/10/100001} {\bibfield
  {journal} {\bibinfo  {journal} {Chin. Phys.}\ }\textbf {\bibinfo {volume}
  {C40}},\ \bibinfo {pages} {100001} (\bibinfo {year} {2016})}\BibitemShut
  {NoStop}%
\bibitem [{\citenamefont {Alba}\ \emph {et~al.}(2017)\citenamefont {Alba} \emph
  {et~al.}}]{Alba:2017mqu}%
  \BibitemOpen
  \bibfield  {author} {\bibinfo {author} {\bibfnamefont {P.}~\bibnamefont
  {Alba}} \emph {et~al.},\ }\href {\doibase 10.1103/PhysRevD.96.034517}
  {\bibfield  {journal} {\bibinfo  {journal} {Phys. Rev.}\ }\textbf {\bibinfo
  {volume} {D96}},\ \bibinfo {pages} {034517} (\bibinfo {year} {2017})},\
  \Eprint {http://arxiv.org/abs/1702.01113} {arXiv:1702.01113 [hep-lat]}
  \BibitemShut {NoStop}%
\bibitem [{\citenamefont {Floerchinger}\ and\ \citenamefont
  {Martinez}(2015)}]{Floerchinger:2015efa}%
  \BibitemOpen
  \bibfield  {author} {\bibinfo {author} {\bibfnamefont {S.}~\bibnamefont
  {Floerchinger}}\ and\ \bibinfo {author} {\bibfnamefont {M.}~\bibnamefont
  {Martinez}},\ }\href {\doibase 10.1103/PhysRevC.92.064906} {\bibfield
  {journal} {\bibinfo  {journal} {Phys. Rev.}\ }\textbf {\bibinfo {volume}
  {C92}},\ \bibinfo {pages} {064906} (\bibinfo {year} {2015})},\ \Eprint
  {http://arxiv.org/abs/1507.05569} {arXiv:1507.05569 [nucl-th]} \BibitemShut
  {NoStop}%
\bibitem [{\citenamefont {Stephanov}(2009)}]{Stephanov:2008qz}%
  \BibitemOpen
  \bibfield  {author} {\bibinfo {author} {\bibfnamefont {M.~A.}\ \bibnamefont
  {Stephanov}},\ }\href {\doibase 10.1103/PhysRevLett.102.032301} {\bibfield
  {journal} {\bibinfo  {journal} {Phys. Rev. Lett.}\ }\textbf {\bibinfo
  {volume} {102}},\ \bibinfo {pages} {032301} (\bibinfo {year} {2009})},\
  \Eprint {http://arxiv.org/abs/0809.3450} {arXiv:0809.3450 [hep-ph]}
  \BibitemShut {NoStop}%
\bibitem [{\citenamefont {Athanasiou}\ \emph {et~al.}(2010)\citenamefont
  {Athanasiou}, \citenamefont {Rajagopal},\ and\ \citenamefont
  {Stephanov}}]{Athanasiou:2010kw}%
  \BibitemOpen
  \bibfield  {author} {\bibinfo {author} {\bibfnamefont {C.}~\bibnamefont
  {Athanasiou}}, \bibinfo {author} {\bibfnamefont {K.}~\bibnamefont
  {Rajagopal}}, \ and\ \bibinfo {author} {\bibfnamefont {M.}~\bibnamefont
  {Stephanov}},\ }\href {\doibase 10.1103/PhysRevD.82.074008} {\bibfield
  {journal} {\bibinfo  {journal} {Phys. Rev.}\ }\textbf {\bibinfo {volume}
  {D82}},\ \bibinfo {pages} {074008} (\bibinfo {year} {2010})},\ \Eprint
  {http://arxiv.org/abs/1006.4636} {arXiv:1006.4636 [hep-ph]} \BibitemShut
  {NoStop}%
\bibitem [{\citenamefont {Stephanov}(2011)}]{Stephanov:2011pb}%
  \BibitemOpen
  \bibfield  {author} {\bibinfo {author} {\bibfnamefont {M.~A.}\ \bibnamefont
  {Stephanov}},\ }\href {\doibase 10.1103/PhysRevLett.107.052301} {\bibfield
  {journal} {\bibinfo  {journal} {Phys. Rev. Lett.}\ }\textbf {\bibinfo
  {volume} {107}},\ \bibinfo {pages} {052301} (\bibinfo {year} {2011})},\
  \Eprint {http://arxiv.org/abs/1104.1627} {arXiv:1104.1627 [hep-ph]}
  \BibitemShut {NoStop}%
\bibitem [{\citenamefont {Ling}\ and\ \citenamefont
  {Stephanov}(2016)}]{Ling:2015yau}%
  \BibitemOpen
  \bibfield  {author} {\bibinfo {author} {\bibfnamefont {B.}~\bibnamefont
  {Ling}}\ and\ \bibinfo {author} {\bibfnamefont {M.~A.}\ \bibnamefont
  {Stephanov}},\ }\href {\doibase 10.1103/PhysRevC.93.034915} {\bibfield
  {journal} {\bibinfo  {journal} {Phys. Rev.}\ }\textbf {\bibinfo {volume}
  {C93}},\ \bibinfo {pages} {034915} (\bibinfo {year} {2016})},\ \Eprint
  {http://arxiv.org/abs/1512.09125} {arXiv:1512.09125 [nucl-th]} \BibitemShut
  {NoStop}%
\bibitem [{\citenamefont {Brewer}\ \emph {et~al.}(2018)\citenamefont {Brewer},
  \citenamefont {Mukherjee}, \citenamefont {Rajagopal},\ and\ \citenamefont
  {Yin}}]{Brewer:2018abr}%
  \BibitemOpen
  \bibfield  {author} {\bibinfo {author} {\bibfnamefont {J.}~\bibnamefont
  {Brewer}}, \bibinfo {author} {\bibfnamefont {S.}~\bibnamefont {Mukherjee}},
  \bibinfo {author} {\bibfnamefont {K.}~\bibnamefont {Rajagopal}}, \ and\
  \bibinfo {author} {\bibfnamefont {Y.}~\bibnamefont {Yin}},\ }\href@noop {} {\
   (\bibinfo {year} {2018})},\ \Eprint {http://arxiv.org/abs/1804.10215}
  {arXiv:1804.10215 [hep-ph]} \BibitemShut {NoStop}%
\bibitem [{\citenamefont {Adamczyk}\ \emph {et~al.}(2014)\citenamefont
  {Adamczyk} \emph {et~al.}}]{Adamczyk:2013dal}%
  \BibitemOpen
  \bibfield  {author} {\bibinfo {author} {\bibfnamefont {L.}~\bibnamefont
  {Adamczyk}} \emph {et~al.} (\bibinfo {collaboration} {STAR}),\ }\href
  {\doibase 10.1103/PhysRevLett.112.032302} {\bibfield  {journal} {\bibinfo
  {journal} {Phys. Rev. Lett.}\ }\textbf {\bibinfo {volume} {112}},\ \bibinfo
  {pages} {032302} (\bibinfo {year} {2014})},\ \Eprint
  {http://arxiv.org/abs/1309.5681} {arXiv:1309.5681 [nucl-ex]} \BibitemShut
  {NoStop}%
\bibitem [{\citenamefont {Luo}(2015)}]{Luo:2015ewa}%
  \BibitemOpen
  \bibfield  {author} {\bibinfo {author} {\bibfnamefont {X.}~\bibnamefont
  {Luo}} (\bibinfo {collaboration} {STAR}),\ }\bibfield  {booktitle} {\emph
  {\bibinfo {booktitle} {{Proceedings, 9th International Workshop on Critical
  Point and Onset of Deconfinement (CPOD 2014): Bielefeld, Germany, November
  17-21, 2014}}},\ }\href@noop {} {\bibfield  {journal} {\bibinfo  {journal}
  {PoS}\ }\textbf {\bibinfo {volume} {CPOD2014}},\ \bibinfo {pages} {019}
  (\bibinfo {year} {2015})},\ \Eprint {http://arxiv.org/abs/1503.02558}
  {arXiv:1503.02558 [nucl-ex]} \BibitemShut {NoStop}%
\bibitem [{\citenamefont {Luo}(2016)}]{Luo:2015doi}%
  \BibitemOpen
  \bibfield  {author} {\bibinfo {author} {\bibfnamefont {X.}~\bibnamefont
  {Luo}},\ }\bibfield  {booktitle} {\emph {\bibinfo {booktitle} {{Proceedings,
  25th International Conference on Ultra-Relativistic Nucleus-Nucleus
  Collisions (Quark Matter 2015): Kobe, Japan, September 27-October 3,
  2015}}},\ }\href {\doibase 10.1016/j.nuclphysa.2016.03.025} {\bibfield
  {journal} {\bibinfo  {journal} {Nucl. Phys.}\ }\textbf {\bibinfo {volume}
  {A956}},\ \bibinfo {pages} {75} (\bibinfo {year} {2016})},\ \Eprint
  {http://arxiv.org/abs/1512.09215} {arXiv:1512.09215 [nucl-ex]} \BibitemShut
  {NoStop}%
\bibitem [{\citenamefont {Akiba}\ \emph {et~al.}(2015)\citenamefont {Akiba}
  \emph {et~al.}}]{Akiba:2015jwa}%
  \BibitemOpen
  \bibfield  {author} {\bibinfo {author} {\bibfnamefont {Y.}~\bibnamefont
  {Akiba}} \emph {et~al.},\ }\href@noop {} {\  (\bibinfo {year} {2015})},\
  \Eprint {http://arxiv.org/abs/1502.02730} {arXiv:1502.02730 [nucl-ex]}
  \BibitemShut {NoStop}%
\bibitem [{\citenamefont {Aprahamian}\ \emph {et~al.}()\citenamefont
  {Aprahamian} \emph {et~al.}}]{Geesaman:2015fha}%
  \BibitemOpen
  \bibfield  {author} {\bibinfo {author} {\bibfnamefont {A.}~\bibnamefont
  {Aprahamian}} \emph {et~al.},\ }\href@noop {} {\enquote {\bibinfo {title}
  {{Reaching for the horizon: The 2015 long range plan for nuclear science}},}\
  }\BibitemShut {NoStop}%
\bibitem [{\citenamefont {Randrup}(2009)}]{Randrup:2009gp}%
  \BibitemOpen
  \bibfield  {author} {\bibinfo {author} {\bibfnamefont {J.}~\bibnamefont
  {Randrup}},\ }\href {\doibase 10.1103/PhysRevC.79.054911} {\bibfield
  {journal} {\bibinfo  {journal} {Phys. Rev.}\ }\textbf {\bibinfo {volume}
  {C79}},\ \bibinfo {pages} {054911} (\bibinfo {year} {2009})},\ \Eprint
  {http://arxiv.org/abs/0903.4736} {arXiv:0903.4736 [nucl-th]} \BibitemShut
  {NoStop}%
\bibitem [{\citenamefont {Randrup}(2010)}]{Randrup:2010ax}%
  \BibitemOpen
  \bibfield  {author} {\bibinfo {author} {\bibfnamefont {J.}~\bibnamefont
  {Randrup}},\ }\href {\doibase 10.1103/PhysRevC.82.034902} {\bibfield
  {journal} {\bibinfo  {journal} {Phys. Rev.}\ }\textbf {\bibinfo {volume}
  {C82}},\ \bibinfo {pages} {034902} (\bibinfo {year} {2010})},\ \Eprint
  {http://arxiv.org/abs/1007.1448} {arXiv:1007.1448 [nucl-th]} \BibitemShut
  {NoStop}%
\bibitem [{\citenamefont {Steinheimer}\ and\ \citenamefont
  {Randrup}(2012)}]{Steinheimer:2012gc}%
  \BibitemOpen
  \bibfield  {author} {\bibinfo {author} {\bibfnamefont {J.}~\bibnamefont
  {Steinheimer}}\ and\ \bibinfo {author} {\bibfnamefont {J.}~\bibnamefont
  {Randrup}},\ }\href {\doibase 10.1103/PhysRevLett.109.212301} {\bibfield
  {journal} {\bibinfo  {journal} {Phys. Rev. Lett.}\ }\textbf {\bibinfo
  {volume} {109}},\ \bibinfo {pages} {212301} (\bibinfo {year} {2012})},\
  \Eprint {http://arxiv.org/abs/1209.2462} {arXiv:1209.2462 [nucl-th]}
  \BibitemShut {NoStop}%
\bibitem [{\citenamefont {Steinheimer}\ and\ \citenamefont
  {Randrup}(2013)}]{Steinheimer:2013gla}%
  \BibitemOpen
  \bibfield  {author} {\bibinfo {author} {\bibfnamefont {J.}~\bibnamefont
  {Steinheimer}}\ and\ \bibinfo {author} {\bibfnamefont {J.}~\bibnamefont
  {Randrup}},\ }\href {\doibase 10.1103/PhysRevC.87.054903} {\bibfield
  {journal} {\bibinfo  {journal} {Phys. Rev.}\ }\textbf {\bibinfo {volume}
  {C87}},\ \bibinfo {pages} {054903} (\bibinfo {year} {2013})},\ \Eprint
  {http://arxiv.org/abs/1302.2956} {arXiv:1302.2956 [nucl-th]} \BibitemShut
  {NoStop}%
\bibitem [{\citenamefont {Steinheimer}\ \emph {et~al.}(2014)\citenamefont
  {Steinheimer}, \citenamefont {Randrup},\ and\ \citenamefont
  {Koch}}]{Steinheimer:2013xxa}%
  \BibitemOpen
  \bibfield  {author} {\bibinfo {author} {\bibfnamefont {J.}~\bibnamefont
  {Steinheimer}}, \bibinfo {author} {\bibfnamefont {J.}~\bibnamefont
  {Randrup}}, \ and\ \bibinfo {author} {\bibfnamefont {V.}~\bibnamefont
  {Koch}},\ }\href {\doibase 10.1103/PhysRevC.89.034901} {\bibfield  {journal}
  {\bibinfo  {journal} {Phys. Rev.}\ }\textbf {\bibinfo {volume} {C89}},\
  \bibinfo {pages} {034901} (\bibinfo {year} {2014})},\ \Eprint
  {http://arxiv.org/abs/1311.0999} {arXiv:1311.0999 [nucl-th]} \BibitemShut
  {NoStop}%
\end{thebibliography}%

\end{document}